\documentclass{aa}
\usepackage{hyperref}
\usepackage{graphicx}
\usepackage{tikz}
\usepackage{txfonts}
\usepackage{xcolor}

\makeatletter
\DeclareRobustCommand{\HI}{%
  \mbox{H\check@mathfonts\fontsize\sf@size\z@\selectfont I}%
}
\makeatother

\makeatletter
\DeclareRobustCommand{\Htwo}{%
  H$_2$}%

\makeatother

\DeclareTextFontCommand{\todo}{\bfseries\color{red}}
\titlerunning{Galaxy edges in the Fornax Cluster}

\begin{document}

   %\title{Dwarf galaxies in the Fornax Cluster: Fifty percent smaller and denser at the edge compared to the field}

   %\title{The impact of environment on size: Fornax Cluster galaxies are fifty percent smaller and denser at the edge compared to the field}

   \title{The impact of environment on size: Galaxies are 50\% smaller in the Fornax Cluster compared to the field}

   %\title{The impact of environment on size: Galaxy edges are fifty percent smaller and denser in the Fornax Cluster compared to the field}
    
   \author{Nushkia Chamba
          \thanks{email: nushkia.chamba@astro.su.se}\textsuperscript{1}
          \and
          Matthew Hayes\textsuperscript{1} 
          \and 
          The LSST Dark Energy Science Collaboration
          }
   \institute{\textsuperscript{1} The Oskar Klein Centre, Department of Astronomy, Stockholm University, AlbaNova, SE-10691 Stockholm, Sweden}

   \date{Received XXX; accepted YYY}

% \abstract{}{}{}{}{} 
% 5 {} token are mandatory
 
  \abstract
    {Size is a fundamental parameter for measuring the growth of galaxies and the role of the environment on their evolution. However, the conventional size definitions used for this purpose are often biased and miss the diffuse, outermost signatures of galaxy growth, including star formation and gas accretion. This issue is addressed by examining low surface brightness truncations or galaxy ``edges'' as a physically motivated tracer of size based on star formation thresholds. Our total sample consists of $\sim900$ galaxies with stellar masses ranging from $10^5 M_{\odot} < M_{\star} < 10^{11} M_{\odot}$. This sample of nearby cluster, group satellite and nearly isolated field galaxies was compiled using multi-band imaging from the Fornax Deep Survey, deep IAC Stripe 82 and Dark Energy Camera Legacy Surveys. We find that the edge radii scale as $R_{\rm edge} \propto M_{\star}^{0.42}$ with a very small intrinsic scatter ($\sim 0.07$\,dex). The scatter is driven by the morphology and environment of galaxies. In both the cluster and field, early-type dwarfs are systematically smaller by $\sim20\%$ than the late-types. However, compared to the field galaxies in the Fornax cluster are the most impacted. At a fixed stellar mass, edges in the cluster can be found at $\sim$ 50\% smaller radii and the average stellar surface density at the edges is a factor of two higher $\sim 1\,M_{\odot}$/pc$^2$. Our findings support the rapid removal of loosely bound neutral hydrogen (\HI{}) in hot, crowded environments which truncates galaxies outside-in earlier, preventing the formation of more extended sizes and lower density edges. Our results highlight the importance of deep imaging surveys to study the low surface brightness imprints of the large scale structure and environment on galaxy evolution.} 
   

   \maketitle
%
%-------------------------------------------------------------------

\section{Introduction}
\label{sect:intro}

Galaxies grow over cosmic time from star formation fed by cold gas accretion, drawn from a reservoir supplied by the surrounding ``intergalactic medium'' and  neighbouring galaxies. The number and frequency of accretion events 
%are set by the host's dark matter halo \citep{2018wechsler} and consequently 
shapes the visible properties of present-day galaxies such as their stellar mass, type (morphology), colour and size. Key insights into the underlying physics of galaxy growth in this scenario are obtained from correlations (``scaling laws'') among these fundamental and physical parameters of galaxies over scales ranging from low mass dwarfs to massive giants. Most of these properties are correlated with the host ``environment'' when defined as a local density using the number of surrounding galaxies over a fixed volume  \citep[see][]{2012muldrew}). For example the well established ``morphology--density'' relation \citep[][]{1980dressler, 2007blanton} shows that redder, elliptical galaxies are more likely found in high density clusters. The environmental impact on size growth, however, has remained a mystery.\par 

\begin{figure}
    \centering
    \includegraphics[width=0.49\textwidth]{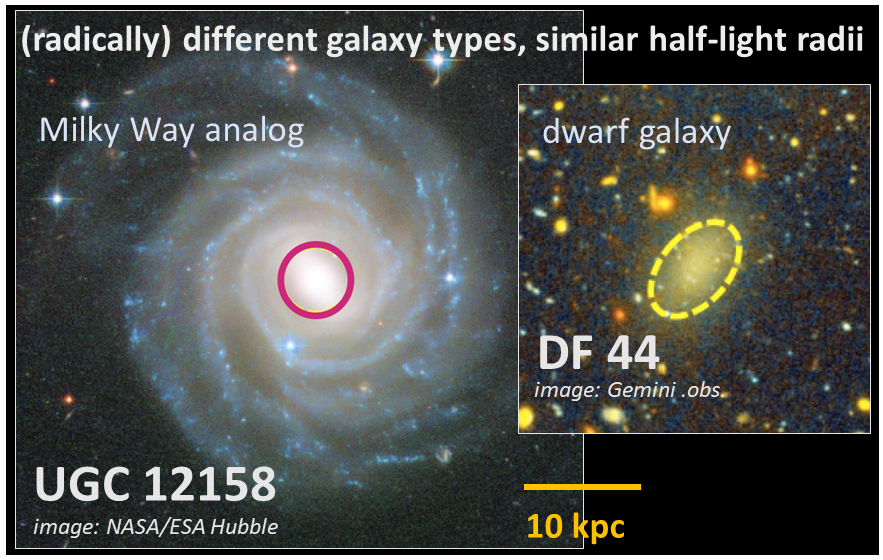}
    \caption{Visualising the problem with using effective (or half-light) radii as a size definition. A massive Milky Way-like galaxy (\textit{left}) and an iconic low surface brightness, low-mass ``dwarf`` galaxy \citep[DF44;][]{2016vanDokkum} (\textit{right}), frequently thought to be as extended as the Milky Way, are shown to the same physical scale using images of similar sensitivity. In spite of the $\sim$~4 orders-of-magnitude mass disparity, both of these galaxies have comparable half-light radius ($r_{\rm e}$; colored contours), partly explaining the large scatter in the $r_{\rm e}-$mass relation and demonstrating that conventional measures of light distribution can be deceptive characterisations of total galaxy extension \citep{2020ctk}. $r_{\rm e}$ may thus not be ideal to trace the environmental impact on size and could explain the lack of consensus in the literature.}
    \label{fig:df44_mw}
\end{figure}

The most common parameter used to examine this issue is the effective or half-light radius. The effective radius is defined at the location which encloses half the total light of a galaxy \citep[$r_{\rm e}$;][]{1948dev}. When plotted against stellar mass ($M_{\star}$), the resulting intrinsic scatter in the $r_{\rm e}-M_{\star}$ plane is large \citep[$\sim 0.3-0.5$\,dex at low redshift;][]{2003shen}, consequently showing little to no distinction for galaxies across varied environments and stellar mass \citep[Fig. \ref{fig:df44_mw}; see also][]{2013huertas, 2017kuchner}. Including the $r_{\rm e}$ of dwarf galaxies with $M_{\star} < 10^9\,M_{\odot}$ which are known to be the most affected by the environment \citep{2017elmegreen} does not significantly change these findings. \par 

These results are puzzling, given that many other physical properties of a galaxy are fundamentally affected by the environment. The large scatter is also problematic for the cosmological predictions of the dark matter halo spin, angular momentum and mass which set the size of a galaxy. The main issues are that the scatter in the $r_{\rm e}-M_{\star}$ plane is difficult to reproduce and the stellar-to-halo mass relation in the dwarf regime from different models diverge. The divergence makes the determination of the minimum halo mass where galaxies can form stars highly model dependant \citep[see][]{2018wechsler}. \par 

Chamba et al. (2020) has shown that the large scatter in the $r_{\rm e}-M_{\star}$ plane can be explained by the fact that $r_{\rm e}$ is insensitive to the total extent of galaxies. This is evident in the comparison between massive galaxies and low mass, diffuse, low surface brightness (LSB) dwarfs (see Fig. \ref{fig:df44_mw} here and their Fig. 5).
\emph{Consequently, various galaxies can have similar $r_{\rm e}$ but fundamentally different masses} \citep[see also the reviews by][]{2019graham, 2020chamba}.  This fact is due to $r_{\rm e}$'s dependence on how concentrated the light is, further complicating the interpretation of the scaling relation \citep[][]{2021donofrio}. 
Decades of research on the use of $r_{\rm e}$ alone to address the problem of the environment on galaxy size have produced ambiguous results \citep[see][]{2022behroozi}, and its scatter issues are unresolved. \par 
Fortunately, $r_{\rm e}$ is not the only radius parameter at our disposal. The development of deep imaging techniques and surveys allow for a more accurate characterisation of the low surface brightness outskirts of galaxies \citep[][see Fig. 11 in \cite{2022euclidcollab} for a comparison between current and future image/survey depth from space and ground based efforts since the Sloan Digital Sky Survey \citep{2000york} era]{1978malin, 1993bland-hawthorn, 2000calcaneo,  2005mihos, 2010martinez, 2011duc, 2014abraham, 2015vegas, 2016fliri, 2016higgs, 2019rich, 2020bilek, 2020annibali, 2021trujillo, 2023ossa-fuentes}. Taking advantage of such data, it is possible to propose a new, physically motivated definition of galaxy size which is more representative of the boundaries of galaxies. Based on theoretical \citep[e.g.][]{2004schaye} and observational evidence \citep[in the ultra-violet, see][]{2019cristina, 2023martinez}, \cite{2020tck} defined the size of a galaxy \emph{as the radial location of a star formation threshold.} This definition is physically motivated because it is directly linked to one of the main growth channels of galaxies, i.e. in situ star formation and the existence of a gas (or stellar) density threshold below which stars are unable to form in a galaxy. \par

The currently used tracer or \emph{proxy} for the \cite{2020tck} size definition which is suitable for general purpose, multi-band optical imaging surveys including the Sloan Digital Sky Survey \citep[SDSS;][]{2000york}, is the truncation radii. The truncation is the radius corresponding to an outermost change in slope or cut-off in surface brightness or surface stellar mass density radial profile in a galaxy, discovered by \cite{1979vanderkruit}. For operative reasons, \citet{2020tck} firstly used a fixed surface stellar density of 1\,$M_{\odot}/$pc$^2$ to measure galaxy size, motivated by the typical density of truncations in Milky Way-like galaxies \citep{2019cristina}. The physical properties of the truncation feature such as its colour and stellar surface density, however, depends on the total stellar mass and morphology of galaxies \citep{2022chamba}. This finding is a reflection of the varied evolutionary histories of galaxies and led \cite{2022chamba} to define the ``edges'' of galaxies more generally as \emph{the outermost radial location where a significant drop in (either past or ongoing) in situ star formation is found in a galaxy due to the existence of a star formation threshold}.  This edge definition can be used as a physically motivated size measure similar to \cite{2020tck}. The impact of the environment on this edge size measure is, however, still unknown and is the main scientific interest of this paper. In Sect. \ref{sect:field-edges}, we shall discuss in more detail the properties of the edges found in \cite{2022chamba}. \par 

In contrast to these new size definitions, it is interesting to point out that $r_e$ was historically fixed to enclose half the total light and not another fraction by \citet{1948dev} mainly to operatively parameterise the light profiles of elliptical galaxies \citep[see also][]{2019graham}. There is no evidence in the archives which suggests that the definition itself is linked to the physics of how galaxies grow in size. The choice of $r_e$ was probably motivated due to the difficulty of measuring galaxy extensions in the past. \par 
In fact, similar arguments can be made about size measures based on fixed surface brightness levels \citep{1936redman, 1958holmberg}. Those levels were pragmatically chosen to be near the typical depth of the (shallower) photographic plates used. With this method, as much as possible of the total galaxy light detected was enclosed, however, only on a per image or observation basis \citep[see the historical review in][]{2020chamba}. \par 
Although pragmatically justified in the past, these historical choices or measurements are arbitrary when considering that current images are far more homogeneous and deeper, clearly showing us that the past measures may in fact be biased against the galaxy outskirts we are capable of detecting now. Our goal is thus to make use of the wide availability of deep images and measure a galaxy size which is based on one of their main formation channels and independent of a prior, arbitrarily selected light fraction or surface brightness level.\par 

Using this physical approach, the \cite{2020tck} and \cite{2022chamba} measurements are a better representation of the total extensions or boundaries of galaxies in a wide stellar mass range compared to $r_{\rm e}$, overcoming the issues visualised in Fig. \ref{fig:df44_mw}. Remarkably, when the measure is used to define the size--stellar mass relation, the intrinsic scatter is very small, $\sim$ 0.06\,dex which is 1/3 of that resulting from $r_{\rm e}$ or other size metrics. These improvements compared to previous size measures make the edge a promising approach to address the environmental problem on galaxy growth. %While the traditional measures of galaxy size have been defined in the optical regime \citep{1936redman, 1948dev, 1958holmberg}, we remark that a physical galaxy size measure need not necessarily be restricted to any specific wavelength. One may certainly improve upon or propose a much better tracer for the \cite{2020tck} and \cite{2022chamba} size definition. \par 

%Ideal datasets for such an endeavour include all-sky, deep, high resolution imaging in the ultra-violet (UV) or H$_{\alpha}$ \citep[e.g.][]{2006koopmann, 2011hunter, 2016herrmann, 2019lei, 2022rampazzo} as well as neutral atomic  (\HI{}) in combination with molecular hydrogen (\Htwo{}) \citep[e.g.][]{2007vanderkruit, 2011toribio, 2016bluedisk, 2022morokuma, 2023watkins}. The UV or H$_{\alpha}$ are more sensitive traces of ongoing or recent star formation in galaxies while \HI{} and \Htwo{} indicate the availability of gas to form stars. These observations may be used for comparison with the optical stellar distribution in galaxies.  Unfortunately, however, such datasets are currently not freely available for all or large numbers of galaxies across varied environments and stellar mass. We are thus currently limited to general purpose optical surveys to measure the sizes needed for any statistical analysis on galaxy size growth and its dependence on the environment. 

In this paper, we use the deep, multi-band optical imaging from the Fornax Deep Survey \citep{2017venhola, 2019iodice} and DESI Legacy Imaging Surveys \citep{2019dey} to measure the sizes of galaxies in three different environments: the Fornax galaxy cluster (a high density environment), confirmed satellite galaxies in nearby $< 40$\,Mpc Milky Way analogues as well as those in the field, i.e. galaxies outside clusters or comparatively in lower density environments and located at distances $< 100$\,Mpc. \par 
While we particularly focus on dwarf galaxies because they are the most impacted by the environment \citep[e.g.][]{2017elmegreen}, the more massive galaxies in Fornax and our satellite galaxy samples are also considered for completeness. Instead of using $r_{\rm e}$, we compare their edge properties and resulting edge radii--mass relations to address the impact of the environment on size. These findings are compared with those traditionally resulting from the use of the effective radius. \par  

The paper is organised as follows.
Sect. \ref{sect:field-edges} summarises the properties of edges and the associated size-mass relation outside clusters (or in field environments) particularly for dwarf galaxies. Sect. \ref{sect:data} describes the catalogues and data used for this work for the different samples considered. Sect. \ref{sect:methods} provides the details of the image processing, radial profile creation and edge identification procedure. The results are described in Sect. \ref{sect:results} and finally discussed in Sect. \ref{sect:discussion}. Throughout this work, we assume a flat $\Lambda$-CDM cosmology with $H_{0} = 70$ and $\Omega=0.3$.

\section{Characteristics of edges in the nearby Universe}
\label{sect:field-edges}

\begin{figure*}[h!]
    \centering
    \includegraphics[width=\textwidth]{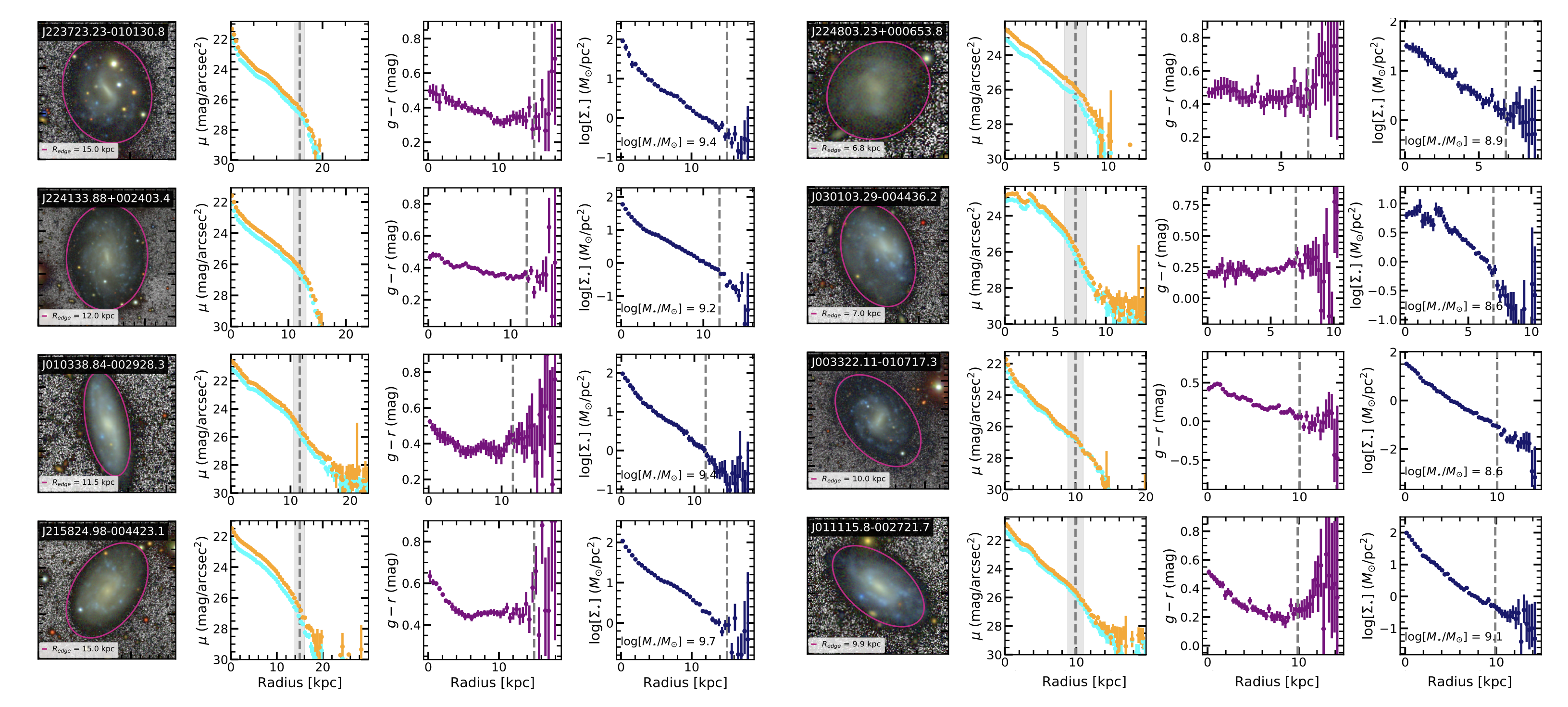}
    \caption{Radial profiles of galaxies with stellar masses $\sim10^9\,M_{\odot}$ from the \citet{2022chamba} field sample. \textit{Left to right}: RGB composite image from IAC Stripe 82 with a grey background to highlight the low surface brightness edge location (pink ellipse). Surface brightness profiles in $r$-band (orange) and $g$-band (cyan), $g-r$ colour and surface stellar mass density profiles. The location of the detected edges are shown as vertical dashed lines and the shaded grey areas are the typical uncertainties associated to repeated measurements by \cite{2022chamba} ($\sim 0.04$\,dex).} %\todo{TO DO: Plot the location of the effective radii.}}
    
    \label{fig:sdss-examples}
\end{figure*}

As mentioned in the Introduction (Sect. \ref{sect:intro}), \cite{2022chamba} defined the ``edge'' of a galaxy as \emph{the outermost radial location where a significant drop in (either past or ongoing) in situ star formation is found in a galaxy due to the existence of a star formation threshold}.  In this section, we summarise the currently known characteristics of galaxy edges in the nearby Universe (redshift $z<0.09$) studied in \citet{2022chamba} which we later use as a reference sample in our analysis. \par 

%Throughout this paper, we define a ``field'' sample as that which is not selected by environment but either by the stellar mass or completeness limits of the surveys used. \par

%\textbf{Field sample:} \citet{2022chamba} used a sample of 624 field galaxies from publicly available catalogues \citep{2010nair, 2013maraston}, selected by imposing the Stripe 82 footprint within the SDSS. These galaxies were studied due to the availability of deep Stripe 82 imaging \citep[two magnitudes deeper than SDSS;][]{2018s82rectified} required to develop the criteria for locating the low surface brightness edge in a wide variety of galaxies. The sample consisted of early-type (elliptical), late-type (spiral) and dwarf galaxies, collectively with stellar masses between $10^7-10^{12}$\,$M_{\odot}$. \par 

\textbf{Criteria to identify the edge:} As a proxy for the location of the edge, \citet{2022chamba} used the location of the truncation or the outermost change in slope in the radial surface brightness, colour and stellar mass surface density profiles of the field galaxies. This feature has previously been connected to star formation thresholds where in situ star formation is cut-off \citep[e.g.][]{2004schaye, 2019cristina, 2022simon} and is therefore currently the strongest signature of the edge detectable in optical imaging. We use the same signature in this work. \par 

We visualise the locations of galaxy edges for a number of example field galaxies with stellar masses $\sim10^9\,M_{\odot}$ from the \citet{2022chamba} sample (the median mass of the range considered in that work) in Fig. \ref{fig:sdss-examples}. These galaxies are categorised as ``dwarfs'' in stellar mass. The location of the detected edges are shown as vertical dashed lines in each profile panel. The surface brightness of the edge typically corresponds to $\sim 26.5-27\,$ mag/arcsec$^2$ in the SDSS $g$-band (see their Appendix B which shows a detailed analysis on the effect of image depth on edge detection). \par 
The shaded grey areas in the panels are the typical uncertainties associated to repeated identifications ($\sim 0.04$\,dex). The repeated identifications were performed by the authors using the same data and profiles. The quoted value is the typical deviation between the repeatedly located edge radii. The technical details on how radial profiles are derived and corrected for inclination is described under the Methodology (Sect. \ref{sect:methods}).\par 
Notice how the colour profile beyond the located edges can either turn redder or bluer. As discussed in \cite{2022chamba}, for dwarf galaxies the different colour profile shapes are indicative of either inside-out or outside-in star formation \citep[see also the work by][]{2016herrmann}. The \cite{2022chamba} criteria for more massive galaxies, however, is slightly different, given that these galaxies have significantly varied evolutionary pathways. \par 
Apart from the change in slope, late-type or spiral galaxies often have a ``$U$''--shape colour profile \citep{2008bakos, 2008azzollini}. \cite{2022chamba} has shown that the reddening of the colour profile signifies both a drop in in-situ star formation when traced using the truncation in ultraviolet observations as well as the stellar halo component or migrated stars from the inner disc (see their Fig. 2). \par 

In the case of early-type or elliptical galaxies, the most clearest criteria for the edge reported in \cite{2022chamba} is in the colour profile where there is a very rapid drop towards bluer colours (slopes compatible with $\sim -5$ in some extreme cases). These bluer colours potentially represent a more recently accreted stellar component which makes up the outer envelope in these galaxies. Depending on the morphology in the total sample considered here, the above features were collectively used when identifying the edges.

\textbf{Edge size-mass properties:} The edges of the 180 dwarfs from the \cite{2022chamba} sample have an average stellar surface density of 0.6$\,M_{\odot}$/pc$^2$ and their sizes follow the relation $R_{\rm edge} \sim M_{\star}^{1/3}$ with an observed scatter of $\sim 0.09\,$dex. The average density of edges in the full sample including the more massive galaxies is $\sim 1\,M_{\odot}$/pc$^2$, similar to the value found for nearby Milky Way-like galaxies \citep{2019cristina}. The elliptical galaxies have even higher edge  densities $> 3\,M_{\odot}$/pc$^2$. The lower edge density in dwarfs compared to the massive galaxies is consistent with the fact that dwarfs are less efficient in forming stars \citep{2012huang} and massive ellipticals are known to be highly efficient in their star formation at earlier times \cite[e.g.][]{2015jaskot}. \par 
Additionally, late-type bluer, younger galaxies were found to be larger in size at a fixed stellar mass compared to redder, older ones. These bluer galaxies are in the range between Sb-Scd morphological types while the redder ones are S0/Sa types \citep[see][]{2020tck}. This result was not definitive for galaxies with $M_{\star} \leq 10^9\,M_{\odot}$, given that the morphological types of these lower mass galaxies studied in \citet{2022chamba} are not available. Nevertheless, this colour (age) bifurcation in the size-mass relation potentially indicates that the older, redder galaxies reached their present-day size at an earlier epoch compared to younger, bluer ones. In this work, we will analyse the environmental influence on the edge locations, densities, galaxy colours and morphology for comparison with the above results from \cite{2022chamba}.\par 

\section{Data and Sample selection}
\label{sect:data}
%%%Sample overview, of each sample, depth of the data. Put it in a table. 

We select a volume limited sample of dwarf galaxies in order to probe a wide range in stellar mass and environmental conditions. We classify galaxies as `cluster' using their membership in the Fornax Cluster, as `satellite' if the galaxies are in a low density group environment but confirmed to be associated to a nearby host  and as `nearly isolated' for all other galaxies in groups or in isolated environments. Throughout this work, we use `field' as a reference to galaxies outside a cluster environment. In our study, the field includes our satellite and nearly isolated dwarf samples.  We describe the catalogues and data used in Table \ref{tab:sample} and the following subsections.

\begin{table*}[]
\begin{center}
\caption{Data and Sample Overview}

\begin{tabular}{c c c c c c}
%\hline
\textbf{Sample} & \textbf{No. of galaxies} &  \textbf{Stellar mass}  & \textbf{Distance} & \textbf{Imaging data/survey}  & \textbf{Depth} ($\mu_r$; \\ %& \textbf{Smallest}\\ 
\textbf{\& environment} & \textbf{selected (out of)} & \textbf{range ($M_{\odot}$)} & \textbf{range (Mpc)} & \textbf{and quality ($g, r$-band)}  &   \textbf{mag/arcsec$^2$)} \\ % & \textbf{scale (pc)} \\ %over scales of 10$\times$10,\ arcsec$^2$  \\ 
\hline \\
\centering 
Cluster & 415 (582) & $10^5-10^{11}$ & 20 & VST/Fornax Deep Survey (1) & 29.7 \\ %\hline
Group \& satellite & 235 (306) & $10^{5.5}-10^{10}$ & $4-50$ & DECaLs/SAGA (2) \& ELVES (3) & 28.4  \\ %\hline
LV/Nearly isolated  & 64 (96) & $10^{5.5}-10^{9}$  & $2-10$ & DECaLs/Updated Nearby Galaxies (4) &  28.4  \\ %\hline
C22/Nearly isolated  & 180 (624) & $10^{7}-10^{10}$ & $50-77$ & IAC Stripe 82/\cite{2022chamba} (5) & 28.5  \\ \\
\textit{Total Sample} & 894 (1608) & $10^5-10^{11}$ & $2-77$ & All surveys have FWHM$_{PSF}\sim 1\,''$ & $28-30$  \\ \\ \hline 
\end{tabular}
\label{tab:sample}
\end{center}
\vspace{-2.5mm}
\textbf{Notes:} In the selection of these samples, we require three conditions. First, the galaxy's axis ratio is required to be $<0.3$ to limit the effect of the galaxy's inclination on our mass estimates (see Sect. \ref{sect:methods}). Second, the galaxies should be entirely within the footprints of the surveys/data used. Third, we require the absence of overlapping or nearby bright stars around the galaxy of interest. The $3\sigma$ depths of the data are estimated over 10$\times$10\,arcsec$^2$ areas. The uncertainty in these values are $\sim 1$\,\%. All data used are well within the limit needed to detect edges, typically found at $\mu_g \sim 26-27$\,mag/arcsec$^2$. We also point out that the imaging quality, defined as the FWHM of the measured point-spread function (PSF), of the different surveys used are very similar $\sim 1\,''$. At the median distance of the total sample studied here $\sim34$\,Mpc, the smallest structures resolvable with our compiled imaging dataset is $\sim 165$\,pc. Therefore, the compiled dataset is suitable for this study based on both depth and spatial resolution. (1) \cite{2018venhola, 2021su}, (2) \cite{2017geha, 2021yao}, (3) \cite{2021carlsten, 2022carlsten}, (4) \cite{2013karachentsev} and (5) compiled from \cite{2016fliri, 2018s82rectified, 2010nair, 2013maraston}. 
\end{table*}

\subsection{Cluster environment}

\begin{figure}[h]
    \centering
    \includegraphics[width=0.49\textwidth]{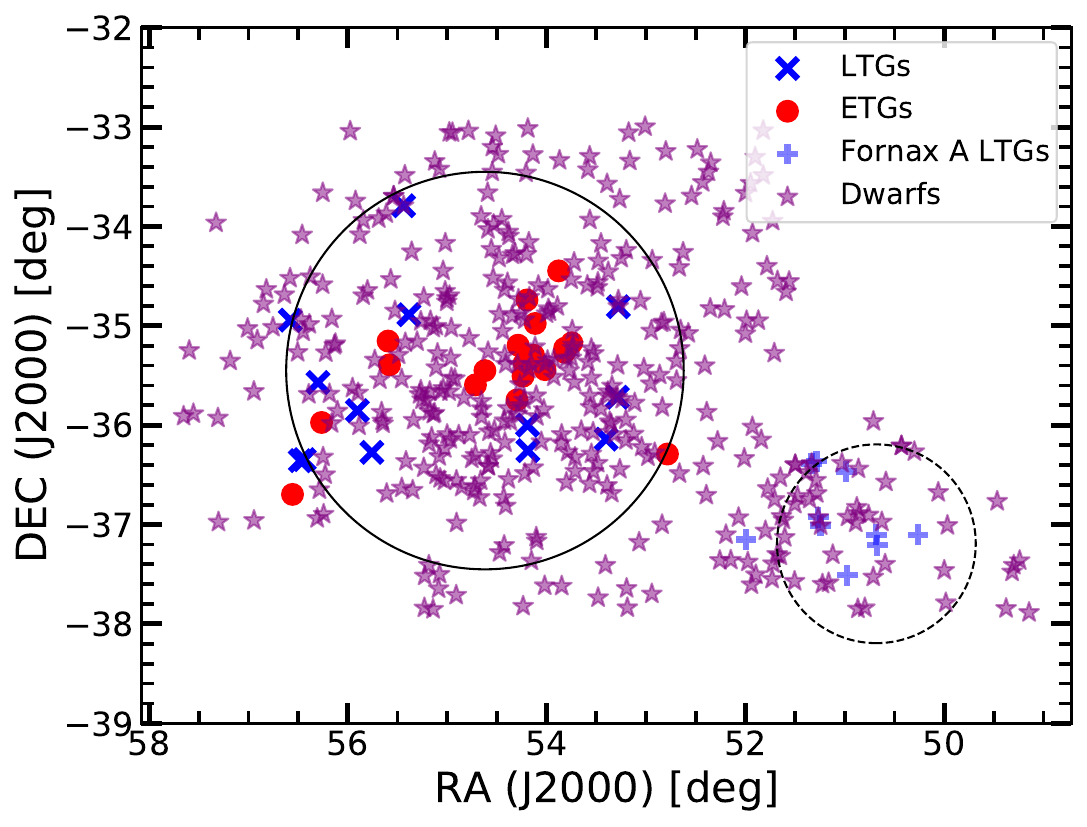}
    \caption{Schematic of the Fornax Cluster and Fornax A sub-group from the Fornax Deep Survey. The positions of dwarfs (purple stars), late-type and early-type galaxies (blue crosses and red dots, respectively) catalogued by \cite{2018venhola} are plotted. The virial radius of the cluster (black circle with radius $\sim 2$\,deg) and Fornax A sub-group (black, dashed circle with radius $\sim 1$\,deg). The late-types in the subgroup are plotted as blue pluses for distinction.}
    \label{fig:FDS-footprint}
\end{figure}

Fornax is the second most massive cluster within 20 Mpc and has a virialized mass of $7\times10^{13}\,M_{\odot}$ \citep{2001drinkwatera}. The cluster has been well-studied using multi-band surveys from the ultra-violet to radio wavelengths \citep[e.g.][]{2005martin, 2013davies, 2018eigenthaler, 2018venhola, 2019zabel, 2021loni, 2023fornax_meerkat}. The most up-to-date catalogues of the cluster include 795 low-mass galaxies with $M_{\star} \lesssim 10^9\,M_{\odot}$. These galaxies include 265 low surface brightness dwarfs \citep{2018venhola, 2021su, 2022venhola}. An additional 42 more luminous early- and late-type galaxies are found in the main cluster and Fornax A sub-group \citep[ETGs and LTGS, respectively;]{2019iodice, 2019raj, 2020raj}, spanning a total stellar mass range between $10^5-10^{11}$\,$M_{\odot}$. The locations of all the galaxies from the \cite{2021su} catalogue are shown in Fig. \ref{fig:FDS-footprint}.

We make use of the deep $g$ and $r$-band optical images from the Very Large Telescope (VLT) Survey Telescope as part of the Fornax Deep Survey \citep[FDS;][]{2017venhola, 2019iodice, 2020peletier}. Given our goal to search for edges in a large sample of galaxies and features which are typically of low surface brightness $\sim 25-27\,$mag/arcsec$^2$ in the SDSS $r$-band \citep[see also Fig. B2 in][]{2022chamba}, the combined depth and pixel scale of the FDS imaging ($\sim 30$\,mag/arcsec$^2$ over boxes of 10$\times$10 arcsec$^2$ scales) provides an ideal dataset to characterise edges in a high density environment. We use the 582 Fornax cluster galaxies catalogued in \citet{2021su} in this work. As in previous work \citep[e.g.][]{2019raj, 2020raj}, we assume these galaxies are located at 20\,Mpc. We do not make any distinction between the `cluster' and `Fornax A' sub-group in this work unless specifically mentioned within the text. \par 
%We also point out that the inner break radii of the massive Fornax late-type galaxies in FDS have been previously characterised in \cite{2019raj, 2020raj} but the features identified here are different as they are the outermost edge. The break radii reported in Raj et al. (2019, 2020) are located at $0.3-0.7\,R_{\rm edge}$ and at a compatible location only in one galaxy FCC013. There could be potential physical connections between the break and edge radii and may have a common origin as suggested in \cite{2019cristina}, however, exploring this connection is beyond the scope of the current work and we leave it for future investigations. %However, where available, we provide a comparison in the Discussion.

\subsection{Group and satellite environment}

For comparison with the Fornax galaxies, we also study the Stage II SAGA satellite sample of 127 galaxies around 36 Milky Way analogues out to 40\,Mpc \citep{2017geha, 2021yao}. The sample is spectroscopically confirmed and predominantly star forming with stellar masses ranging from $10^6-10^{10}$\,$M_{\odot}$, the majority of which have a stellar mass of $\sim 10^8\,M_{\odot}$. Public deep photometric imaging catalogues from the Dark Energy Camera Legacy Survey \citep[DECaLs;][]{2019dey} are also available for the sample. Similar to FDS, we use the $g$ and $r$-band images of DECaLs which have a depth of $\sim 28$\,mag/arcsec$^2$ in the $g$-band, sufficient for the detection of edges. \par 

To increase our sample in the lower mass regime $< 10^7\,M_{\odot}$, we combine the SAGA sample with the ELVES satellites belonging to Milky Way mass hosts within the local volume \citep[$<10$\,Mpc;][]{2021carlsten, 2022carlsten}. The stellar mass range of ELVES galaxies are between $10^{5.5}-10^{8.5}\,M_{\odot}$. In our work, we only use a sub-sample of the ELVES satellites with the criteria that their distances are confirmed and public DECaLs imaging are available at their locations. These two criteria leads to a sample of 179 ELVES satellite galaxies. Our total satellite galaxy sample from SAGA and ELVES combined consists of 306 galaxies. 

\subsection{A nearly isolated environment}

%How field is defined? Not selected by environment, only by mass and completeness limit of the surveys. All the rest. Contains the broad galaxy population. 

For a nearly isolated sample, we use the edge measurements made in \cite{2022chamba} for a wide selection of galaxies within a stellar mass range of $10^7-10^{12}\,M_{\odot}$ using the deep IAC Stripe 82 survey \citep{2016fliri, 2018s82rectified}. This sample consists of 624 galaxies taken from a combination of the \cite{2010nair} and  \cite{2013maraston} catalogues. 180 of these galaxies are categorised in that work as dwarfs according to their stellar mass $< 10^{10}\,M_{\odot}$. The morphological classifications of these `dwarfs' are not available. However, the classifications of the late-type (spiral) and early-type (elliptical) galaxies are available from the \cite{2010nair} catalogue. Only galaxies within the foot print of the SDSS Stripe 82 were studied. 

%As already discussed in Sect. \ref{sect:field-edges}, the galaxies in this sample with stellar mass $10^{10}\,M_{\odot}$ can be considered to be nearly isolated. \par 
%$\sim$31\% of our dwarf sample have their centres within 1\,arcsec of the luminosity weighted group centres used in \citet{2007yang}. In other words, they are considered the most massive galaxy in their respective group. The nearest group matched with the rest of the sample a

To gain insight on the surrounding environment of these galaxies, we cross match the \cite{2022chamba} sample with the \citet{2007yang} SDSS group and cluster catalogues\footnote{\protect\url{https://gax.sjtu.edu.cn/data/Group.html}}. Unfortunately, we are able to find definitive group associations only for 79-83 dwarfs out of the 180 studied in \citet{2022chamba} as their central coordinates are at least within 1-2\,arcsec of those in the \citet{2007yang} galaxy catalogue. More recent catalogues on the SDSS galaxies do not include the Stripe 82 footprint \citep[see e.g.][]{2017lim}. However, as this sub-sample covers the full stellar mass range of the \citet{2022chamba} dwarfs, it is still useful to analyse them. We find that $\sim$75\% of the sub-sample in \citet{2007yang} are reported as the most massive and single member of their respective group. In other words, they are isolated. \par 
For the rest of the sample, we find that the median angular separation between the dwarf and a nearest galaxy or group from \citet{2007yang} is 223\,arcsec. Upon further investigation, these galaxies or groups are located at significantly different redshifts, with our sample being nearer at a median $z \sim 0.016\pm0.002$ and the groups at $\sim 0.081\pm0.04$. 84\% of these groups are composed of single member galaxies. According to \citet{2007yang}, such galaxies are defined as isolated. Therefore, these results allow us to conclude that the \citet{2022chamba} dwarfs are either isolated or in low-mass groups not characterised in \citet{2007yang}. In other words, \emph{the \citet{2022chamba} dwarfs are nearly isolated or in the field.}

%$sim 26.5$\,kpc. 

%This value is comparable to the size of a Milky Way-mass galaxy. 

%We find that $\sim$22\% of the galaxies with stellar masses $\geq 10^8\,M_{\odot}$ are considered the most massive galaxy in their respective group/cluster.  %which have stellar masses $\geq 10^{9.5}$\,$M_{\odot}$.
%However, we point out that group/cluster masses were only available in \cite{2007yang} for those galaxies where $M_{\star} > 10^{10}\,M_{\odot}$ in the \citet{2022chamba} sample. The lower mass galaxies in \cite{2007yang} are generally considered as either isolated or group members. 
%\textit{We may thus conclude that the field sample studied in \citet{2022chamba} consists of galaxies in groups and clusters for those above stellar masses 10$^{10}\,M_{\star}$ and either small groups or isolated environments for the lower mass galaxies.}

%should the physical interpreation of this previous work be included in the Intro or here?

We additionally use a lower mass, Local Volume (LV) sample of dwarfs studied in \cite{2021carlsten} which overlaps in stellar mass and distance with ELVES. These galaxies were selected as a sub-sample from the Updated Nearby Galaxy Catalogue \citep{2013karachentsev}, however, we only use galaxies that have DECaLs imaging. This criteria results in a sample of 96 galaxies in addition to the 180 low mass galaxies in the \cite{2022chamba} sub-sample. \par
\begin{figure}
    \centering
    \includegraphics[width=0.49\textwidth]{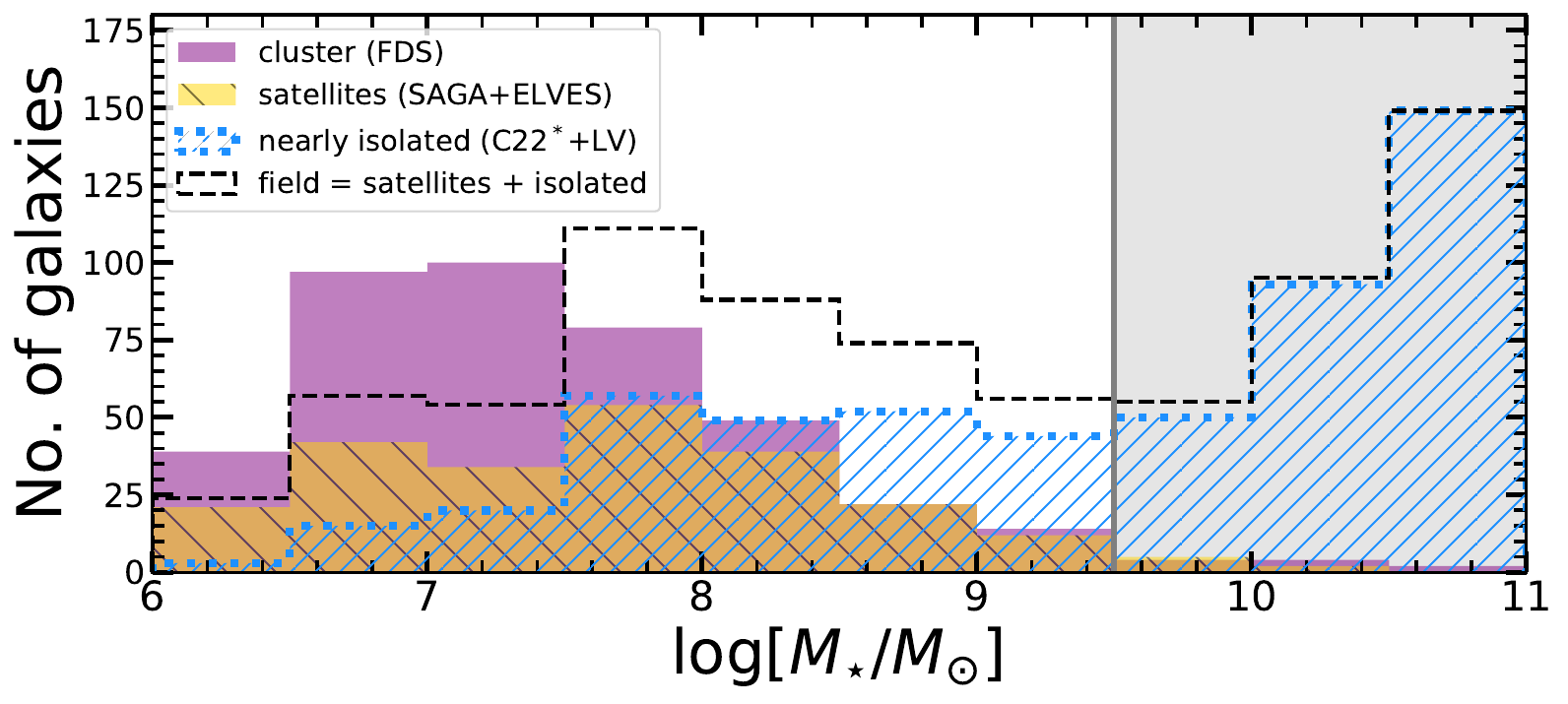}
    \caption{Distribution of the number of galaxies in stellar mass bins of width 0.5\,dex in our final Fornax cluster (purple), satellite (yellow, right hatched) and nearly isolated (blue, dotted and left hatched) samples. The field sample combining the satellites and isolated galaxies are plotted in dashed black lines. C22* in this figure indicates that we have included the \citet{2022chamba} LTG and ETG sample here only to show the lack of cluster galaxies beyond $M_{\star} > 10^{9.5}\,M_{\odot}$ (grey region).}
    \label{fig:mass_dist}
\end{figure}

 
\citet{2013karachentsev} also estimates tidal indices for these galaxies to describe and quantify their local environment. The index is computed by ranking the nearest neighbours `n' according to the strength of their tidal force on the galaxy, estimated simply by $F_{tidal} \sim M_n/D_n^3$ where $M_n$ is the mass of the neighbour and $D_n$ is the distance between the galaxy and the neighbour. Negative values of this index correspond to isolated galaxies. Of the 96 field galaxies we are able to consider here, 78\% have negative tidal indices.\footnote{We do not remove galaxies with positive tidal indices as it is undesirable to lower our sample size in the low mass regime even more (see Fig. \ref{fig:mass_dist}).} Therefore, when combining the \cite{2022chamba} and \cite{2013karachentsev} sub-samples, the majority of these galaxies can be considered nearly isolated.

\par \bigskip 

\noindent \textbf{Total sample:} The total sample of galaxies (Fornax, SAGA, ELVES and field) considered in this work amounts to 1608. Following \cite{2020tck}, after removing galaxies that have an inclination or axis ratio $< 0.3$ to avoid projection effects in the computation of the stellar mass (Sect. \ref{sect:methods}) and those significantly impacted by neighbouring galaxies (overlapping or due to bright stars), the final sample analysed amounts to 894 galaxies, 17 of which are Fornax massive early- or late-type galaxies. \par 
Table \ref{tab:sample} summarises the data and final sample used. Fig. \ref{fig:mass_dist} shows the distribution of the number of galaxies in our cluster, field and satellite samples in stellar mass bins of 0.5\,dex. While our total isolated and satellite samples consist of about half the number of Fornax cluster dwarfs in the lowest stellar mass bins $M_{\star} < 10^{7.5}\,M_{\odot}$ and nearly double at the higher mass bins, we have sufficient galaxies to derive and compare the distributions of these galaxies in the size scaling relations. \par 

\subsection{Characterisation by morphology and distance}
Morphological classification of the dwarf galaxies in this study is only available for the LV, ELVES and FDS sub-samples. These samples are classified as either `early'- or `late'-type galaxies based on visual inspection. In the case of LV and ELVES, \citet{2022carlsten} categorise `dwarfs with smooth, regular, and generally low surface brightness morphology' as early-type satellites. Late-types are those with `clear star forming regions, blue clumps or dust-lanes'. \par 

The FDS galaxies are classified similarly. However, \citet{2018venhola} additionally divide the early-type sample as either `smooth' red galaxies without `clearly distinguishable bars or spiral-arms' or those with structure. The former category includes `giant early-type galaxies with no clear structure, nucleated and non-nucleated dwarf ellipticals'. The latter category also consists of red galaxies but they have sub-structure such as a `bulge, bar or spiral arms...includ[ing] the S0s and dEs with prominent disk features'. \citet{2017venhola}'s late-type category includes `spirals, blue compact dwarfs and dwarf irregular galaxies'. In our analysis, we treat galaxies with or without sub-structure in FDS as belonging to the same category, broadly defined as early- or late-type.

%FDS
%Smooth early-type: Galaxies that have a smooth red ap- pearance, and do not have structures like clearly distinguish- able bars or spiral arms. If a dwarf galaxy has an unresolved point-like nucleus, it will be classified into this group as well. This group therefore includes giant early-type galaxies with no clear structure, and nucleated and non-nucleated dwarf ellipticals.
%– Early-type with structure: Galaxies that are red and have no star-forming clumps, but have structures such as bulge, bar, or spiral arms. They are not well modeled by a single Sérsic function. This group includes S0s and dEs with promi- nent disk features.
%– Late-type: Galaxies that are blue, and have star-forming clumps. This group includes spirals, blue compact dwarfs and dwarf irregular galaxies.

%ELVES
%Dwarfs with smooth, regular, and generally low surface bright- ness morphology are classed as early-types. On the other hand, dwarfs with clear star-forming regions, blue clumps, dust-lanes, or any other complications in their surface brightness profile are classed as late-type

Morphological classification is not available for the SAGA and \citet{2022chamba} samples. However, in the case of SAGA, \citet{2021yao} used a condition on the equivalent width of the H$_{\alpha}$ line such that EW(H$_{\alpha}) < 2 \textup{\AA}$ to tag galaxies as `quenched'. The galaxy is considered `star forming' in other cases. Using the EW(H$_{\alpha}$) to make this distinction is motivated as the H$_{\alpha}$ line reflects the presence of star formation within the last 5\,Myr \citep[e.g.][]{2021flores}. \par 
In the visual comparison between ELVES and SAGA by \citet{2022carlsten}, the above criteria for quenched satellites is shown to be compatible with that of the red ultra-violet (UV)-optical colours of the ELVES early-type classification scheme (see their Fig. 10). However, they identify three exceptional cases in their visual comparison. Relevant to our final sample are two of these galaxies which are visibly blue and irregular but with no detected H$\alpha$ emission: SAGA galaxy 1237661852013166857 (RAdeg=206.99,  DEdeg=43.5872) and DES-203016260 (RAdeg=322.9709, DEdeg=-43.6539). Their stellar masses are log$_{10} M_{\star}$ = 7.9 and 7.7, respectively. \par 

Apart from these two galaxies, SAGA's satellites listed as star forming also have a similar colour to the ELVES late-type galaxies. For this reason, in our analysis, we treat the SAGA quenched sample as `early-type' and star-forming as `late-type'. Classifying the two exceptional galaxies as late-type do not change our results significantly. \par 

For completeness, we make a similar classification criteria for the \citet{2022chamba} sample. We obtain the EW(H$_{\alpha}$) from the computations made by the Portsmouth Group for SDSS/BOSS galaxies \footnote{\protect\url{https://data.sdss.org/datamodel/files/BOSS_GALAXY_REDUX/GALAXY_VERSION/portsmouth_emlinekin_full.html}} produced by \citet{2013thomas}. We impose the same condition as in SAGA to separate quenched and star forming galaxies in this sample. Similar to the SAGA analysis by \citet{2022carlsten}, we treat the quenched sample as early-type and star forming as late-type. \par 
The number of galaxies in our total sample considered as either early or late-type for each environment are listed in Table 
\ref{tab:morph}. A few remarks are in order. First, five of the FDS early-types are from the sample studied in \citet{2019iodice} which are located in regions of the cluster not dominated by intra-cluster light (ICL). These early-types have stellar masses $10^{9}\,M_{\odot} < M_{\star} < 10^{10}\,M_{\odot}$.  As we require a sky background estimation in our radial profile derivation (Sect. \ref{sect:methods}), including galaxies within the ICL regions would specifically require a more careful treatment of this component or ad hoc methods such as that shown in  \citet{2019iodice}. Therefore, for this paper we only focus on galaxies without ICL in their background.  \par 
Second, in the case of the FDS late-types previously studied in \citet{2019raj, 2020raj}, we able to study 12 galaxies from their sample. These galaxies have stellar masses $> 10^8\,M_{\odot}$. Apart from the ICL issue, we remind the reader that this selection also includes our additional criteria on the inclination of the galaxy and and those impacted by neighbouring galaxies (see the paragraph on `Total Sample' in the previous section). Throughout the paper we refer to our sub-sample from \citet{2019iodice} and \citet{2019raj, 2020raj} collectively as `FDS ETGs+LTGs' and FDS `dwarfs' for the rest of the sample from \citet{2021su}.  Our FDS ETGs+LTGs sample are predominantly LTG. \par 

Finally, as we use the same nearly isolated sample as in ELVES \citep{2021carlsten}, we lack early-type galaxies in this category by construction. For this reason, we focus on comparing the sizes of galaxies in different environments primarily at fixed stellar mass. The segregation by morphology for each sample is also shown and discussed where appropriate. \par
%\textbf{POINT OUT HERE SPECIFICALLY IN FDS THE num. of LTGS and ETGS which are not contaminated by ICL.}

%the quenched SAGA satellites do have similar red UV-optical color to the ELVES early-types to populate the same regions in the ultra-violet (UV)-optical colour to be similar to the early- and late-type

\begin{table}[h]
\caption{Number of early- and late-type galaxies in each environment}
    \begin{center}
    
    \begin{tabular}{c c c}
       \textbf{Environment}  & \textbf{Early-type} & \textbf{Late-type} \\  \hline 
         Cluster & 353 & 64\\
         Group \& satellite & 106 & 129 \\
         Nearly isolated & 26 & 218\\ 
         \textit{Total} & \textit{485} &  \textit{411} \\ \hline 
    \end{tabular}
    \label{tab:morph} 
        \end{center}
        \vspace{-7pt}
\textbf{Notes:} The values listed here are for the galaxies within the full stellar mass range listed in Table \ref{tab:sample}. Visual classifications are only available for galaxies from the FDS (cluster) and ELVES (satellite) samples. The rest of the sample has been categorised using the equivalent widths of the H$\alpha$ line as a proxy for star forming or quenched systems. We use this categorisation as late- or early-type, respectively (see text for details).
\end{table}

\begin{figure}[h!]
    \centering
    \includegraphics[width=0.49\textwidth]{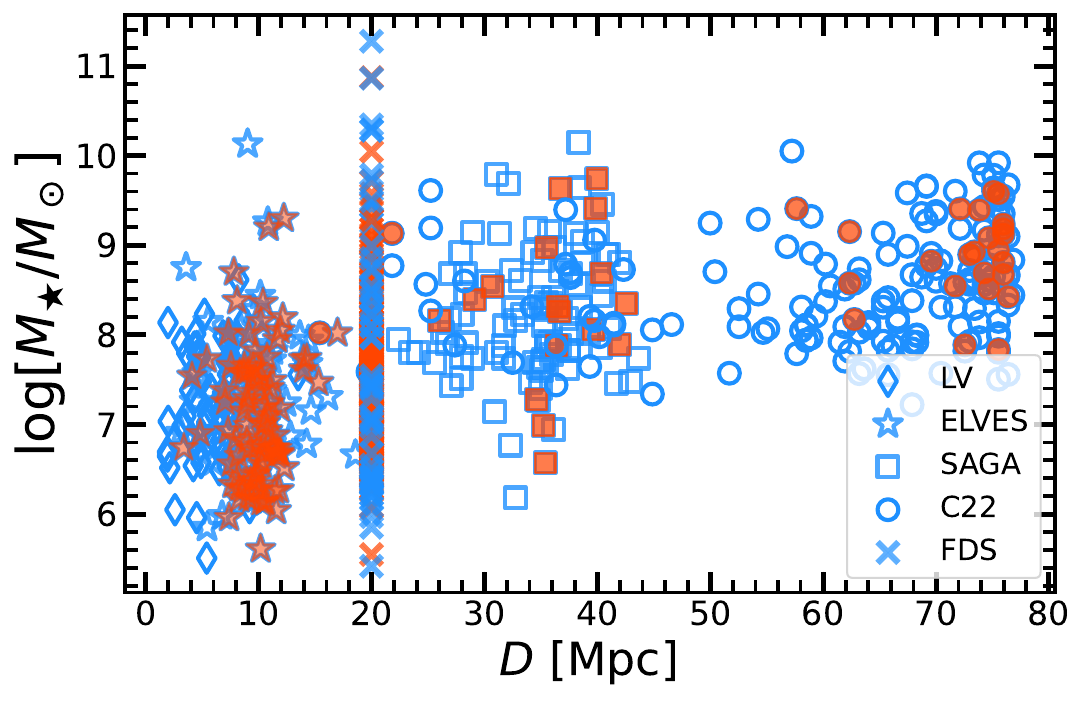}
    \caption{Distribution of the sample listed in Table \ref{tab:sample} in stellar mass as a function of distance. The points are coloured according to their morphology, broadly categorised here as early- (red) or late-type (blue) [see text for details]. The markers in the legend label each sub-sample. The number of galaxies in each type are listed in Table \ref{tab:morph}.}
    \label{fig:mass_distance_plot}
\end{figure}

In Fig. \ref{fig:mass_distance_plot}, we plot the stellar mass distribution of each sample as a function of their distances. We mark the early-type galaxies in orange and late-type in blue. The FDS sample is considered to be at a distance of 20\,Mpc. Limiting our analysis to galaxies out to twice the distance of Fornax would almost completely remove the C22 sub-sample. Making such a distance cut would significantly restrict our comparison of edge properties in Fornax to galaxies with $M_{\star} \lesssim 10^{8.5}\,M_{\odot}$. Above those stellar masses within $D \lesssim 40$, we would have the confirmed SAGA satellites and only $\sim 25$ nearly isolated galaxies from C22. \par

Low surface brightness galaxies such as those catalogued in Fornax \citep{2022venhola} are not considered in our sample due to the difficulty in measuring the distances of similar galaxies in the field or isolated environments \citep[see e.g.][]{2019roman, 2022greene}. Future works specifically targeting low surface brightness galaxies within our lower stellar mass bins will readily complement the relations we report here.  %Throughout this paper, our comparison of the properties of edges between samples and their respective environmental category (Table \ref{tab:sample}) is primarily based on their stellar mass. 

%We also point out that the imaging quality, defined as the FWHM of the measured point-spread function (PSF), of the different surveys used are very similar $\sim 1\,''$. At the median distance of the total sample studied here $\sim34$\,Mpc, the smallest structures resolvable with our compiled imaging dataset is $\sim 165$\,pc. Therefore, the compiled dataset is suitable for this study in both depth and spatial resolution. 

\section{Methods}
\label{sect:methods}

We follow the methodology described in detail by \cite{2022chamba} to derive the surface brightness, colour and surface stellar density radial profiles of the 894 galaxies in our sample. We show the profiles of a number of galaxies with stellar masses $\sim10^9\,M_{\odot}$ from the \cite{2022chamba} sample in Fig. \ref{fig:sdss-examples}. The same methodology is applied to the FDS, SAGA, ELVES and LV galaxies. We summarise the procedures for profile derivation and edge identification for each galaxy below.\par 

%is shown in Fig. \todo{TO DO, \ref{fig:fds_examples}}

%show the effective radius in the radial profiles? 

%TO DO: Summarise how radial profiles are computed (ellipse fitting, background subtraction and masking).
%Summarise visualisation procedure here (see Fig. 1 in \cite{2022chamba} flowchart). Summarise criteria used to detect edges. 
\textbf{Image stamps:} We create an image centered on the galaxy using the $g$ and $r$-band data. The dimensions of the image depends on the diameter of the galaxy. We used three times the Homlberg radius \citep{1958holmberg} as a benchmark to create an image which includes the immediate `local' surroundings of the galaxy. \par 
\textbf{Masking:} We sum the $r$ and $g$-band images and run Sourcerer \citep[formerly Max-Tree Objects;][]{2016mto, 2021haigh}, a robust segmentation tool, for masking extremely diffuse light sources in an image. We create a `master mask' using the summed image to include source detections from both bands in the segmentation map. The associated sourcerer catalogue includes the best fit elliptical parameters (axis-ratio and position angle) using the map for each source. We mask every source in the images except the galaxy of interest and the background when deriving radial profiles. Additional manual masking is used when necessary. \par 

\textbf{Background subtraction:} We define `background' as every pixel in the image undetected by Sourcerer. While the survey images are already background subtracted, the subtracted values are not always representative of the more local background variations in the immediate surroundings of galaxies and are either over- or under-subtracted \cite[see][]{2020tck}.  Therefore, to make a more precise and representative background subtraction to the galaxy of interest, in our default pipeline we make an additional `local' background estimate. We select an elliptical annulus between two and three times outside the segmented region of the galaxy. We visually verify that all light sources within this annulus are masked and as mentioned in the masking step, additional pixels are manually masked if necessary. An average local sky background value is then measured using all the unmasked pixels within this annulus. This value is then subtracted from the images. This method of subtraction is similar to the approach developed in \cite{2006pohlen} as it removes the flat, average asymptotic flux value in the image at least twice the distance beyond the edge of the galaxy. 

\textbf{Radial profiles:} We use the elliptical parameters from the segmentation tool to average flux within bins of concentric elliptical annuli beginning from the centre of the galaxy until the end of the background subtracted image. The number of bins we use  vary with galaxy extension. We first define a parameter '$b_{prof}$' such that $b_{prof} = 2r_{eff}$ where $r_{eff}$ is the effective semi-major axis directly taken from the catalogues we use (see Sect. \ref{sect:data}) in pixel units. The bin size is then computed by dividing the distance from the centre of the galaxy to $3b_{prof}$ i.e. well-beyond the edge radii of our sample, by the number of bins $N_{bin}$. We set $N_{bin} = b_{prof}$ rounded to the nearest integer. This gives us a binsize of 3 pixels in our profiles, corresponding to  $\sim 0.79$ arcsecs in DeCALs and $\sim 0.63$ arcsecs in FDS. \par 
Such a binsize is justified if the edge feature occurs over sub-kpc scales. At the distance of the Fornax Cluster, the bin size is 60.5\,pc. At the median distance of our field sample (populated by SAGA) $\sim 34$\,Mpc, the bin size $\sim 134$\,pc. Given that our FDS data is our deepest dataset with the smaller pixel scale, we also test our profile derivation method using different bin sizes in Appendix \ref{app:profile_bins} for a bright dwarf galaxy in our Fornax sample. \par 

We use fixed elliptical parameters to ensure that the edge we locate is not artificially caused by the use of a different position angle or axis ratio within the galaxy. This step produces the $g$ and $r$ flux radial profiles of the galaxy.\par  
The profiles are then converted to surface brightness ($\mu_g$ and $\mu_r$) following the appropriate calibrations of the datasets considered. The $g-r$ and surface stellar mass density $\Sigma_{\star}$ profiles are also computed. The stellar masses are computed using colour-mass-to-light ratio relations from \cite{2015roediger} as in \citet{2020tck}. \par 

\textbf{Inclination correction:} \cite{2022chamba} only considered low inclination galaxies with axis-ratios $\geq 0.3$ to reduce the impact of the point spread function (PSF) and projection effects on the stellar mass estimation. This axis-ratio limit was carefully chosen by considering the inclination correction necessary for estimating the stellar mass using the disk models developed in \cite{2020tck} for galaxies with varied thickness (see their Fig. 2). For galaxies with higher axis ratios, the inclination correction was found to be independent of thickness. We make the same correction on the radial profiles derived in this work for consistent comparison with the \cite{2022chamba} measurements. Table 1 in \cite{2020tck} list the fitted parameters we use to make this correction.  

%The correction takes the form:
% \begin{equation}
%     \todo{EQUATION}
% \end{equation}

\textbf{Galactic extinction:} We additionally correct the radial profiles for Galactic extinction using the $A_g$ and $A_r$ values computed using the NED calculator.\footnote{\protect\url{https://ned.ipac.caltech.edu/forms/calculator.html}}

\textbf{Identifying the edge}: As summarised in Sect. \ref{sect:field-edges}, the signature of the edge is identified as a change in slope in the outer radial profiles of galaxies. In particular, for field dwarf galaxies, \citet{2022chamba} reported that the most visible signature is that the colour beyond the edge in the profile may either increase or decrease towards redder or bluer colours, respectively. This shape of the colour profile can be understood in terms of inside-out or outside-in galaxy evolution, depending on whether older or younger stellar populations are found in the outskirts \citep{2016herrmann, 2017denija}. \par 
The pipeline to search for and identify the edge signature is drawn as a flowchart in Fig. 1 of \cite{2022chamba}. Here we summarise the main steps as follows. We first visually search for similar signatures where the colour drops or rises towards bluer or redder colours in the derived $g-r$ colour profile of the galaxies. We then visually check whether the galaxy is elliptically symmetric at the chosen radial location ($R_{\rm edge}$) by overplotting the elliptical contour at the edge on the RGB image of the galaxy to ensure that the truncation is not created by an artifact in the data or a neighbouring source that was unmasked. \par 
If unmasked sources are found, a second masking is performed at the potential contaminant and the radial profiles are re-derived. We search again for the edge in the same manner described above on the new profile. We report edges only in cases where all potential contaminants that we are able to visually verify are masked. \par 

In many cases studied by \citet{2022chamba} the edge is also visible in the surface brightness and density profiles. While multiple signatures of a change in slope in all three profiles is further confirmation of the edge feature reported here, we point out that in NC's experience the most obvious signature in the majority of the dwarf galaxies studied in this and previous work is in the $g-r$ colour profile. There are two potential reasons for this outcome. \par 

The first is that in surveys like SDSS where the shape of the PSF in the $g$ and $r$ band are similar, \cite{2022chamba} has argued that the PSF effect of broadening a galaxy's $g-r$ radial profile is roughly cancelled out to first order, at least within the locations where edges are found. From this technical perspective, the inflection point in the colour profile is the more robust signature of the edge compared to the ones in the surface brightness profile. While we do not explore deconvolution techniques on our profiles, dedicated effort is underway within the LIGHTS collaboration \citep{2021trujillo} to exactly account for these effects on ultra-deep profiles (G. Golini in prep.). \par 
The second reason is that a surface brightness profile in a single band cannot be used as an indicator of the varying stellar populations dominating the structure of galaxies at different radii. Since we are interested in locating a \emph{physical} feature in the galaxy outskirts that can mark the `edge'' of the bulk of the stellar material formed in situ, the colour of the galaxy is an important observable to make this task possible. \par 
In fact, in the recent article by \citet{2023fernandez} who use machine learning to automate the identification of galaxy edges using previously analysed \textit{Hubble} images \citep[see][]{2023buitrago}, the authors use different combinations of colour images to define the edge and show that the algorithm is able to recover the identifications made by expert observers. As the use of machine learning is beyond the scope of this paper, we refer the interested reader to that work for details on their training dataset and algorithm. For the above reasons, we are convinced that our criteria for identifying edges using the colour profile \emph{in addition to the surface brightness and stellar surface density profiles} is more reliable compared to only using surface brightness or stellar mass density profiles \citep[see also][]{2016herrmann, 2023sanchez}.  In Appendix \ref{app:asymmetry} we show how our profile derivation and edge identification method performs for four cases in the \citet{2019raj} sample which are clearly asymmetric. \par 

%If the edge is located in multiple profiles, we report an average value and include this in t
%\todo{Discuss something about robustness?} 

\textbf{Effective radii:} In addition to $R_{\rm edge}$, we also measure the effective radii ($r_e$) of each galaxy. Instead of computing the average flux in the elliptical annuli as before, we compute the sum in each annulus and create a light growth curve. The curve is cut-off at the surface brightness limit of data used (see Table \ref{tab:sample}) to measure the total light (or apparent magnitude) belonging to the galaxy in each band. $r_e$ is then defined as the radius enclosing half the light of the galaxy. In Appendix \ref{app:reff_difference}, we show that the scaling relations and results using the modelled radii values from \citet{2021su} for Fornax galaxies do not change our main conclusions. \par 
\textbf{Colour and age:} We use the global $g-r$ colour of the galaxy to estimate its age from the E-MILES models \citep{2012vazdekis}. The model predictions for the SDSS filters using Kroupa Universal isochrones\footnote{\protect\url{http://research.iac.es/proyecto/miles/pages/photometric-predictions-based-on-e-miles-seds.php}} are provided for seven specific metallicity values ranging from -2.32 $<=$ [M/H] $<=$ 0.22. As all our measurements account for the outskirts of galaxies, we select an average metallicity value which is representative of the stellar populations in this regime. For simplicity, we adopt a fixed metallicity of [M/H] = -0.71 across the full stellar mass range. This low metallicity value accounts for both, the metal-poor stellar populations found in the faint outskirts of nearby galaxies \citep{2017denija, 2021neumann} as well as in observations and simulations of dwarf galaxies \citep[see e.g.][]{2023cardona}. \par 
In Appendix \ref{app:metallicity}, we additionally explore our results using alternative metallicity values from E-MILES which follow the \citet{2013kirby} stellar--mass metallicity relation for dwarf galaxies. While the estimated age values for individual galaxies are different as expected, the colour and age bifurcation we report in this work remains unchanged. \par 

%We do not assume a stellar mass-metallicity relation as the relations are often computed using the brightest central regions of galaxies \citep{2023dominguez} while  

%and have not yet been extensively examined for galaxies with $M_{\star} < 10^8\,M_{\odot}$.

%However, \cite{2022chamba} additionally used solar metallicity and showed that it does not significantly change the age bifurcation they found in the size--stellar mass relation. For all these reason, and because we are more interested in the low mass dwarf galaxies in our sample, we adopt a fixed metalicity value of [M/H] = -0.71 in this work.  \par 
\textbf{Stellar mass and surface density:} The surface stellar density at the edge ($\Sigma_{\star}(R_{\rm edge})$) is obtained from the surface density radial profiles via linear interpolation. The total stellar masses $M_{\star}$ are computed by integrating the surface stellar density profile out to the surface brightness limit of data used (Table \ref{tab:sample}). \par 
All the average or mean values in edge locations or edge densities reported are computed by 3$\sigma$-clipping the distribution of interest. Using 5$\sigma$-clipped values do not significantly change our results. 
%using colour-mass-to-light ratio relations from \cite{2015roediger} as in \citep{2020tck} and 

\textbf{Uncertainty estimation:} We compute the uncertainty in the $R_{\rm edge}$ and $\Sigma_{\star}(R_{\rm edge})$ measurements due to background subtraction and stellar mass estimation following the approach described in \cite{2020tck, 2020chamba} and summarised in \cite{2022chamba}. We perform this task in order to quantify the movement of $R_{\rm edge}$ in the size--stellar mass plane due to both uncertainties and the edge surface density--stellar mass plane in the case of $\Sigma_{\star}(R_{\rm edge})$.  \par 
Each galaxy is parameterised by a pair of values ($R_{\rm edge}$, $\Sigma_{\star, edge}$). Fixing $R_{\rm edge}$, for each band the surface brightness radial profile is moved up or down by a random value drawn from a Gaussian distribution defined using the previously measured dispersion value of the image background. \par 
The steps to compute the colour and stellar mass surface density profiles are then followed using the `moved' surface brightness profiles and we  infer the stellar surface mass density at the edge location in the moved profile, i.e. $\Sigma^{move}_{\star}(R_{\rm edge}) = \Sigma^{move}_{\star, edge}$ . Similarly, we fix the density at the edge and track how the associated edge location changes in the moved profiles. In other words, $R^{move}_{\rm edge}$ is taken as the value where $\Sigma^{move}_{\star}(R^{move}_{\rm edge}) = \Sigma_{\star, edge}$. \par 
We repeat this process $1000$ times. At each step, we record ($R^{move}_{\rm edge}, \Sigma^{move}_{\star, edge}$) and measure the differences ($\delta_{R}, \delta_{\Sigma}$)  between each `moved' value and our original ($R_{\rm edge}$, $\Sigma_{\star, edge}$) pair. Finally, we compute the dispersion of the distributions $\sigma_{\delta_R}$ and $\sigma_{\delta_{\Sigma}}$ and use these dispersion values as the uncertainty in our measurements due to background subtraction.\par

In the case of the uncertainty due to stellar mass, we follow the same procedure outlined above except we obtain random values to move the profiles using a Gaussian distribution with a fixed dispersion value of 0.25\,dex. This value was chosen because it is the typical uncertainty in stellar masses when computed using mass-to-light vs. colour relations \citep{2015roediger}. \par 

\textbf{Scaling relations:} We use a Huber regression which is robust towards outliers to estimate the best fit lines and dispersion values in the  scaling relations as in \citet{2022chamba}. To estimate how each of the background and stellar mass uncertainties for each galaxy contributes to the `global' dispersion of the size--stellar mass scaling relation $\sigma_{back}$ and $\sigma_{mass}$, we use the fact that the best-fit relation to our observations has a non-negligible dispersion ($\sigma_{obs}$; Table \ref{table:bestfit}). \par 
For simplicity, we assume that the observed dispersion can be expressed as a quadratic sum:

\begin{equation}
    \sigma_{obs}^{2} = \sigma_{int}^{2} + \sigma_{back}^{2} + \sigma_{mass}^{2} + \sigma_{vis}^{2}
\end{equation}

where $\sigma_{int}$ is the intrinsic dispersion of the relation and $\sigma_{vis}$ is the typical uncertainty from visual identifications \citep[0.04\,dex;][]{2022chamba}. We move each observed profile such that the data point ($R_{\rm edge}, M_{\star}$) is exactly on or crosses the best-fit relation and infer ($R_{\rm cross}, M_{\star, \rm cross}$). This process sets $\sigma_{back}^{2} = \sigma_{mass}^{2} = \sigma_{vis}^{2} = 0$ and leaves an artificial `intrinsic' dispersion of the  ($R_{\rm cross}, M_{\star, \rm cross}$) plane which is the total observed dispersion, as that is the typical displacement of our observed data points to cross the best fit line. Let us call this artificial `intrinsic' dispersion as $\sigma_{\Delta}$. We then randomly draw a new $R^{rand}_{\rm edge}$ value from a Gaussian distribution centred on $R_{\rm edge}$ with the associated $\sigma_{\delta_R}$ from the procedure described above and record its movement on the ($R_{\rm cross}, M_{\star, \rm cross}$) plane. In this case, we assume that that final dispersion in this plane is:

\begin{equation}
    \sigma_{\delta \rm cross}^2 = \sigma_{\Delta}^2 + \sigma_{X}^2
\end{equation}

where %$\sigma_{\Delta}$ is the artificial `intrinsic' dispersion of the ($R_{\rm cross}, M_{\star, \rm cross}$) plane we started with because we artificially moved the profiles to derive it.   
$X$ is the uncertainty we seek to estimate, considering which $\sigma_{\delta_R}$ value is used. The estimated uncertainty on the scaling relation due to the background and mass is then separately computed as the difference in quadrature %between the dispersion of the $R^{rand}_{\rm edge}-M_{\star, cross}$ plane and the artificial intrinsic of the best-fit relation, i.e. 
$\sqrt{\sigma_{\Delta}^2 - \sigma_{\delta \rm cross}^2}$. For simplicity, we fixed $\sigma_{\Delta}$ to the total dispersion of the observed relation 0.13\,dex (Table \ref{table:bestfit}) as it is the typical displacement of the profiles in the size--mass plane to cross the best-fit line.  \par 
For comparison, we follow the same procedures as described above to estimate the dispersion in the effective radius--stellar mass relation contributed by the uncertainties in image background subtraction and stellar mass estimation in our measurements. However, here we mark each galaxy using the ($r_e, \Sigma_{\star, eff}$) pair of values where $\Sigma_{\star, eff} = \Sigma_{\star}(r_e)$ is the surface stellar mass density at $r_e$. While we do not analyse the $\Sigma_{\star}(r_e)-M_{\star}$ plane in this work as we focus on the edge properties of galaxies, we use this parameter here for practical purposes in order to obtain $r^{move}_e$ such that $\Sigma^{move}_{\star}(r^{move}_{e}) = \Sigma_{\star, eff}$. We refer the reader to \citet{2021stone} for alternative methods of estimating the intrinsic scatter of scaling relations.

%we fixed the location of the edge and then followed how the radial profiles move by a random quantity prescribed by the dispersion in the measured background value per image and stellar mass estimate in our procedure. In other words, we fol- lowed how the inferred Σ⋆ (Redge ) changes due to our background and mass estimate if we fix Redge and vice versa. The disper- sion in the stellar mass comes from comparing the estimate from Eq. (1) and those published by Maraston et al. (2013)

%The intrinsic uncertainties in the scaling relations are computed using....\todo{TO DO: Try computing intrinsic uncertainties. Because galaxies are already faint this will be difficult and maybe I will overestimate the errors (see Fig. 5 already in the edge relation).}
\section{Results}
\label{sect:results}

\begin{figure*}[h!]
    \centering
    \includegraphics[width=0.49\textwidth]{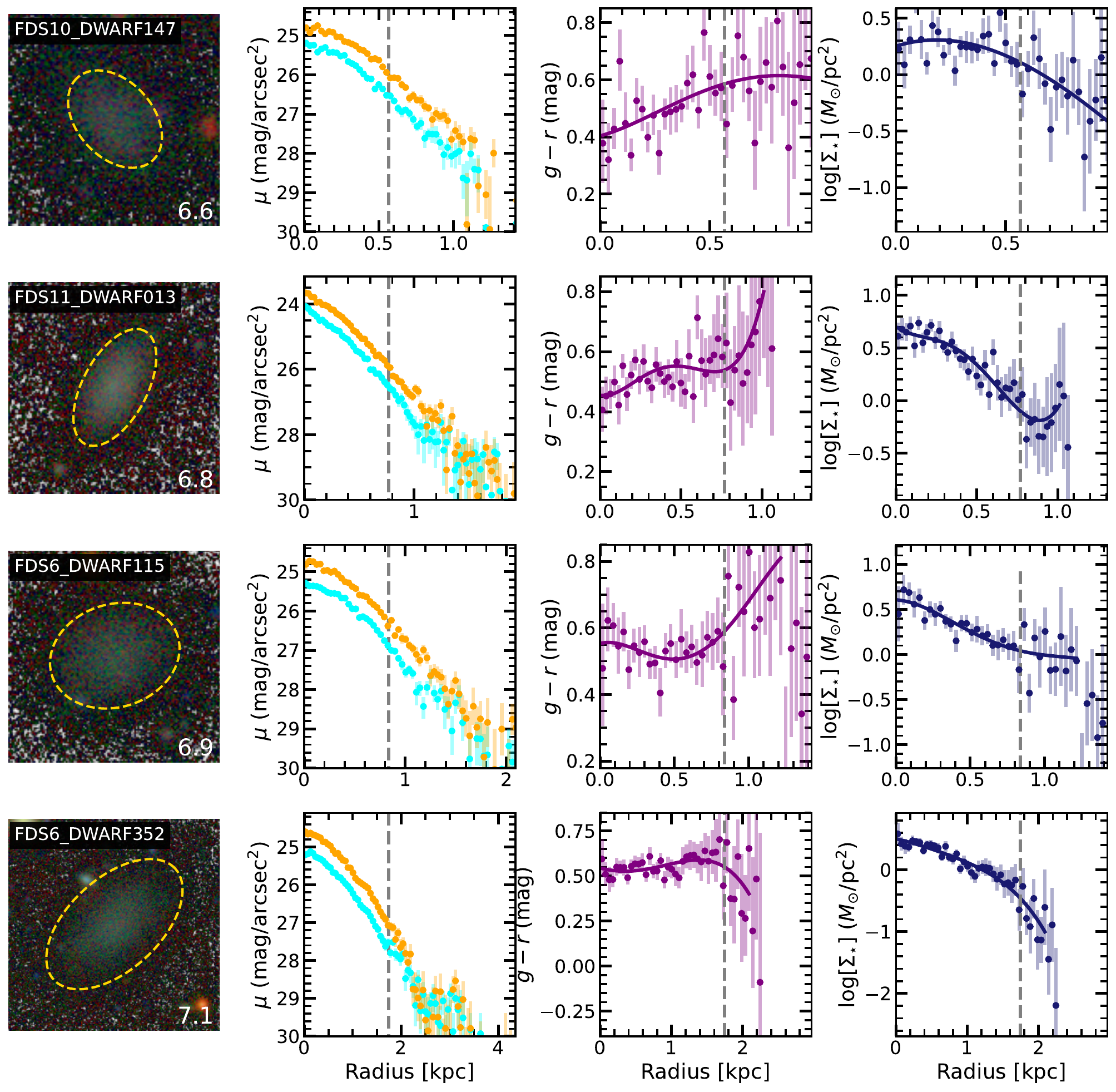}
    \includegraphics[width=0.49\textwidth]{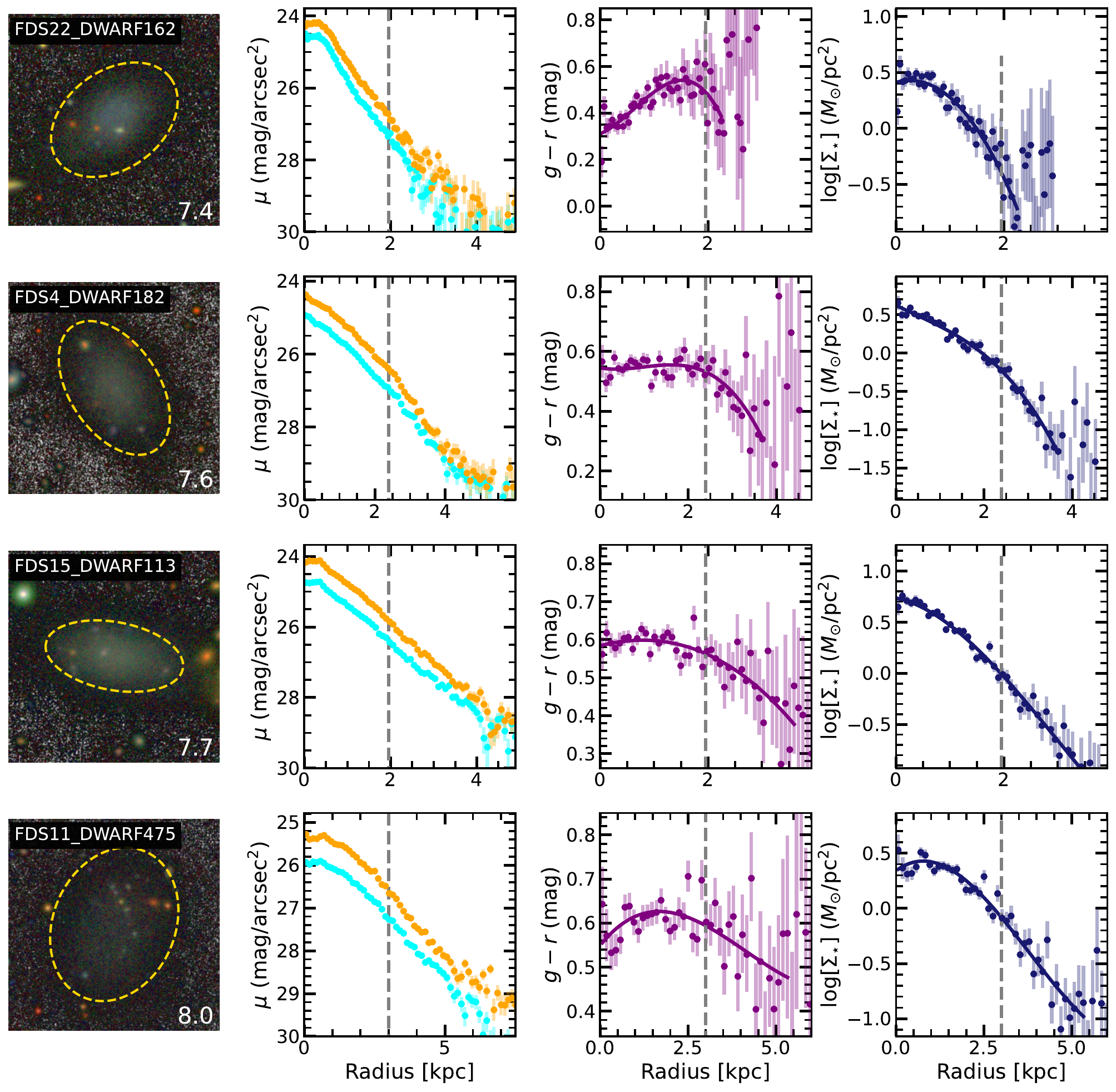}
    \caption{Similar to Fig. \ref{fig:sdss-examples}, example radial profiles of cluster dwarf galaxies with $M_{\star} \lesssim 10^8\,M_{\odot}$. The first panel is the $gri$-colour composite image using FDS images \citep{2018venhola}, overlaid on a grey scale background to highlight the low surface brightness boundaries of these galaxies. White in this scale indicates non-detections or background pixels in the data. FDS names for each galaxy from \citet{2018venhola, 2021su}, the estimated stellar mass in units of log$M_{\star}/M_{\odot}$ (left) and edge locations (dashed ellipses) are indicated in each panel. The second, third and fourth panels are the $g$ (cyan) and $r$-band (orange) surface brightness, $g-r$ and $\Sigma_{\star}$ profiles respectively. The identified edge feature is indicated as a vertical dashed line. To highlight the inflection point we mark as the edge feature in the colour and stellar surface density profiles, we additional plot a least squares polynomial fit of degree four as a solid line in those panels. Galaxies are ordered by increasing stellar mass from the upper to lower panels. }
    \label{fig:fds-examples}
\end{figure*}

\begin{figure*}[h!]
    \centering
    \includegraphics[width=0.49\textwidth]{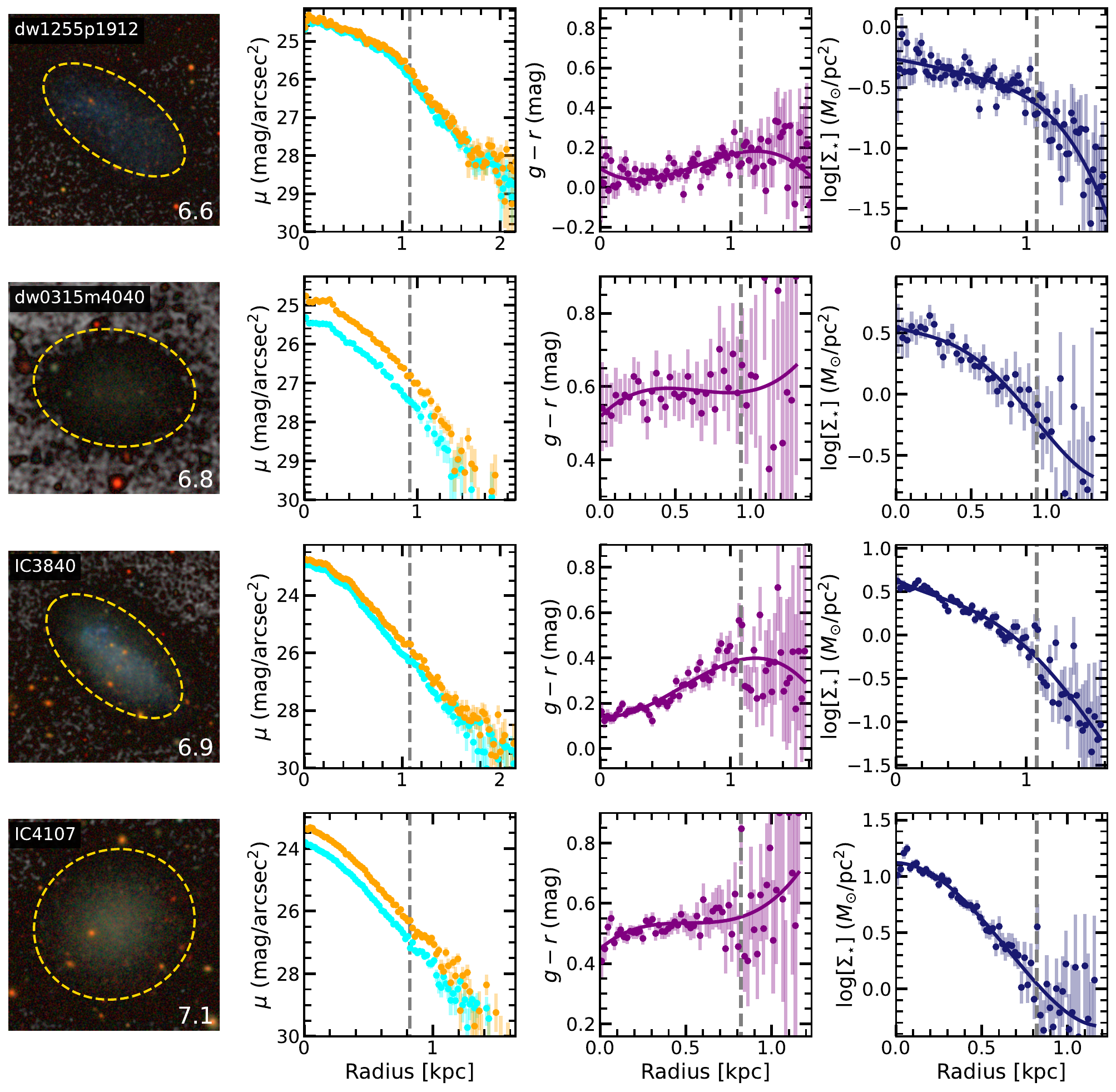}
    \includegraphics[width=0.49\textwidth]{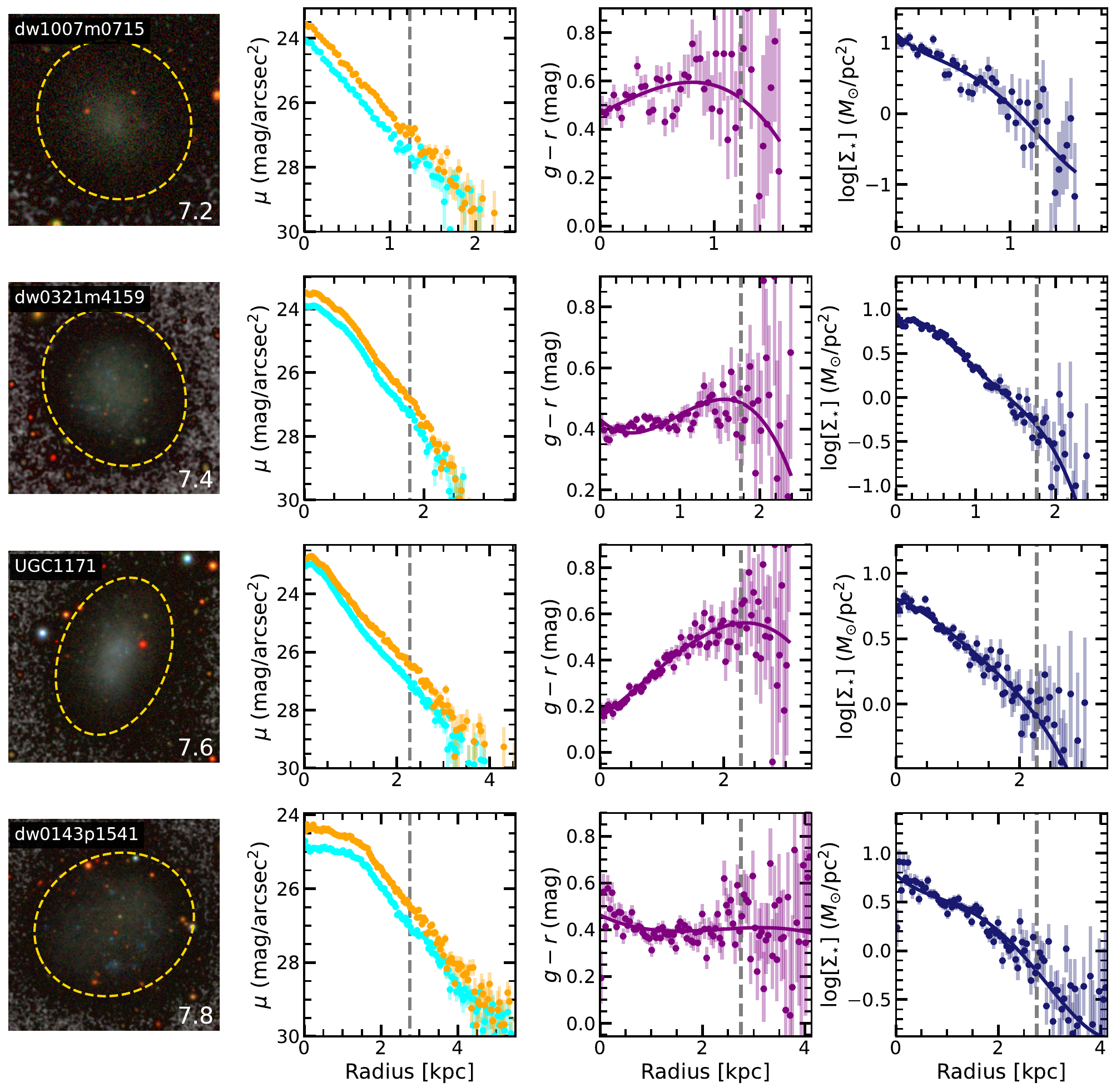}
    \caption{Similar to Fig. \ref{fig:fds-examples} but for satellite galaxies from ELVES \cite{2021carlsten} within DECaLs \citep{2019dey}.}
    \label{fig:elves-examples}
\end{figure*}

\begin{figure*}[h!]
    \centering
    \includegraphics[width=0.49\textwidth]{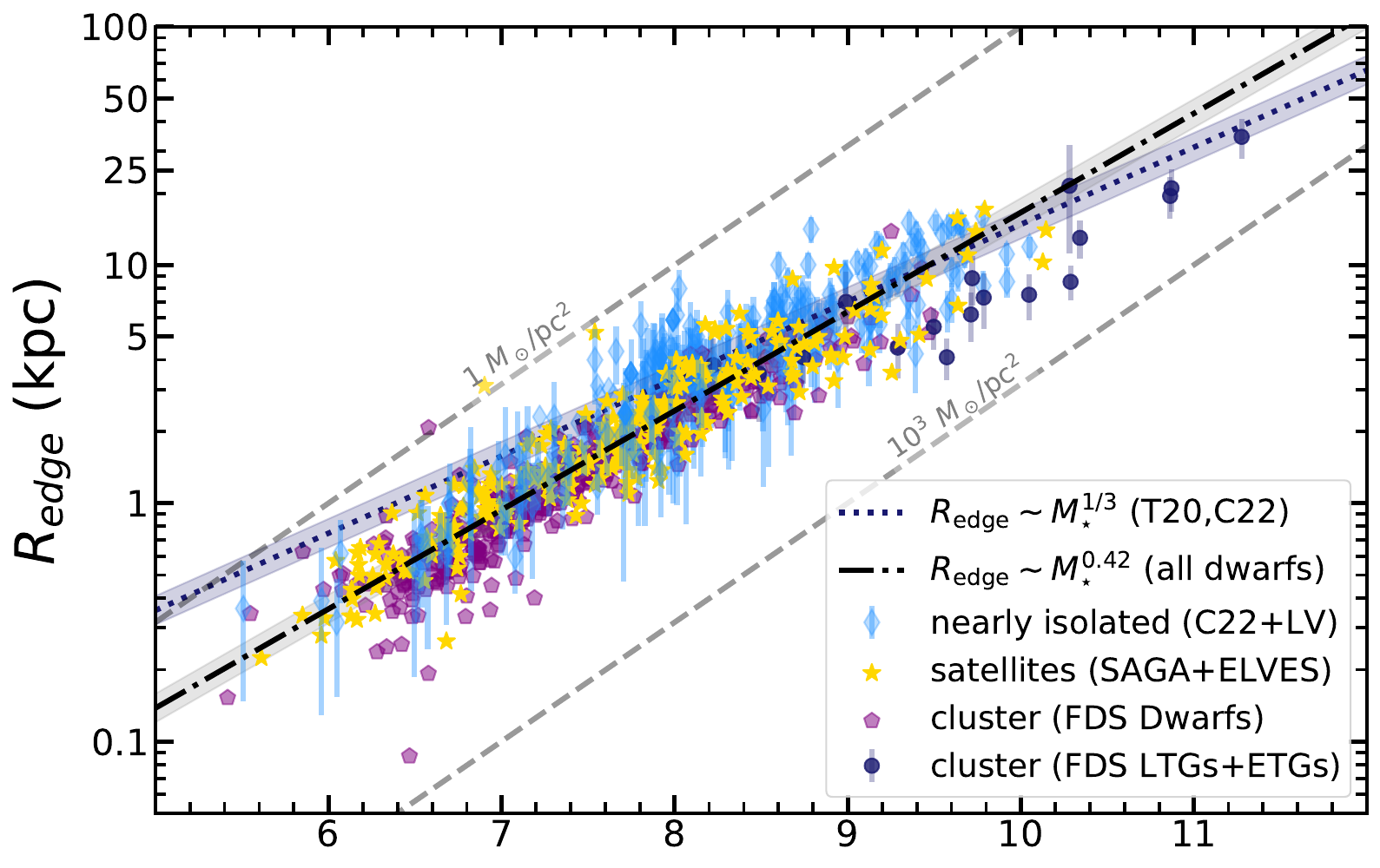}
    \includegraphics[width=0.49\textwidth]{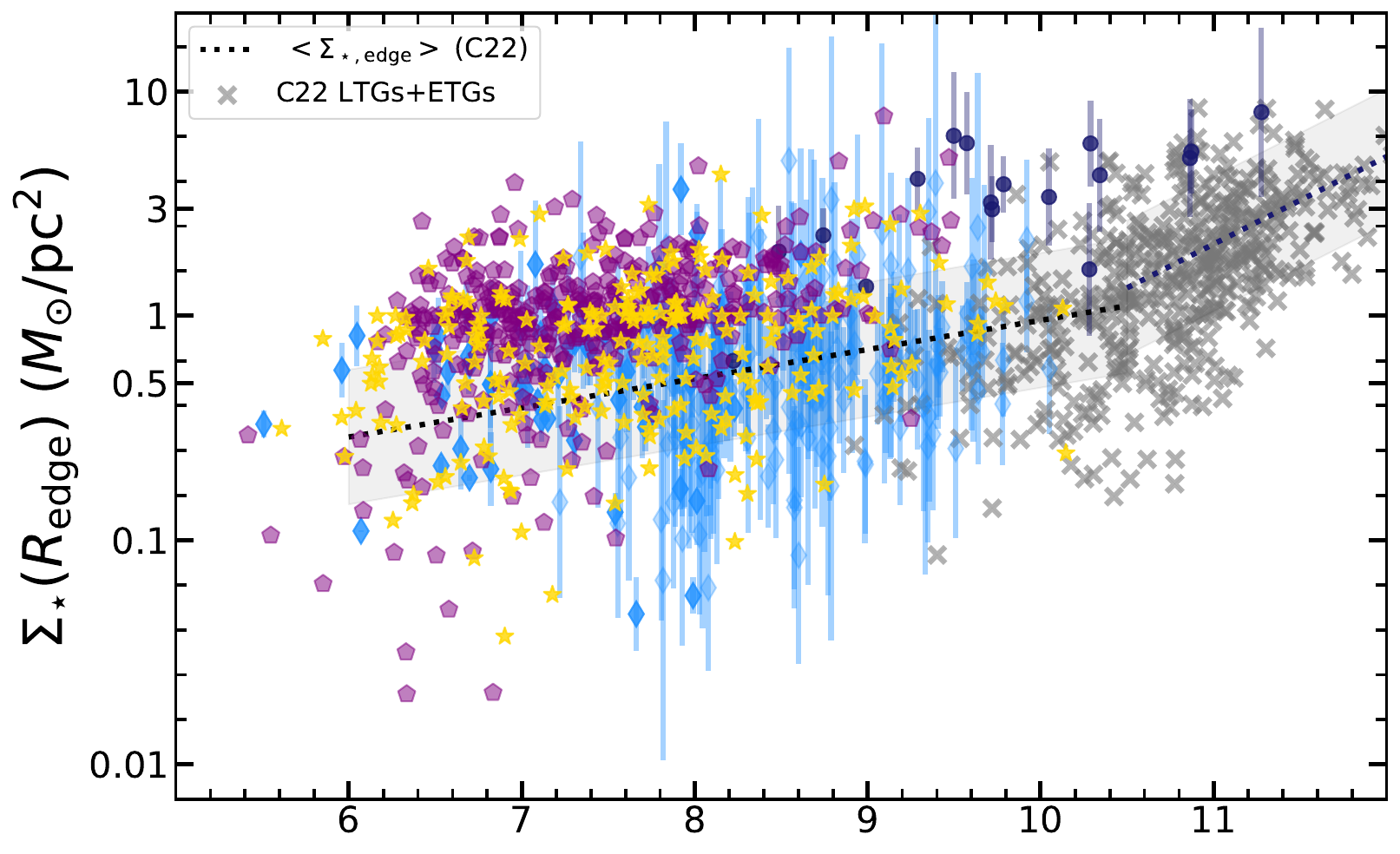}
    \includegraphics[width=0.49\textwidth]{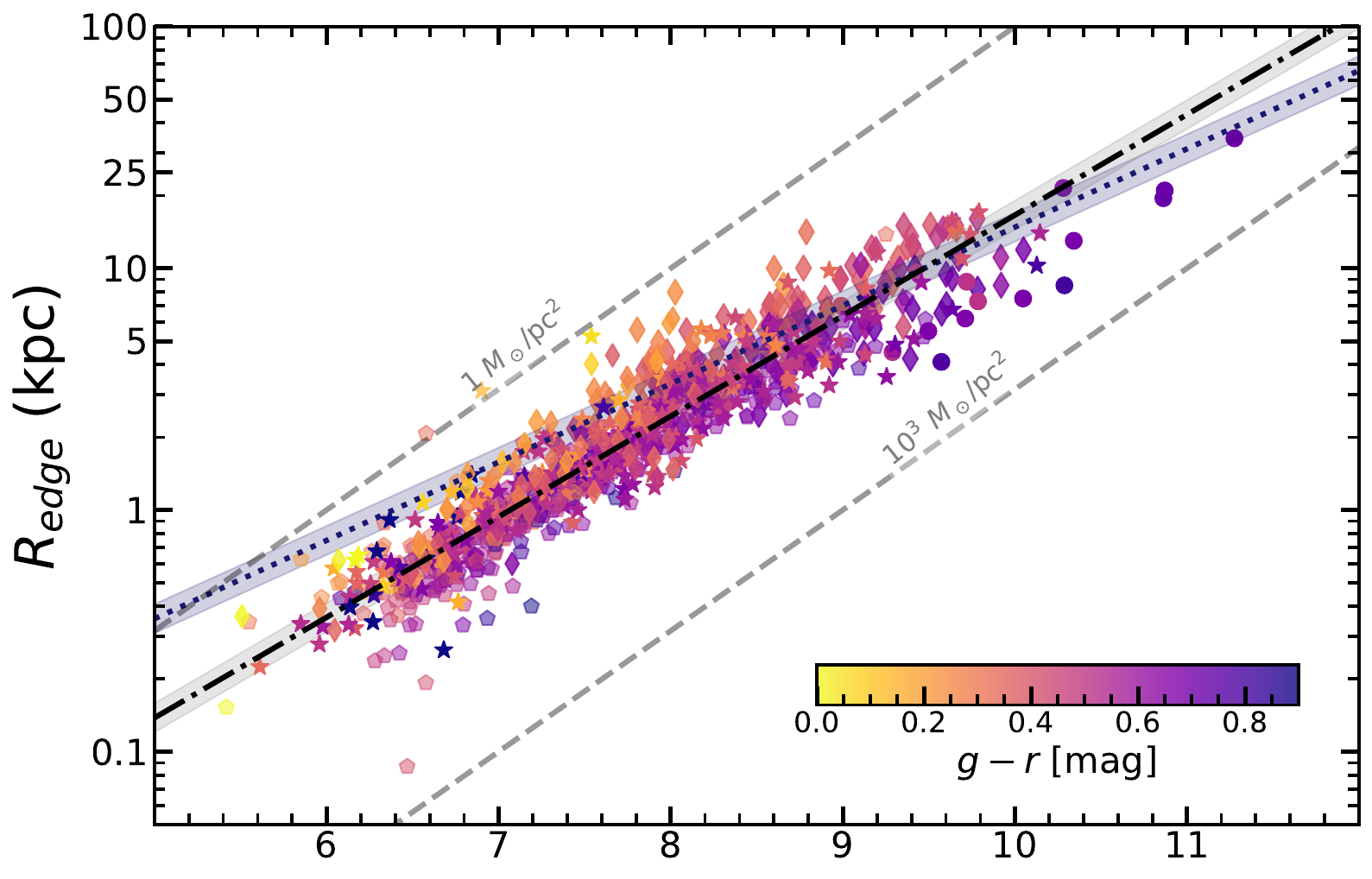}
    \includegraphics[width=0.49\textwidth]{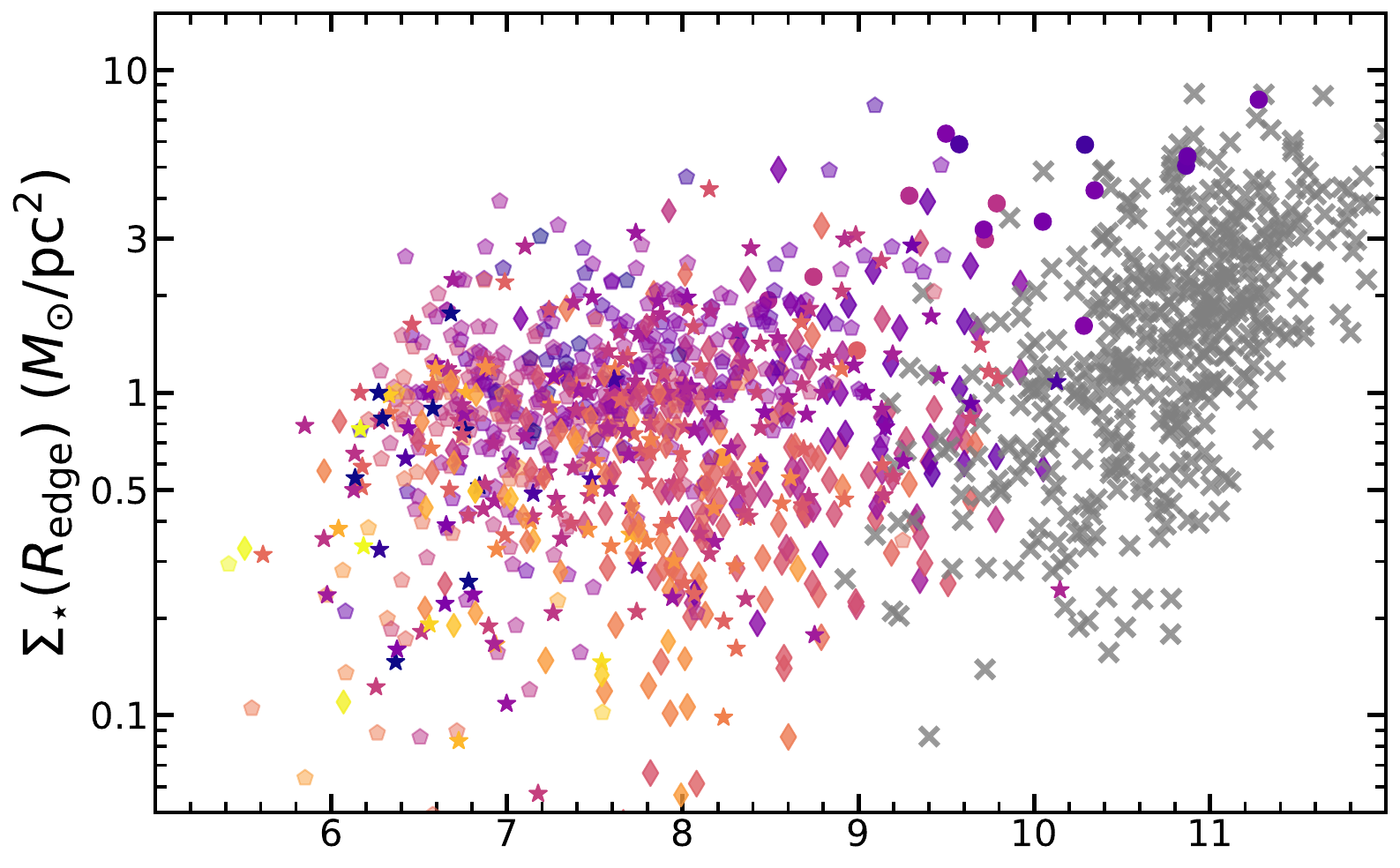}
    \includegraphics[width=0.49\textwidth]{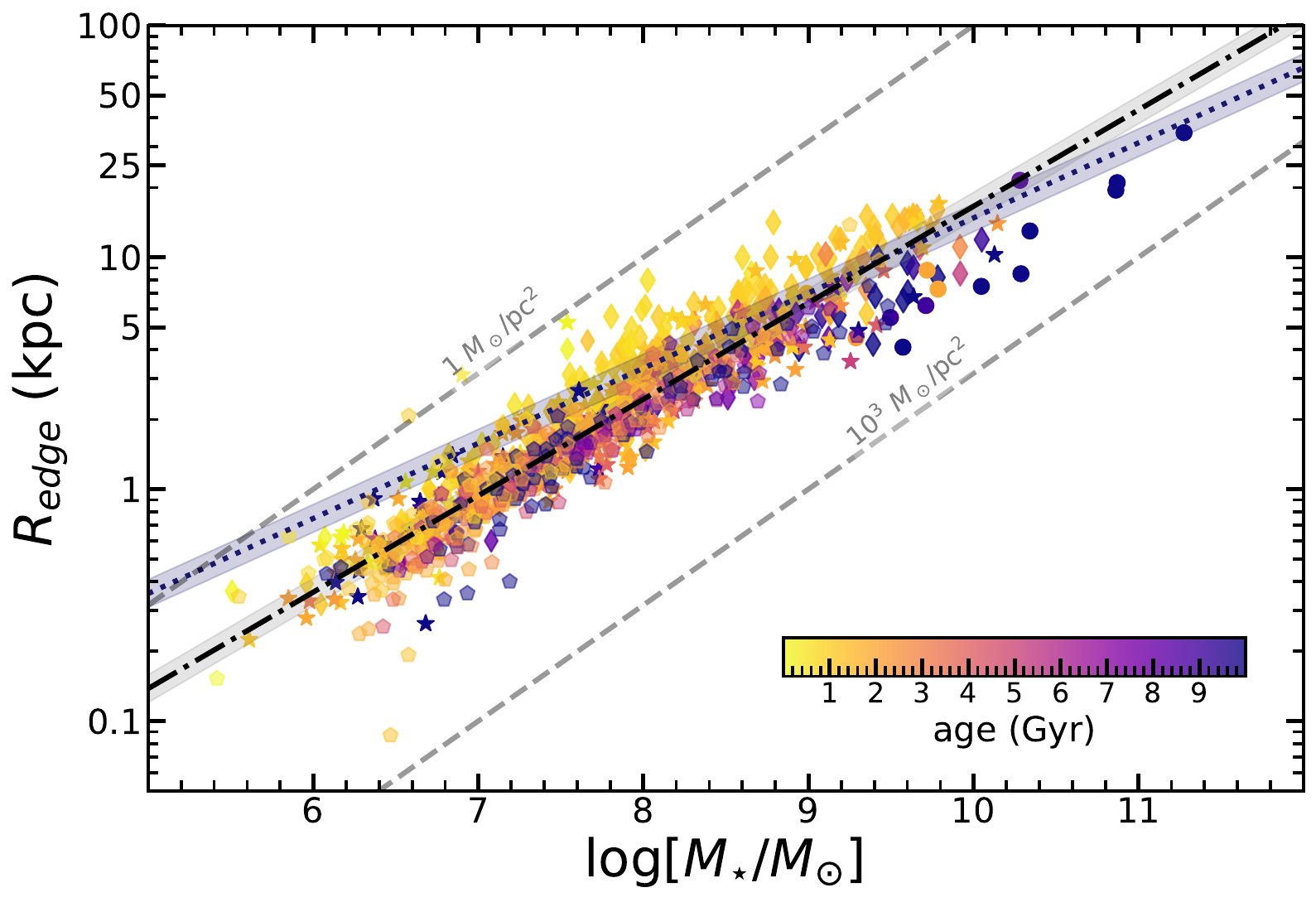}
    \includegraphics[width=0.49\textwidth]{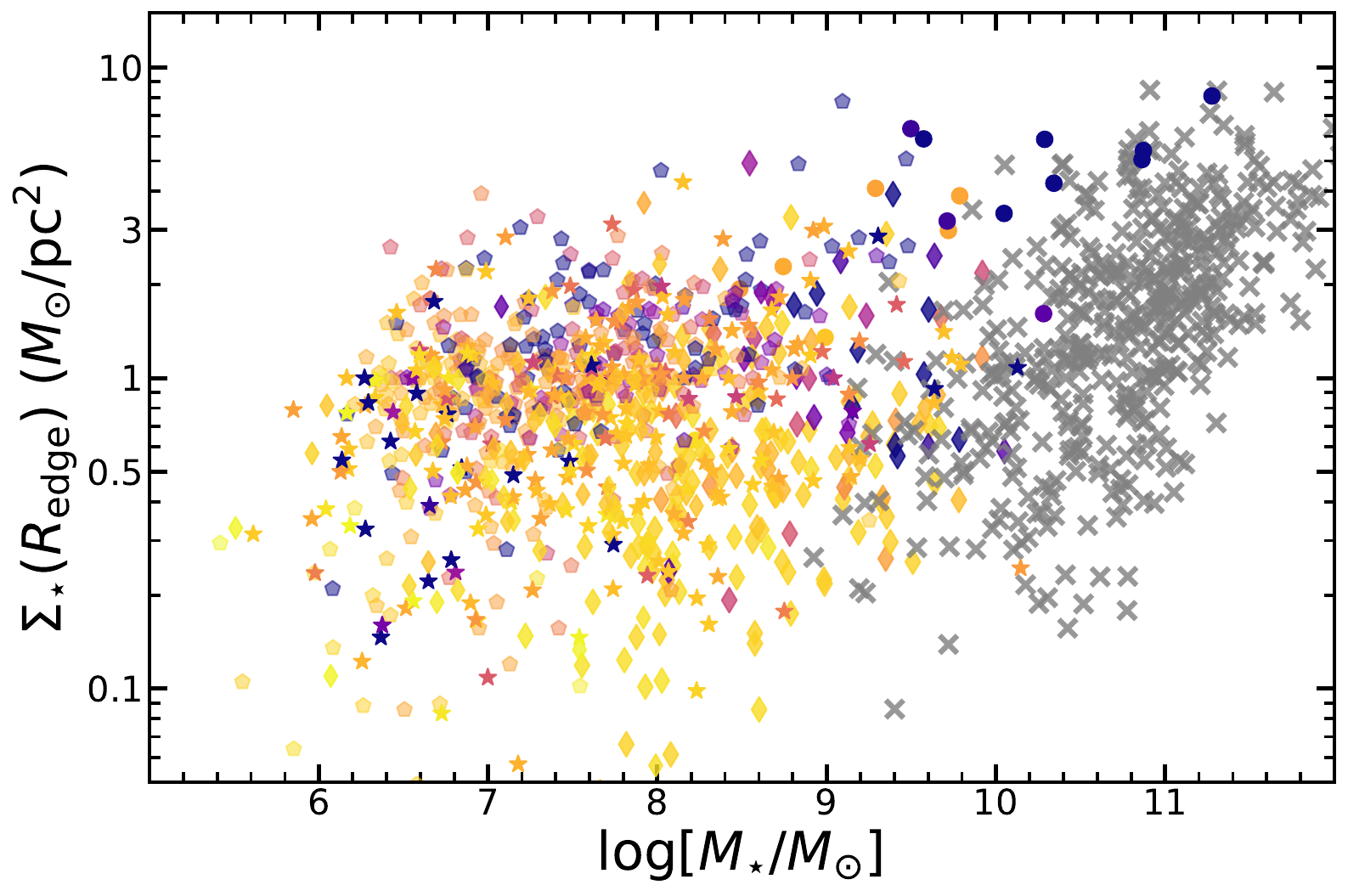}
    
    \caption{Size--stellar mass relation using edge radii (left) and the associated edge density--stellar mass relation (right). For visualisation reasons, we only plot the combined uncertainties due to the background and stellar mass estimation in our measurements (see text for details) for the field and Fornax LTGs/ETGs samples in the upper panels to demonstrate the typical values across the full stellar mass range of the scaling relations. \textit{\textbf{Upper}}: Best-fit laws fitting the labelled dwarf samples:  $R_{\rm edge}\sim M_{\star}^{1/3}$ \citep[reported for galaxies with $M_{\star} > 10^{7}\,M_{\odot}$ in][]{2020tck, 2022chamba} and $R_{\rm edge}\sim M_{\star}^{0.42}$ (this work). We plot the Fornax cluster dwarfs (purple pentagons), SAGA \citep{2021yao} and ELVES \citep{2022carlsten} satellites (yellow stars) and sub-sample \cite{2022chamba} and \cite{2013karachentsev} field galaxies (blue diamonds). For completeness, the radii for Fornax cluster massive late-type and early-type galaxies (black circles) are also shown. The shaded regions show the intrinsic scatter of each relation ($\sim 0.06\,dex$ for edge radii). At a fixed stellar mass, the majority of Fornax galaxies are below the best fit line. The edge density-stellar mass plane is plotted using the same symbols and colours as in the left panels. The piecewise, $\Sigma_{\star, edge}$ best-fit relations as well as the luminous ETGs and LTGs from \cite{2022chamba} (grey crosses) are shown for reference. \textit{\textbf{Middle:}} The same relations are now colour coded according to the global $g-r$ colour of the galaxy. The symbols representing each sub-sample are those used in the upper panels. \textit{\textbf{Lower:}} Similar to the middle panels, but now using the $g-r$ colour to estimate the age of the galaxy from MILES \citep{2012vazdekis} for a fixed metallicity of [M/H] = -0.71 (see text for details).  For reference, the age of the dwarf elliptical galaxies with $M_{\star} > 10^7\,M_{\odot}$ in the Fornax Cluster is $\sim$ 10\,Gyr \citep{2001rakos}. The figure shows that galaxy edges in the Fornax cluster are smaller, denser and redder compared to the satellite and field samples. We use the best fit relations to quantify and represent these results as histograms in Fig. \ref{fig:edge-histograms}.} %Compared to $r_{\rm e}$ where there is no clear distinction between the radii for the different samples given the larger scatter, the edge radii of Fornax galaxies %and \ref{fig:density-histogram}
    %Across the full stellar mass range, almost all the Fornax galaxies lie above the best fit lines. We use the best fit relations to represent these results as histograms in Fig. \ref{fig:edge-histograms}. 
    %This low metallicity value is justified for (nearby) faint galaxy outskirts \citep{2017denija, 2021neumann} and representative of the metal-poor stellar populations found in observations and simulations of dwarfs \citep[see e.g.][]{2023cardona}.
    \label{fig:size-mass-relations}
\end{figure*}

\begin{table*}[h!]
\begin{center}
\caption{Best fit parameters of the radius--stellar mass scaling relations. The estimated dispersion of the scaling relations due to uncertainties in the image background ($\sigma_{back}$) and stellar mass ($\sigma_{mass}$) estimation are also included. These values are used to calculate the intrinsic uncertainty ($\sigma_{int}$) of the scaling relation.}
\begin{tabular}{c c c c c c c c c c}
%\hline
\textbf{Relation} & \textbf{Sample} & \textbf{$M_{\star} (M_{\odot})$} & $\boldsymbol{\beta}$ &  $\boldsymbol{\alpha}$  & $\boldsymbol{\sigma_{obs}}$  & $\boldsymbol{\sigma_{back}}$ & $\boldsymbol{\sigma_{mass}}$ & $\boldsymbol{\sigma_{int}}$ & \textbf{r} \\
\hline \\
\centering 
Edge radii & Cluster & $10^5-10^{11}$ & 0.40$\pm$0.01 & $-$2.88$\pm$0.10 & 0.096 & 0.057 & 0.052 & 0.057  &0.93  \\ 
vs. stellar mass & Group \& satellite & $10^{5.5}-10^{10}$ & 0.38$\pm$0.02 & $-$2.67$\pm$0.12 & 0.12 & 0.064 & 0.066 & 0.077 & 0.94 \\ 
 & LV/Nearly isolated & $10^{5.5}-10^{9}$  & 0.40$\pm$0.05 & $-$2.79$\pm$0.33 & 0.14 & 0.055 & 0.093 & 0.089 & 0.87  \\ 
& C22/Nearly isolated & $10^7-10^{10}$ & 0.32$\pm$0.03 & $-$2.07$\pm$0.24 & 0.12  & 0.052 & 0.079 & 0.073 & 0.85  \\ \\  
$R_{\rm edge} \sim M_{\star}^{\beta}$ & \textit{Total Sample} & $10^5-10^{11}$ & 0.42$\pm$0.01 & $-$2.94$\pm$0.07 & 0.13 & 0.057 & 0.066 & 0.068 &  0.94 \\ \\ \hline \\ 

Effective radii & Cluster & $10^5-10^{11}$ & 0.27$\pm$0.02 & $-$2.11$\pm$0.14 & 0.15 & 0.073 & 0.080 & 0.10 & 0.79\\ 
vs. stellar mass & Group \& satellite & $10^{5.5}-10^{10}$ & 0.27$\pm$0.03 & $-$2.15$\pm$0.18 &  0.19 & 0.074 & 0.076 & 0.16 & 0.78\\ %\hline 
& LV/Nearly isolated & $10^{5.5}-10^9$ & 0.29$\pm$0.07 & $-$2.29$\pm$0.48 & 0.18 & 0.087 & 0.061 & 0.15 & 0.68\\ %\hline
& C22/Nearly isolated & $10^7-10^{10}$ & 0.30$\pm$0.045 & $-$2.41$\pm$0.38 & 0.21  & 0.12 & 0.062 & 0.16 & 0.65 \\ \\ 
$r_{\rm e} \sim M_{\star}^{\beta}$ & \textit{Total Sample} & $10^5-10^{11}$ & 0.28$\pm$0.01 & $-$2.19$\pm$0.10 & 0.18 & 0.087 & 0.074 & 0.14 & 0.80\\ \\ \hline  
\end{tabular} 
\label{table:bestfit}
\end{center}
%\vspace{-2.5mm}
\textbf{Notes:} The relations are of the form $\log_{10}R = \beta\,\log_{10}M_{\star} + \alpha$ where $R$ is the radius of interest. The intrinsic dispersion of the relation is computed using $\sigma_{int} = \sqrt{\sigma_{obs}^2 - \sigma_{back}^2 - \sigma_{mass}^2}$, and $r$ is the Pearson correlation coefficient. If we additionally account for the typical uncertainty in visual identifications of edge radii \citep[0.04\,dex;][]{2022chamba}, then the intrinsic uncertainty in the $R_{\rm edge}-M_{\star}$ relation for the total sample drops to $\sim0.06$\,dex.
\end{table*}

\begin{figure*}[h!]
    \centering
    \includegraphics[width=0.43\textwidth]{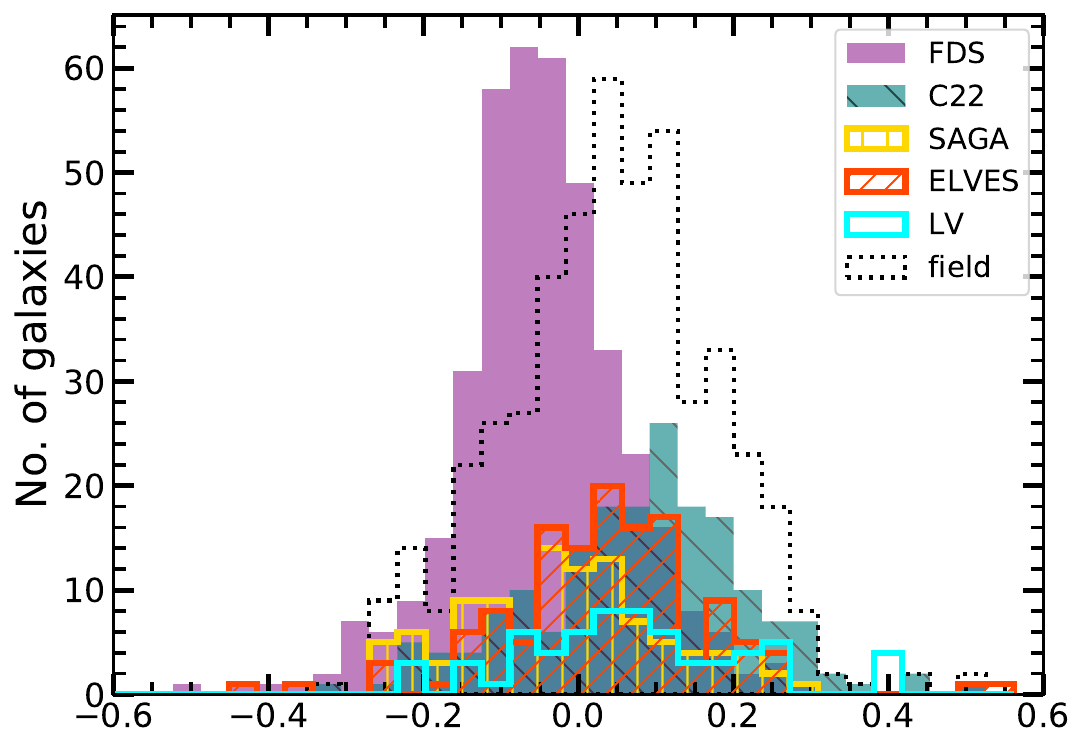}
    \includegraphics[width=0.41\textwidth]{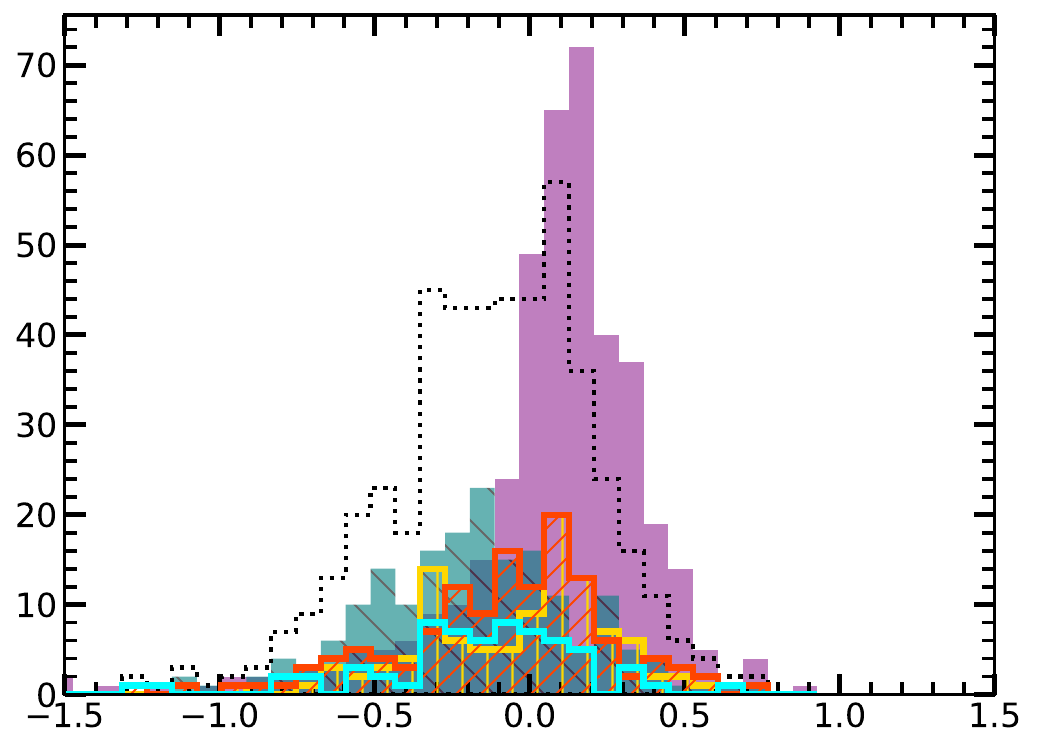}
    \includegraphics[width=0.43\textwidth]{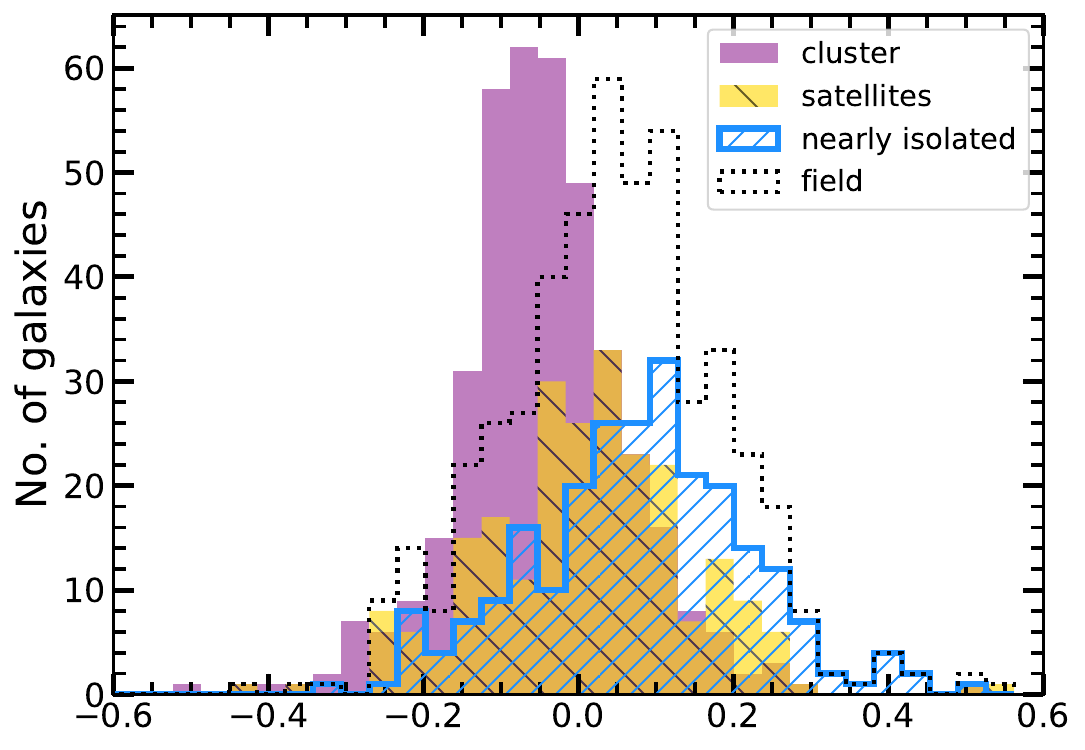}
    \includegraphics[width=0.41\textwidth]{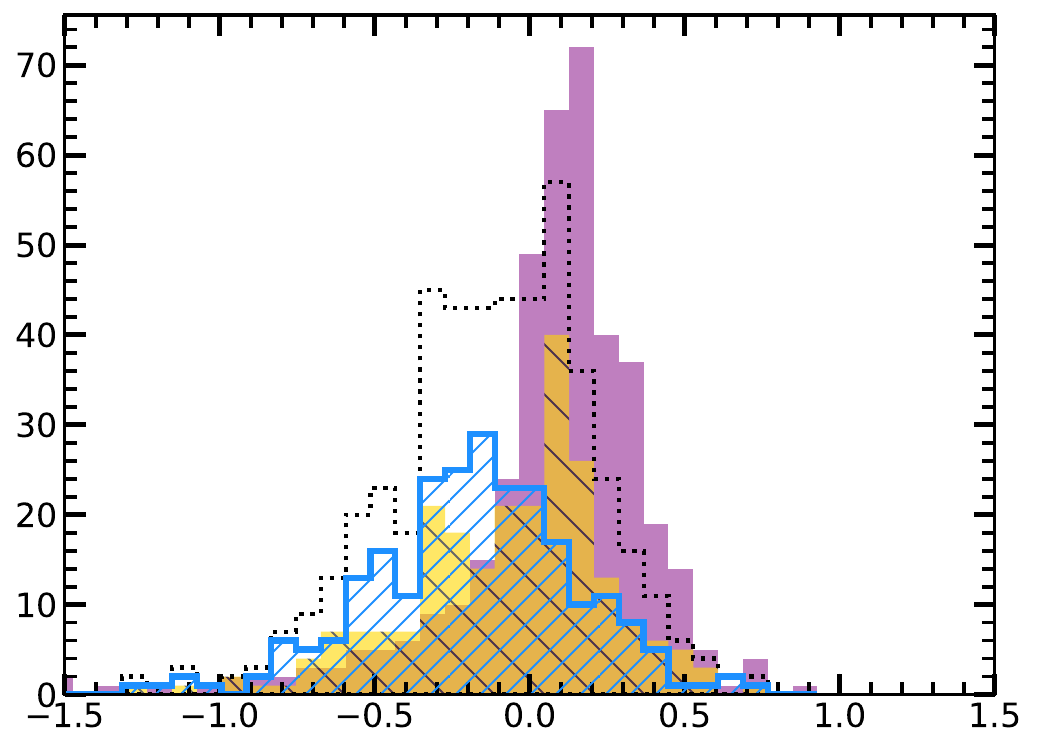}
    \includegraphics[width=0.43\textwidth]{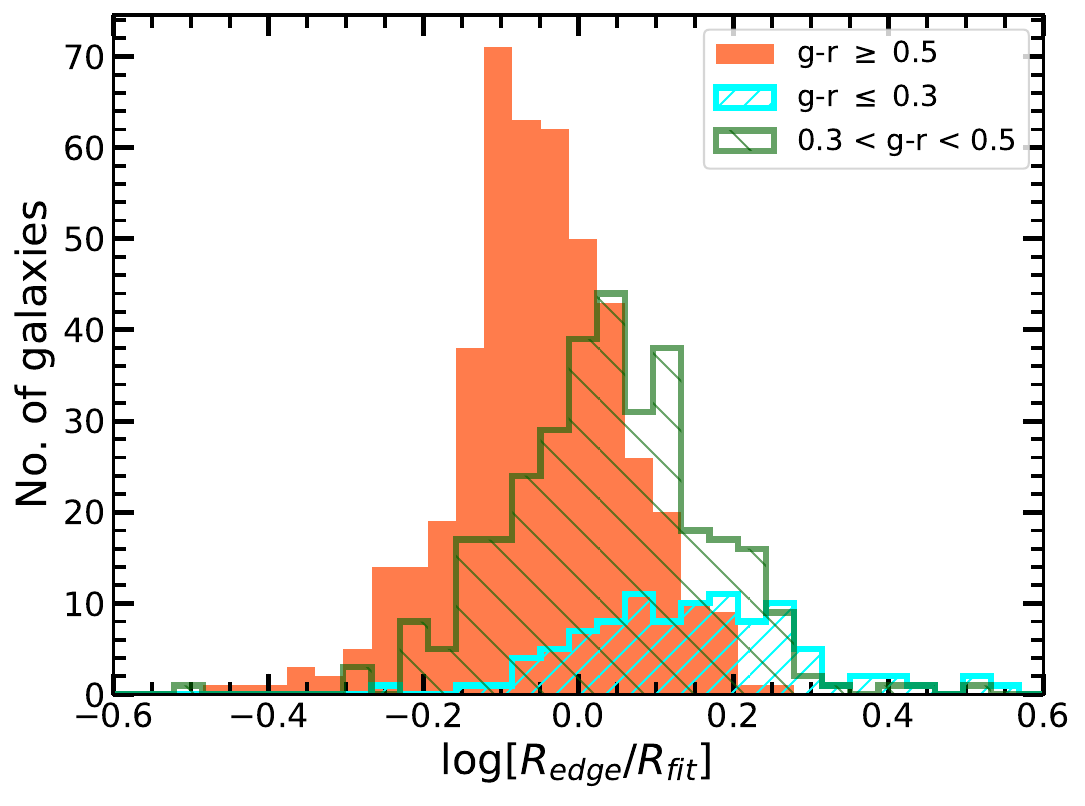}
    \includegraphics[width=0.41\textwidth]{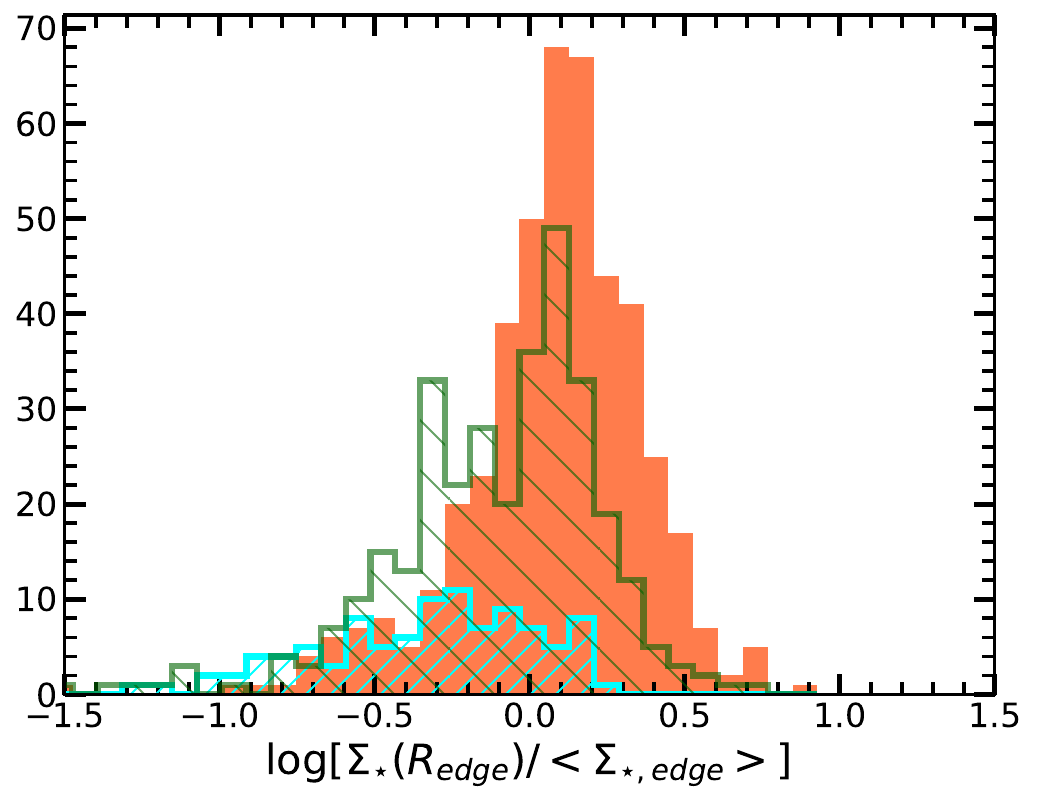}
     \caption{Distribution of edge radii (left) and surface stellar mass densities (right), normalised by the expected value derived from the regression of these measurements against stellar mass ($R_{fit}\sim M_{\star}^{0.4}$ and $<\Sigma_{\star, edge}>$ from \cite{2022chamba} respectively, see Fig. \ref{fig:size-mass-relations}). In the upper and middle panels, Fornax dwarf galaxies are shown in purple. The individual samples considered are separated by the labelled designs (upper). The distribution of the rest of the sample (field + satellites combined) is plotted in black dotted lines.  In the middle panel, we combine the field (\cite{2022chamba} and \citep{2013karachentsev} in blue) and satellite (SAGA \citep{2021yao} and ELVES \citep{2022carlsten} in yellow) samples to further visualise the distinction between their distributions. The lower panels show the distribution of galaxies in the size-- and edge density--stellar mass planes when separated in three bins of $g-r$ colour, showing that redder, older galaxies populate the lower half of the size--mass relation (left) and upper half in the density-stellar mass relation (right).  Quantitatively, the panels show that compared to field galaxies, Fornax cluster galaxies can be up to 50\% smaller and on average, redder, older and two times denser at the edge.}
    \label{fig:edge-histograms}
\end{figure*}

\begin{figure*}[h!]
    \centering
    \includegraphics[width=0.49\textwidth]{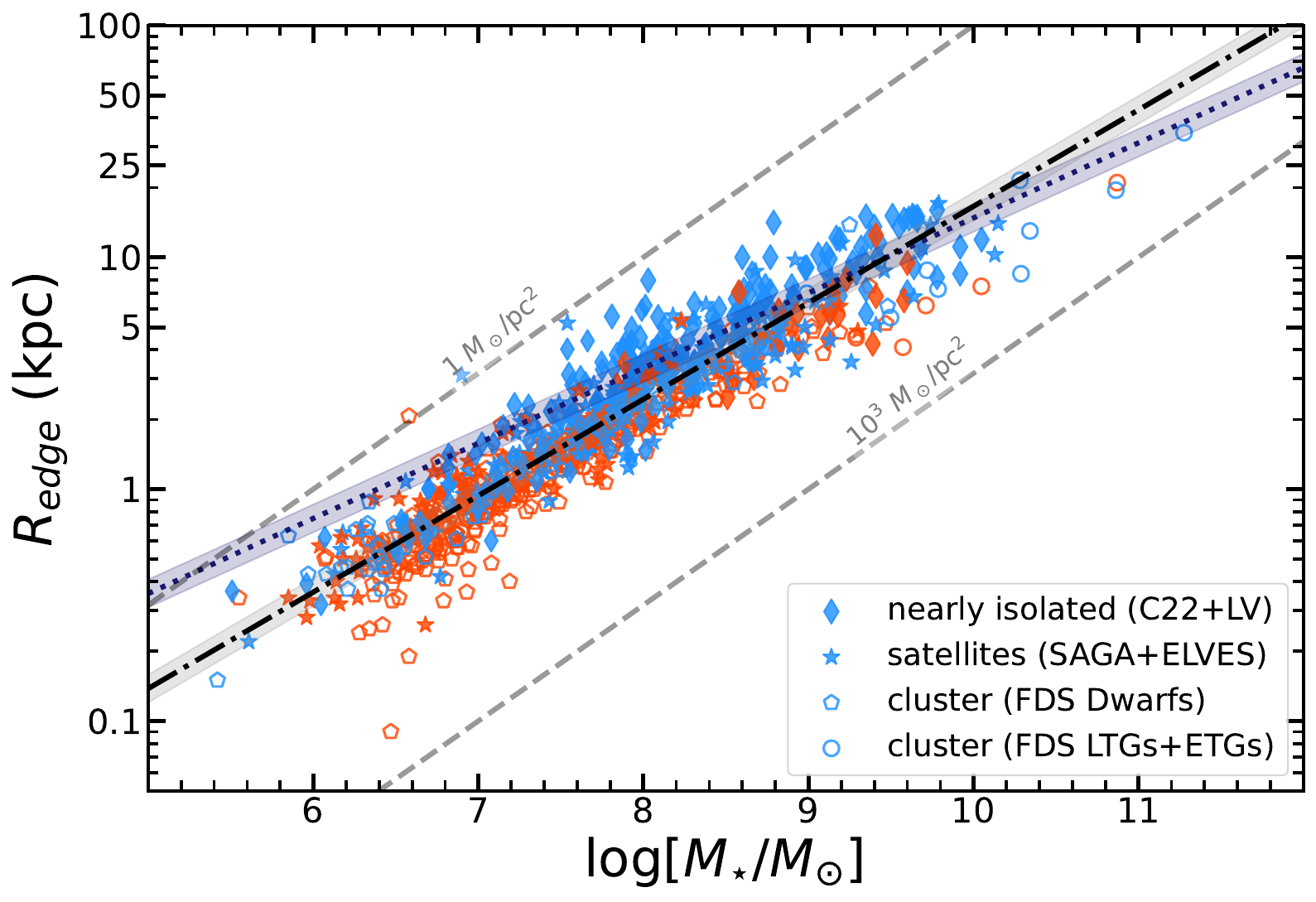}
    \includegraphics[width=0.49\textwidth]{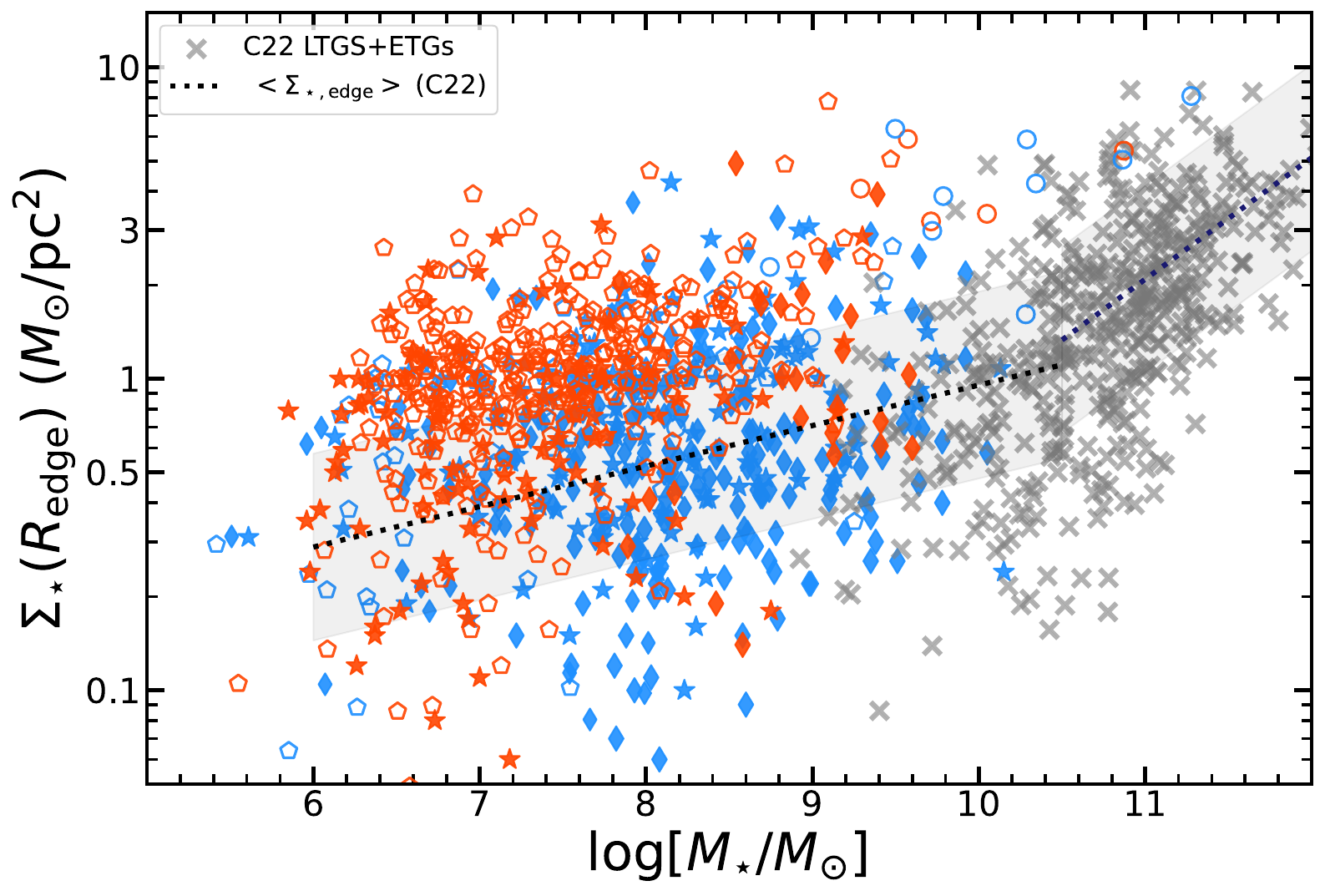}    
    \caption{Scaling relations colour coded by morphology. Early-type galaxies are coloured in red are smaller on average than the late-types in blue across the stellar mass range studied. The dotted and dashed lines are the same as those previously plotted in the upper panels of Fig. \ref{fig:size-mass-relations}.}
    \label{fig:edge-morph}
\end{figure*}

%The best-fit relations for $r_{\rm e}$ are also plotted.

%\todo{TO DO: Remove x-axis label in upper panel. Instead show the histogram distribution of mass in each edge-stellar mass bin above the plot as a separate panel. Do the same on the y-axis. Consider splitting panels in two seperate figures? Maybe Fig. 5 (upper) + Fig 6 and Fig. 5 (lower) + Fig 7?}

\subsection{Radial profiles}

We display the identified edges over plotted on the $gri$-colour composite images and radial profiles derived in this work for a number of galaxies in our sample in Fig. \ref{fig:elves-examples} and \ref{fig:fds-examples}. To highlight the inflection point we mark as the edge feature, we additionally plot a least squares polynomial fit of degree four in the colour and stellar surface density profile panels as a solid line. Using a polynomial of degree three does not significantly change these fits. We only plot them in these panels for visualisation purposes. We do not use these fits for any of our measurements. While we show these data only for a handful of galaxies as in Fig. \ref{fig:sdss-examples}, we make the profiles and measurements made in this work publicly available as catalogues associated with this publication. 

%Colour composite images of more galaxies are shown in the Appendix Figs. \ref{fig:fds_examples} and \ref{fig:elves_examples}. \par 

\subsection{Scaling relations}
The resulting size--stellar mass relation using the edge locations are shown in the left panel of Fig. \ref{fig:size-mass-relations}. The left panel of the same figure shows the associated edge density-stellar mass relation. The best fit relations in each panel are also shown along with their intrinsic dispersion ($\sim$ 0.06\,dex for the edge radii) in a lighter shade. The samples considered in this work are labelled in the legend: Fornax \citep[][dwarfs in purple pentagons and early- and late-type galaxies in dark blue circles]{2021su}, field \citep[][in blue diamonds]{2013karachentsev, 2022chamba} and satellites \citep[][in yellow stars]{2021yao, 2022carlsten}. The same relations  are colour coded using the global $g-r$ colour of the galaxy in the middle panels) and estimated age (see Sect. \ref{sect:methods}) in the lower panels. All the best fit relations used and quoted in this work are provided in Table \ref{table:bestfit}. \par 

\noindent The main features of these relations are: \\

\textbf{Edge radii-mass:} First, all the dwarfs with stellar masses $<10^{7.5}\,M_{\odot}$ collectively deviate from the 1/3 slope reported in \cite{2020tck} and \cite{2022chamba} for field galaxies. We find that the galaxies follow a size--mass relation with a higher slope of 0.42$\pm$0.01. \par 
Second, the observed scatter in the relation is 0.13\,dex across the full mass range analysed. After accounting for the uncertainties due to background subtraction and stellar mass estimation as listed in Table \ref{table:bestfit}, the dispersion of the relation reduces to 0.068\,dex. Additionally removing the typical uncertainty from repeated identifications found in previous work \citep[$\sim $0.04\,dex;][]{2022chamba} lowers this value to $\sim$0.055\,dex, consistent with the findings from \cite{2020tck}. \par 
Third, on average, the Fornax dwarfs are 0.04 dex below the mean edge--mass relation, while the rest of the sample is 0.05 dex above the relation. This deviation is highlighted in the histograms plotted in upper and middle panels of Fig. \ref{fig:edge-histograms}. The lower panel separates the total sample in three bins of the global $g-r$ colour of the galaxies: $g-r < 0.3$, $0.3 < g-r < 0.5$ and $g-r > 0.5$. \par 
Fig. \ref{fig:edge-histograms} shows that Fornax galaxies are on average 20\% smaller compared to galaxies outside a cluster. \textit{If we consider only the nearly isolated sample, then Fornax cluster galaxies are on average 25\% smaller. When considering the group satellites, the Fornax dwarf galaxies are 13\% smaller than this sample. Within a 1$\sigma$ interval, Fornax galaxies can be up to 50-60\% smaller in edge radius than those in the field} (Fig. \ref{fig:edge-histograms}, middle panel). In this sense, the satellite galaxies have intermediate sizes compared to the Fornax cluster and nearly isolated galaxies.  \par 

Fourth, the lower panels of Fig. \ref{fig:size-mass-relations} and Fig. \ref{fig:edge-histograms}, show that \emph{at a fixed stellar mass,  younger and bluer galaxies are larger}. In Fig. \ref{fig:edge-morph}, we plot the same relations as in Fig. \ref{fig:size-mass-relations} colour coded by morphology. We find that this colour bifurcation is a reflection of morphology. In other words, bluer, late-type galaxies are larger than the redder, early types. This result is similar to that found in \cite{2022chamba} for the late-type spiral galaxies (Sect. \ref{sect:field-edges}) but here we have extended it to even lower mass galaxies $M_{\star} < 10^9\,M_{\odot}$.

%we colour code the galaxies in the edge-mass relation according to the global $g-r$ colour (middle) and estimated age (lower, see Sect. \ref{sect:methods}). 
%Fifth,  \par  (Fig. \ref{fig:edge-histograms}, left panels). 

Fifth, the five Fornax massive ETGs studied here are below the best fit $M^{1/3}$ line. All except one of the massive LTGs with $M_{\star} > 10^9\,M_{\odot}$ are also below this line. This result implies smaller sizes relative to the field sample studied in \cite{2022chamba}.  \par
%, suggesting that Fornax dwarf galaxies reached their present-day size earlier than those in the field.

\textbf{Edge surface stellar mass density-mass:} In relation to the above results is the edge surface stellar mass density -- stellar mass relation, first published in \cite{2022chamba}. We plot the values measured in this work for the galaxies analysed here in the right panels of Fig. \ref{fig:size-mass-relations}. The average edge density--mass relations from \citep{2022chamba} for their field sample as well as the measurements for the more massive, luminous galaxies are over-plotted for reference. The same colour and age bifurcation we report in the edge radii--stellar mass plane is also observed in the edge surface stellar mass density-stellar mass relations (middle panels of Fig. \ref{fig:size-mass-relations}). \par 
We find that the majority of Fornax and satellite galaxies populate the upper half of the field relations. This region in the edge density--stellar mass plane systematically corresponds to redder, older and higher edge densities at a fixed stellar mass compared to the average values found in nearly isolated environments. In other words, \emph{the edges of cluster and satellite galaxies are denser compared to those found to be near isolated. On average, the edges of the Fornax (satellites) galaxies are two (1.5) times denser than in isolation.} \par 
Most of the galaxies with smaller sizes and higher edge densities in Fornax are of early-type morphology (Fig. \ref{fig:edge-morph}.) Our results are therefore a reflection of the different edge properties found in early- and late-type galaxies. However, we further investigate this issue in Sect. \ref{subsect:morph-environment} and demonstrate that the edge radii of late-type galaxies are also impacted by the environment. \par 

The edge densities of the more massive Fornax galaxies are higher than $\sim 3\,M_{\odot}$/pc$^2$ in many cases. If we use the best fit edge density-stellar mass relations from \cite{2022chamba}, the edge density values of the massive Fornax ETGs and LTGs is approximately a factor of four times higher than the mean edge density values of the C22 field galaxies with the same stellar mass. While we cannot perform a statistical analysis on the massive end of our sample in this work, these results suggest that the observation of higher edge densities in the Fornax Cluster occurs across the full mass range considered here. \par \bigskip

%\bigskip 

%\todo{TO DO: Put all slopes as decimals if not exact (including in Figures). Give uncertainties in the slopes. Provide y intercepts in appendix. Investigate standard errors of mean, not of distributions.}

\begin{figure}[h!]
    \centering
    \includegraphics[width=0.5\textwidth]{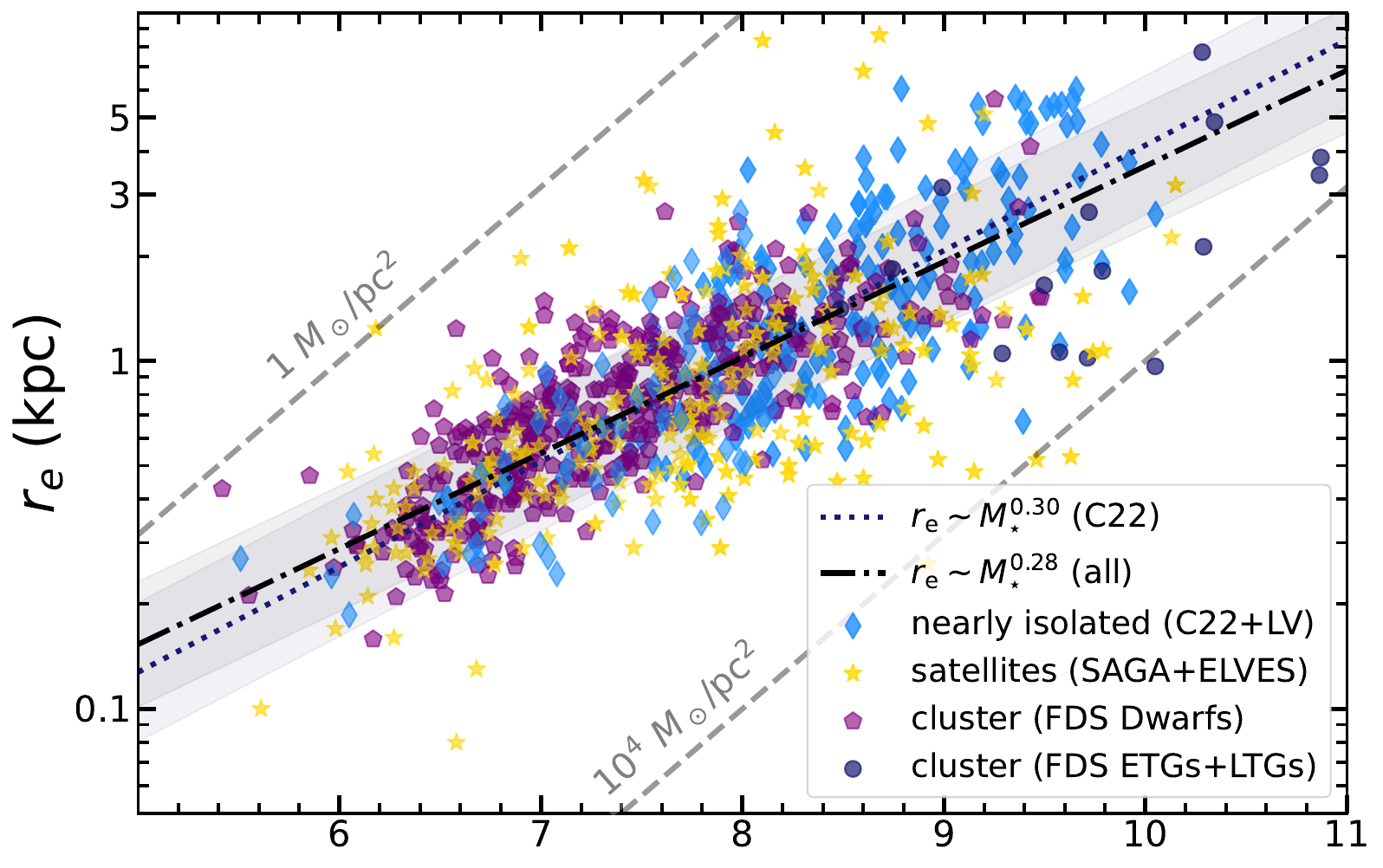}
    \includegraphics[width=0.5\textwidth]{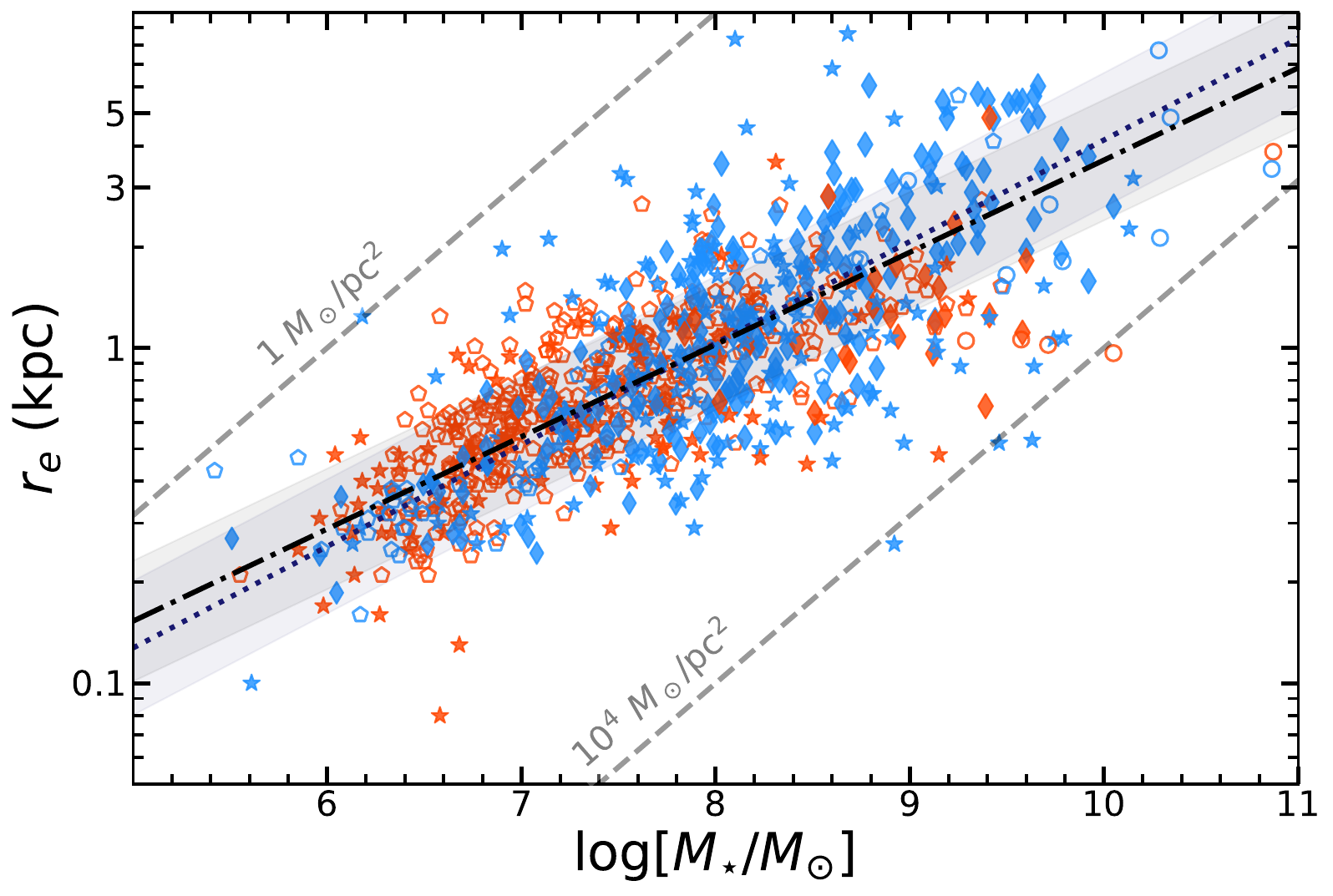}
    \caption{Effective radius--stellar mass plane for galaxies in different environments. \textit{\textbf{Upper:}} The data points are labelled as in Fig. \ref{fig:size-mass-relations}. \textit{\textbf{Lower:}} Similar to Fig. \ref{fig:edge-morph}, we colour code the effective radius -- stellar mass relation using red for the early-types and blue for the late-type galaxies. In both panels the shaded region corresponds to the $0.2$\,dex scatter for reference.}
    \label{fig:effective-radius-mass-plane}
\end{figure}

\textbf{Effective radii-mass:} For comparison with the above results, we show the well-known effective radii-stellar mass scaling relation in Fig. \ref{fig:effective-radius-mass-plane}. The upper panel colour codes the data points according to the environment and the lower panel according to our early and late-type categories as in Fig. \ref{fig:edge-morph}. As before, the best fit relations are also plotted with the 0.2\,dex scatter shaded for reference. \par 

\begin{figure}[h!]
    \centering
    \includegraphics[width=0.45\textwidth]{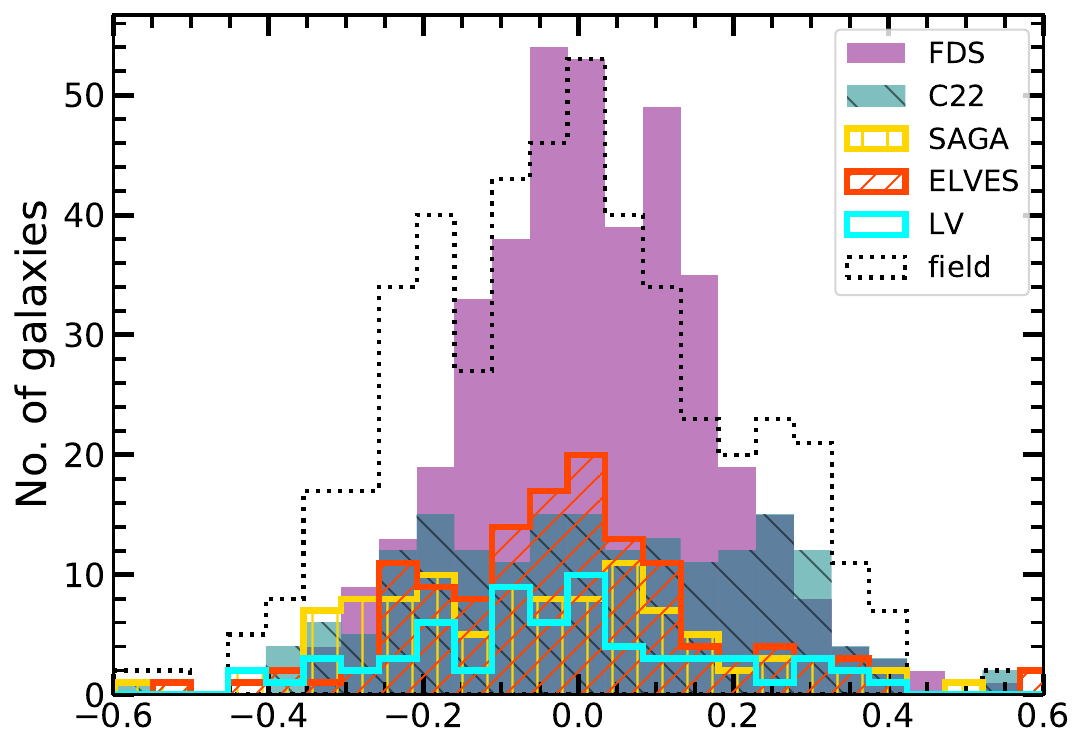}
    \includegraphics[width=0.45\textwidth]{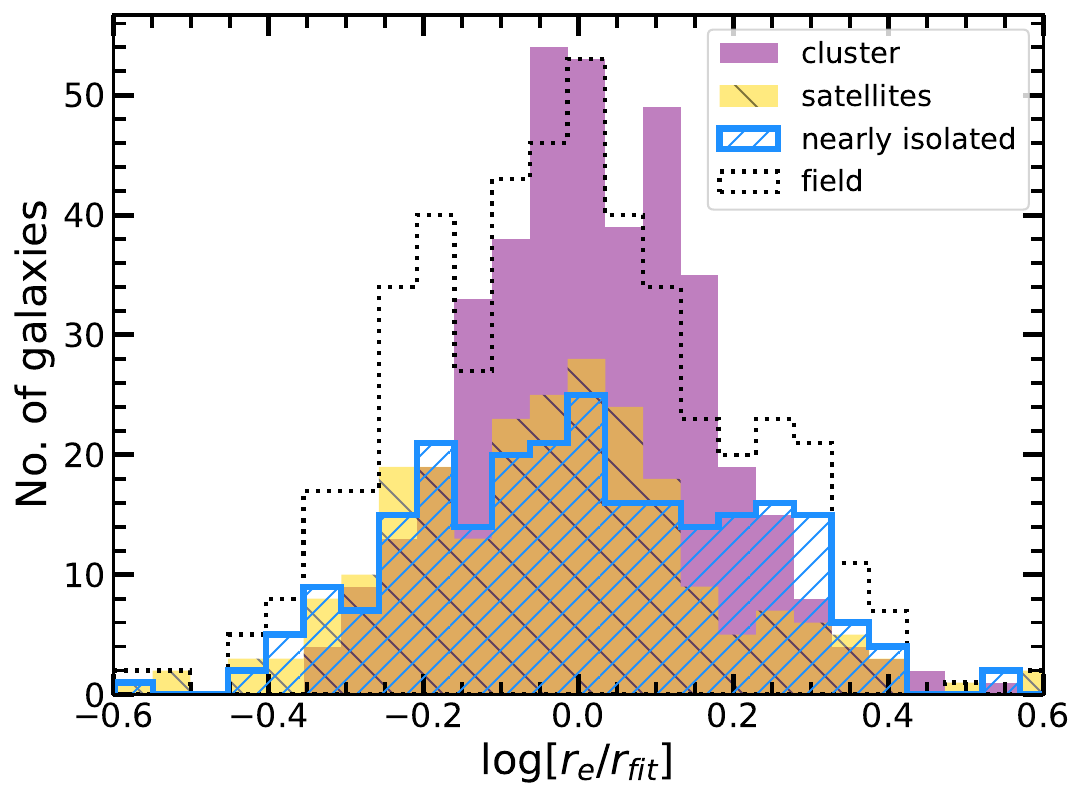}
     \caption{Distribution of half-light radii, normalised by the expected value derived from the regression of $r_{\rm e}$ against stellar mass ($r_{fit} \propto M_{\star}^{0.28}$; see Fig. \ref{fig:effective-radius-mass-plane}). As in Fig. \ref{fig:edge-histograms}, Fornax dwarf galaxies are shown in purple. The upper panels shows the distribution of individual samples separated by the labelled designs as in the left of Fig. \ref{fig:edge-histograms}. The lower panel shows the full satellites (yellow) and the combined nearly isolated + satellites samples (blue). The total field sample is plotted in dotted black lines. There is no obvious distinction between the samples in the effective radius--stellar mass plane.}
    \label{fig:re-histogram}
\end{figure}

First, the slope of the $r_e$--mass plane considering only the field sample is 0.25$\pm0.01$ while with all the dwarfs it is 0.28$\pm0.01$. These values are very similar, with the slope slightly higher when the satellites and Fornax galaxies are included. However, the scatter is double that of the edge--mass relation. \par 
Second, \textit{there is no clear separation of the sub-samples in effective radius--mass plane.} We verify this point by using the histograms shown in Fig \ref{fig:re-histogram}. This figure shows the distribution of radii at a fixed stellar mass for individual sub-samples and combined samples as in Fig. \ref{fig:edge-histograms}. The result further emphasises the fact that the effective radius has previously led to ambiguous results on the impact of the environment on sizes. \par 
Third, although with a larger scatter, similar to the edge radii, almost all the more massive sub-sample of Fornax LTGs and ETGs studied here are below the best fit lines which implies smaller effective radii relative to the field sample studied in \cite{2022chamba}. \par
%Fourth, galaxies with $M_{\star} < 10^7\,M_{\odot}$ and early-type morphology appear to have larger effective radii compared to the late-types. This trend appears to reverse when $M_{\star} > 10^{8.5}\,M_{\odot}$. \par \bigskip 

\subsection{Comparison by morphology and environment}
\label{subsect:morph-environment}

\begin{figure}[h!]
    \centering
    \includegraphics[width=0.491\textwidth]{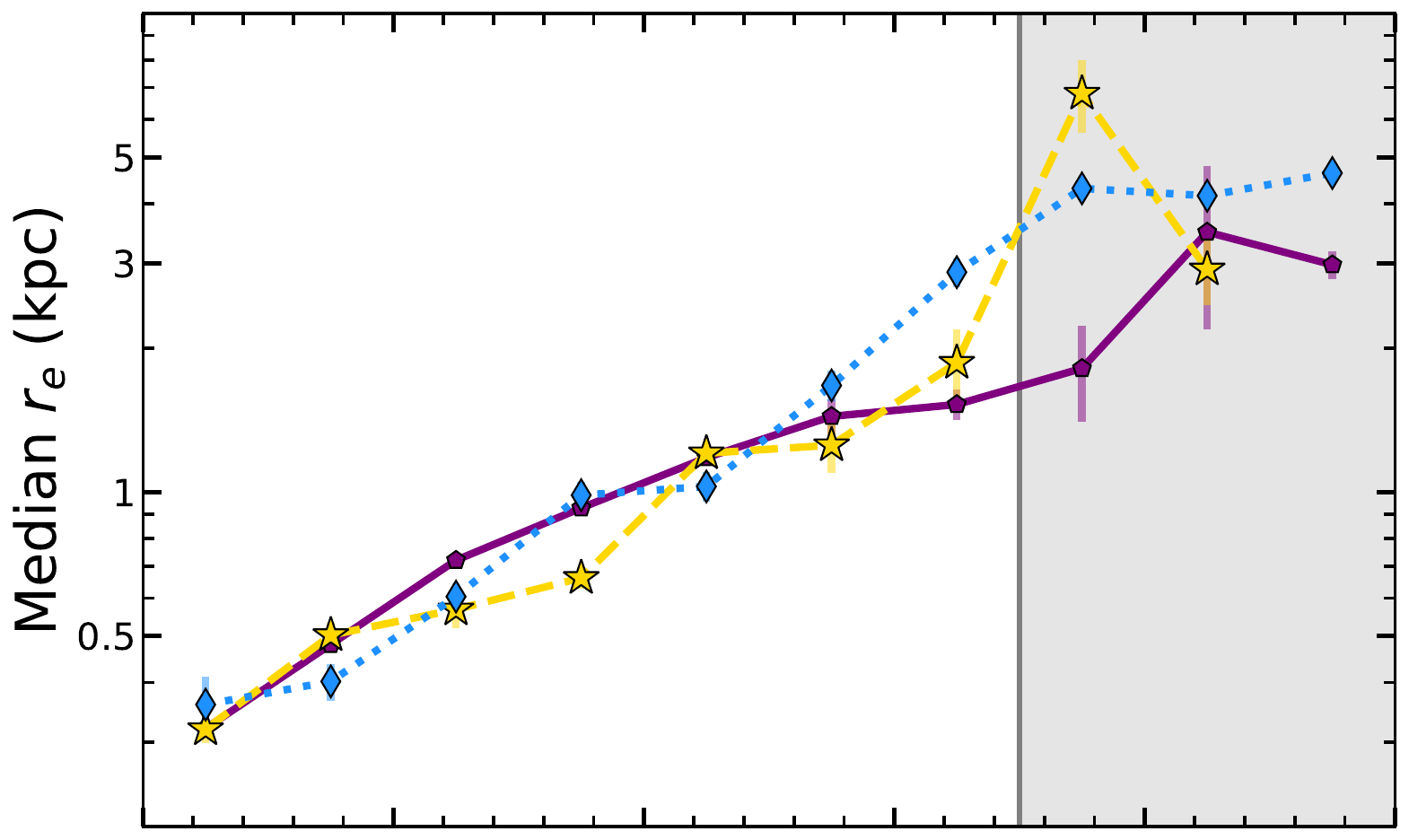}
    \includegraphics[width=0.5\textwidth]{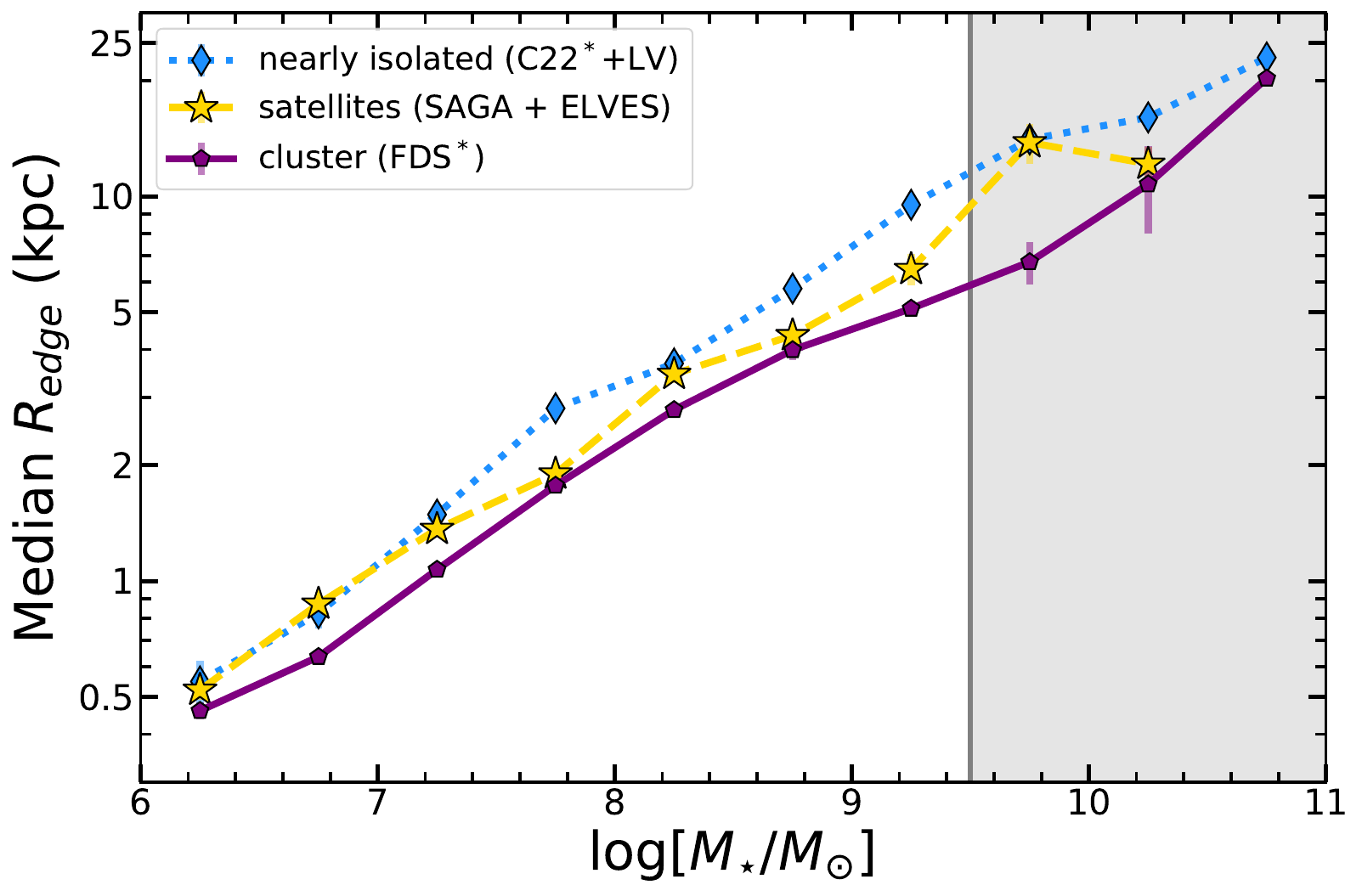}
    \caption{Median effective (upper) and edge radii (lower) in bins of stellar mass. The distribution of the number of galaxies used to compute the median radii in each stellar mass bin was plotted in Fig. \ref{fig:mass_dist}. The shaded grey region corresponds to the stellar mass regime beyond which we lack cluster and satellite galaxies for a meaningful, statistical comparison. However, we compute the median points with the few galaxies in our sample for completeness. The separation between the Fornax cluster and field sample is much more clear in median edge radii: the cluster galaxies are systematically smaller in edge radii across the full stellar mass range considered, which is not reflected by the $r_e$ measurements.}
    \label{fig:comparison-wre-medians}
\end{figure}

\begin{figure}[h!]
    \centering
    \includegraphics[width=0.48\textwidth]{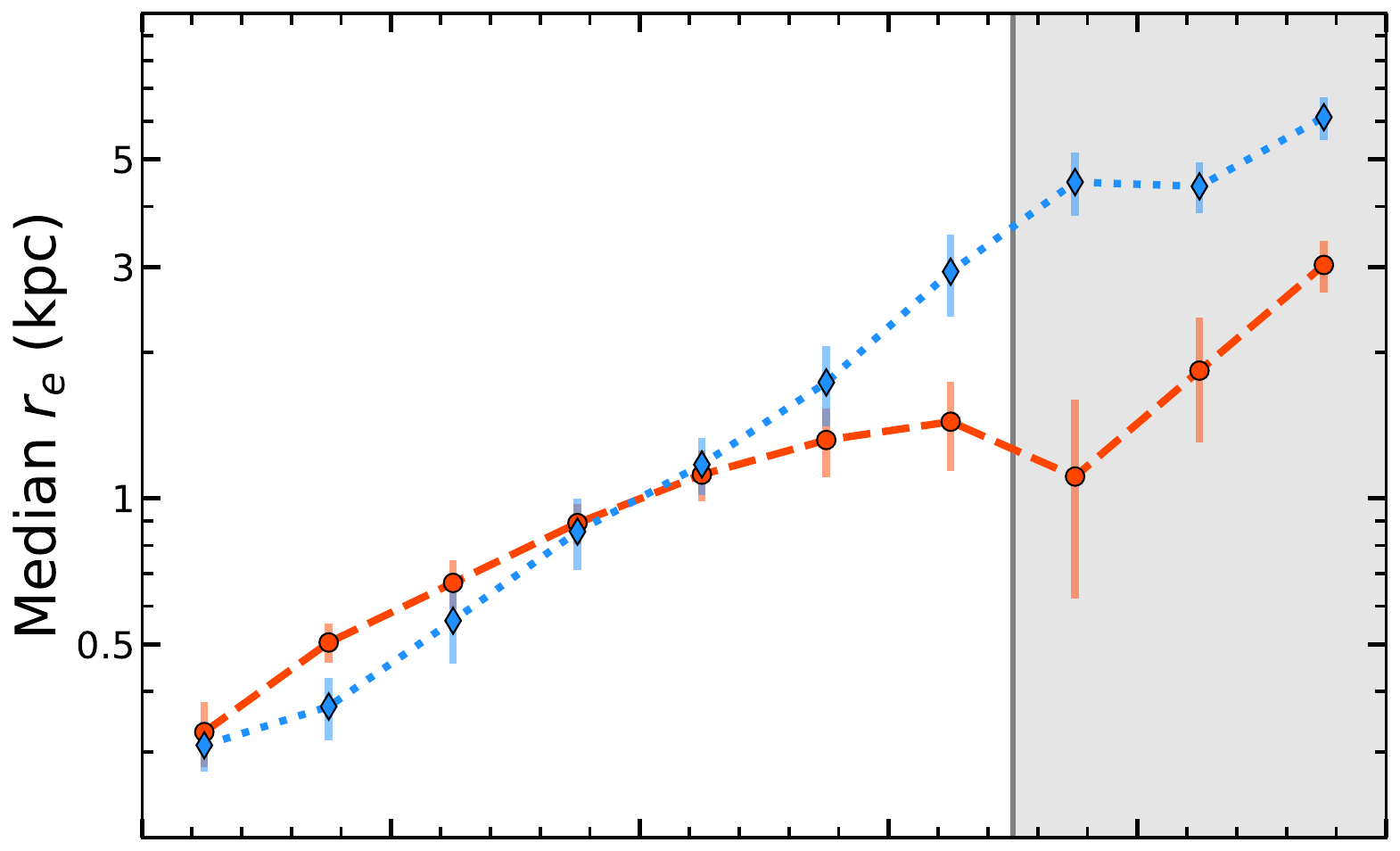}
    \includegraphics[width=0.494\textwidth]{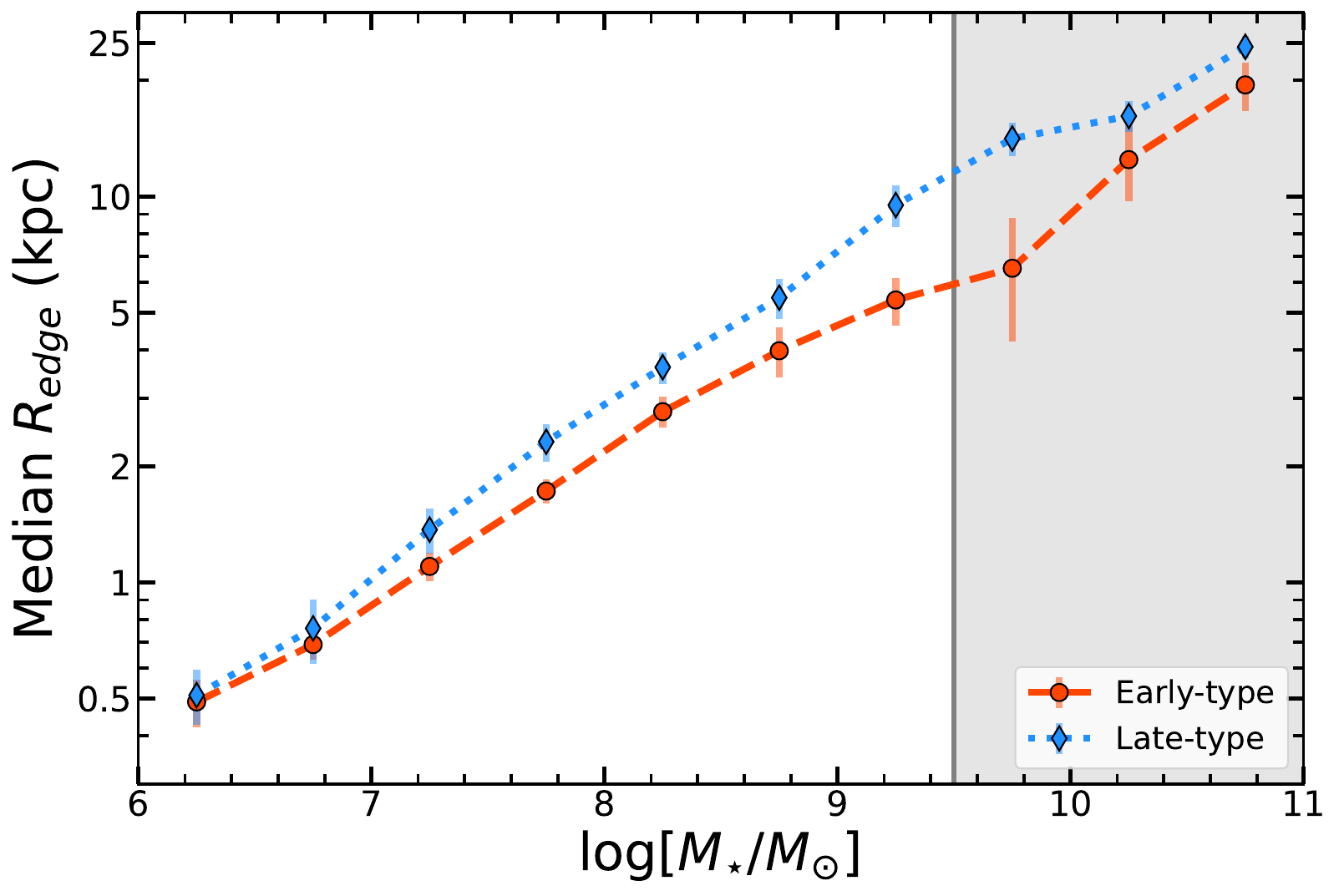}
    \caption{Similar to Fig. \ref{fig:comparison-wre-medians}, the median radii for early and late-type galaxies in our total sample. The error bars correspond to three times the dispersion in the estimated median value. Early type galaxies are larger in effective radius when $M_{\star
    } \lesssim 10^{8}\,M_{\odot}$ compared to the late-types (upper). However, in their edge radii (lower), these galaxies are systematically smaller by $\sim 20\%$ than the late-types across the full stellar mass range.}
    \label{fig:total_median_early_late_types}
\end{figure}

\begin{figure*}[h!]
    \centering
    \includegraphics[width=0.49\textwidth]{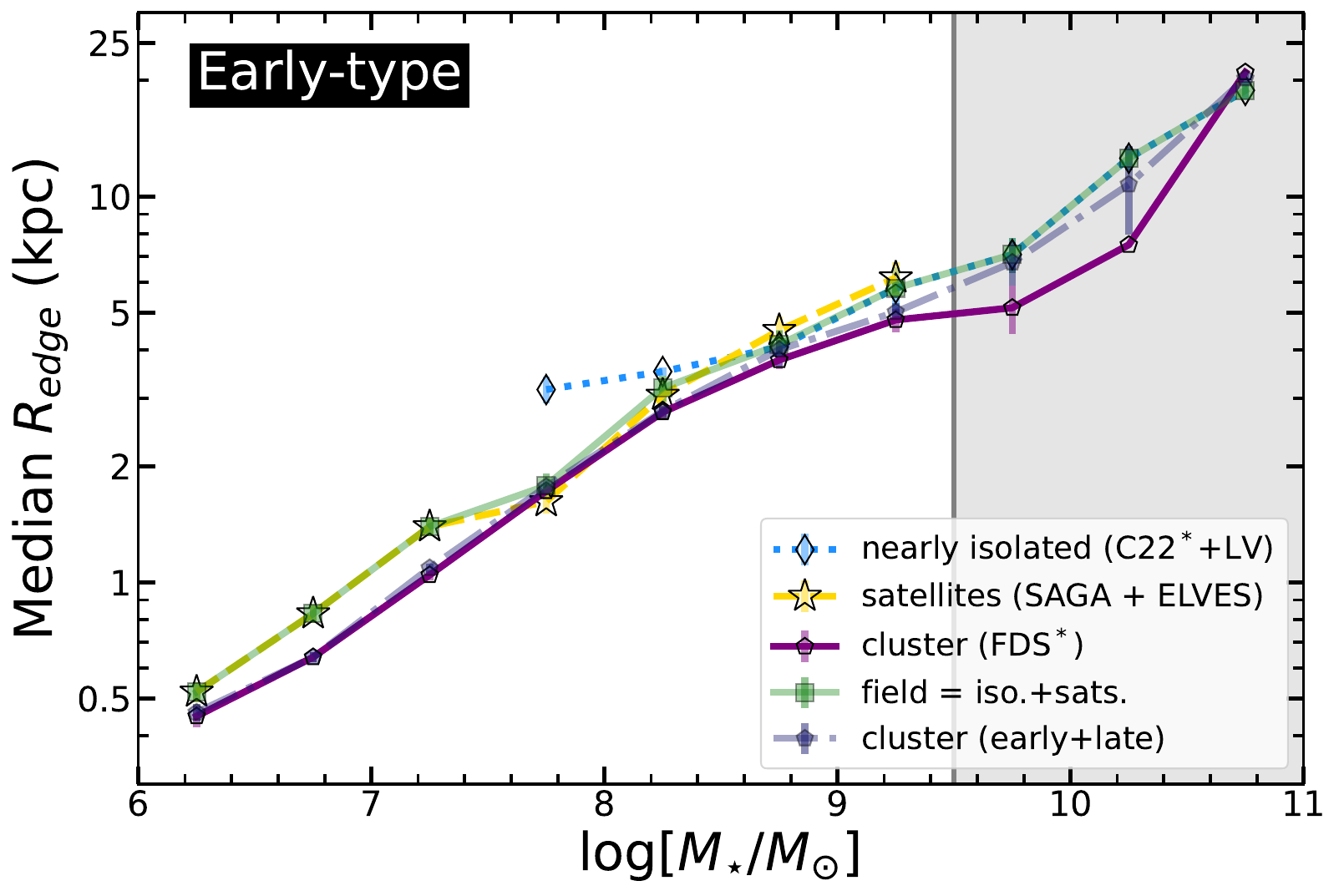}
    \includegraphics[width=0.468\textwidth]{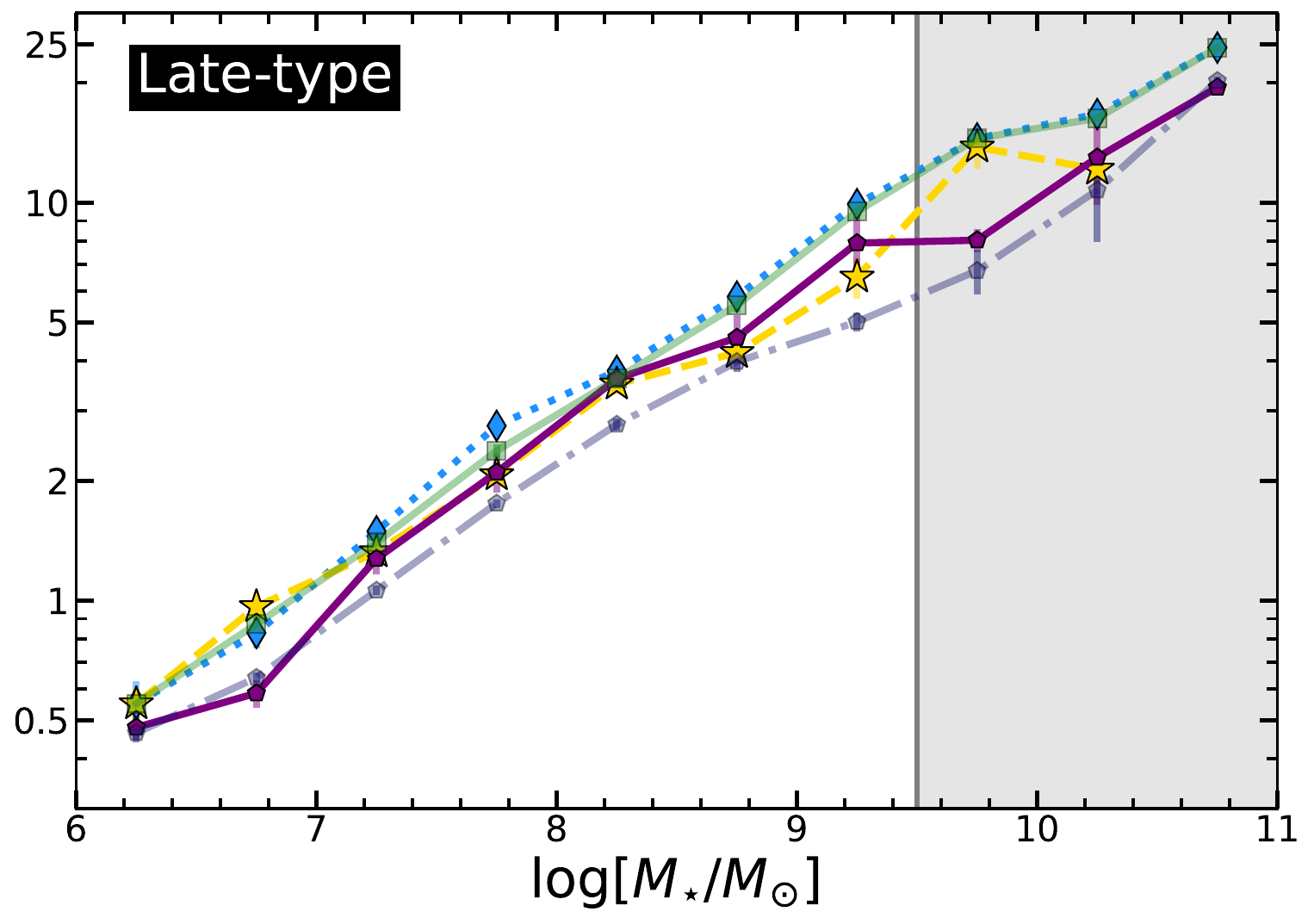}
    \caption{Similar to Fig. \ref{fig:comparison-wre-medians} now separated by morphology and environment. The early-types are plotted on the left and late-types on the right. In both panels we plot those galaxies in the field by combining the nearly isolated and satellite samples (green squares). We also plot the combined line for the cluster using the full sample of early- and late-type galaxies (the same shown in Fig. \ref{fig:comparison-wre-medians}) in the dot-dashed line. Both the early and late-type galaxies are smaller in the Fornax cluster compared to the field. However, the early-type galaxies are even smaller than the late-types, regardless of the environment (Fig. \ref{fig:total_median_early_late_types}).}
    \label{fig:morph_environment}
\end{figure*}

We reaffirm the above results in Fig. \ref{fig:comparison-wre-medians}, \ref{fig:total_median_early_late_types} and \ref{fig:morph_environment} where we plot the median radii in bins of stellar mass to further analyse the differences in our sample as a function of their environment and morphology. In all figures, the median radii are computed in bins of stellar mass with bin size 0.5\,dex and the plotted error bars are the 3$\sigma$ clipped standard error. Our results do not change if we consider three or five $\sigma$ clipped median and error values. The grey shaded region in each panel corresponds to the stellar mass regime where we lose statistics due to the low number of massive galaxies in the satellites and Fornax Cluster samples (see Fig. \ref{fig:mass_dist} and associated text). \par 
In Fig. \ref{fig:comparison-wre-medians} we plot the median effective radius  (upper panel) and the same for the edge radii (lower) for all the Fornax cluster galaxies (purple pentagons), group satellites (yellow stars) and isolated samples (blue diamonds).  We find that when $M_{\star} \lesssim 10^9\,M_{\odot}$, the median $r_e$ of Fornax cluster galaxies are comparable to the isolated and satellite samples. This trend appears to switch at higher stellar masses where the median radii of Fornax galaxies are smaller. We point out, however, that our higher mass data points (grey shaded) suffer from low statistics (see Fig. \ref{fig:mass_dist}). In Appendix \ref{app:reff_difference} we show that our conclusions based on the use of the effective radii to study Fornax galaxies are independent of the method we used to derive them by comparing these relations with that derived from \citet{2021su}. \par

If the large scatter in the $r_e-M_{\star}$ relation could be explained by differences in the sample environment, then it is reasonable to expect that the ratio between the median radii at a fixed stellar mass for different environmentally selected samples would have to be at least of the order of the intrinsic dispersion of the relation, i.e. $10^{\pm\sigma_{int}}$ which is $\sim$0.72 for smaller galaxies below the best fit scaling relation or $\sim$ 1.4 for larger galaxies above the line. We find that the average ratio between the median Fornax and isolated or satellite data points at each bin where $M_{\star} \lesssim 10^9\,M_{\odot}$ is $0.9-1.0\pm0.3$.  These values imply there is no significant differences in the radii of these galaxies.\par

In contrast, the median edge radii is systematically larger for the isolated and satellite samples compared to the Fornax cluster galaxies across the full stellar mass range. The average ratio between the data points at each bin where $M_{\star} \lesssim 10^9\,M_{\odot}$ is 0.7-0.8$\pm$0.12. If the scatter in the edge radii--stellar mass can be explained by the environmental differences in our sub-sample, then the expected difference in their radii using the intrinsic scatter of the relation is $\sim$0.86. Therefore, compared to the $r_e$, we can conclude that the edge radii--stellar mass relation is capable of separating the radii of galaxies in different environments at a fixed stellar mass. \par 

To further analyse which galaxies are driving these trends by morphology, in Fig. \ref{fig:total_median_early_late_types}, we separate our sample in to the early and late-type categories and compute their median radii. While the early-type galaxies are consistently smaller than the late-types across stellar mass in their median $R_{\rm edge}$, the trend is similar in effective radius only for galaxies with $M_{\star} > 10^{8.4}\,M_{\odot}$. This trend reverses for lower mass galaxies as mentioned earlier. \par 
\citet{2021carlsten} have already analysed the effective radius--mass plane in detail in the same stellar mass range as in our work and their results are very similar (see their Figs. 5, 6 and 9 in particular). However, they point out that the galaxies in their stellar mass bin $10^6\,M_{\odot} < M_{\star} < 10^7\,M_{\odot}$  suffer from incompleteness in the surveys due to the limiting surface brightness of the images (their Fig. 2 and discussion therein). As we are using a sub-sample of their survey in this stellar mass regime, this argument also applies here. It is therefore currently difficult to interpret the difference in radii in our lowest stellar mass bins $M_{\star} < 10^7\,M_{\odot}$ physically as we are likely missing the faint, early and late-type galaxies in our sample. Therefore, we do not repeat the same analysis as \citet{2021carlsten} and focus on our edge radii measurements for galaxies. \par 

We repeat the same analysis as that presented in \citet{2022chamba} Appendix B2 to examine the surface brightness near $R_{\rm edge}$ and find that it is indeed the faintest galaxies in our low mass sample ($\mu_r(0) \sim 26\,$mag/arcsec$^2$ and $M_{\star} < 10^7\,M_{\odot}$) that approach the limiting depth of the images. This result further supports our choice for leaving the analysis of edges in LSB galaxies using future deeper datasets (see Sect. \ref{sect:data}). However, given that we find only five LSB-type galaxies are in our sample which have size measurements likely impacted by the surface brightness limit, we do not remove these galaxies in our analysis as doing so do not change our main results. \par 

Nevertheless, in the median edge radii panels, we find that the early-type line is driven by our cluster sample where $M_{\star} < 10^{9.5}\,M_{\odot}$. This point is more clearly visible in Fig. \ref{fig:morph_environment} where we further separate the early and late-types according to their environment. We plot the cluster line from the lower panel of Fig. \ref{fig:comparison-wre-medians} which includes both the early and late-type sample for comparison. While the late-type cluster galaxies are larger than the early-types, these galaxies are still systematically smaller than those found in the isolated environment. The average ratio between the median Fornax and isolated data points at each bin for both the early- and late-types are 0.8$\pm$0.1, i.e. a difference of about $\sim 20\%$. However, both the cluster early and late types are closer to the satellite galaxies with a ratio 0.9$\pm$0.1. \par 

We additionally plot the median lines after grouping the satellite and isolated early- (left) and late-type (right) samples as `field'. Both the early and late-type galaxies are smaller in the cluster compared to the field. The average ratio between the cluster and field median radii for both the early and late-type samples are very similar $\sim$ 0.8$\pm0.1$. Although we caution the reader that our sample in the field lacks early-type, nearly isolated galaxies with $M_{\star} < 10^{7.5}\,M_{\odot}$. Nevertheless, we may conclude that the median edge radii at fixed stellar mass for both the early and late-type galaxies depends on the environment. \par 
In other words, \emph{we find that both the early and late-type galaxies in the cluster environment are systematically smaller compared to similar galaxies in the field.} However, the early-type galaxies are generally even smaller compared to the late-type galaxies in both the cluster and the field. These results complement those we report at fixed stellar mass without considering the differences in morphology. The intrinsic scatter in the edge radii--stellar mass plane is thus driven by both morphology and environment.\par

%\bigskip 

%r_fornax - r_field -0.12234498626491327 0.7544926506007481
% r_fornax - r_sats -0.05760527473958875 0.8757793994308956
% r_fornax - r_outside -0.08901136457819167 0.8146829651927532
% r_fornax - r_extreme -0.26184922613371264 0.5472059032069917
% r_fornax_extreme - r_extreme -0.3615180868117853 0.43499264436570606

%Show the distribution of mass across the relations in histogram. A top histogram of the scaling relation. 

% \begin{figure*}
%     \centering
%     \includegraphics[width=0.9\textwidth]{figs/rho-edge-relation.pdf}
%      \caption{Edge surface stellar density--stellar mass relation. The piecewise, best-fit relations from \cite{2022chamba} are shown for reference. As before, the symbols represent: Fornax cluster dwarfs (purple pentagons), SAGA \citep{2021yao} and ELVES \citep{2022carlsten} satellites (yellow stars) and sub-sample \cite{2022chamba} and \cite{2013karachentsev} field galaxies (blue diamonds). Fornax cluster massive late-type and early-type galaxies (black circles) and the luminous galaxies from \cite{2022chamba} (grey crosses) are shown for completeness. We use the best fit relation to represent these results as histograms in Fig. \ref{fig:density-histogram}.}
%     \label{fig:stellar-density-relation}
% \end{figure*}

%Paperstergis + 2012, Maddox + 2015, MHI/M* 

%\clearpage 

\section{Discussion}
\label{sect:discussion}

%\todo{Summary of results and what the discussion is about} 
Our results indicate that the edges of Fornax cluster galaxies can be up to 50\% smaller and denser compared to field galaxies. Additionally, at a fixed stellar mass in the edge-mass relation, younger and bluer galaxies were found to be larger than older and redder ones. This colour bifurcation is related to morphology where the bluer, late-type galaxies in our sample are larger than the redder, early types, similar to the previous results found in \cite{2022chamba}. The interpretation of these findings are discussed in subsections \ref{dis:physical_processes} and \ref{sub-sec:compare} , considering the physical processes which are dominant in the Fornax Cluster and other environments. 

\subsection{Which physical processes impact the formation of lower density edges?}
\label{dis:physical_processes}

%\subsection{How would tidal forces impact the shape and properties of the edge?}
\textbf{FORNAX CLUSTER} In a crowded cluster environment, three main processes may impact the evolution of galaxies.  As galaxies move through the cluster, their gas can be removed due to the ram pressure from the hot intra-cluster medium \citep{1972gunn}, their stellar and gaseous material may be removed due to tidal forces from multiple galaxy-galaxy interactions and collisions with surrounding galaxies \citep{1984dressler, 1996moore} or from the gravitational potential of the full cluster \citep{2003gnedin}. We explore the impact of each of these processes on galaxy edge properties in the following paragraphs. \par 

There are several observational signatures of ram pressure and tidal forces acting on galaxies. In the case of ram pressure, the signatures include: 1) the observation of relatively gas-poor galaxies in clusters and groups compared to those in isolated environments 2) extended neutral or molecular hydrogen (\HI{} or \Htwo, respectively) tails, together with their alignment with respect to 3) the presence of hot gas or halos traced with X-rays. In the case of tidal forces, observable signatures include the visual deformation of galaxy shapes (e.g. the ``S'' shape, dislocated centers), stellar tidal structures in the outskirts or the dislocation of \HI{} or \Htwo{} with respect to the optical centres of galaxies (similar to ram pressure candidates) because \Htwo{} is more strongly bound to the galaxy and thus require stronger forces for them to be removed. \par 
All of the above signatures have been reported in the Fornax Cluster using optical imaging from FDS \citep{2018venhola, 2022venhola}, X-rays using the RÖntgen SATellite \citep{2002paolillo} or NASA's \emph{Chandra} X-ray Observatory \citep{2005scharf} and observations of molecular \citep{2019zabel, 2022morokuma} and neutral hydrogen gas \citep{2021loni, 2023fornax_meerkat} with the Atacama Large Millimeter/submillimeter Array (ALMA) and Meer-Karoo Array Telescope (MeerKAT), respectively. \par 

However, \cite{2022asencio}'s recent, detailed MCMC and N-body analysis of the susceptibility of tidal forces acting specifically on the FDS dwarf galaxies in the Fornax Cluster provides further clarity on which of these processes might be shaping the optical morphology of the dwarfs. In particular, these authors compare the susceptibility of tidal forces from galaxy-galaxy interactions and from the cluster potential and show that the latter process is the more likely contributor for the visual deformations reported in the morphology of these galaxies by \citet{2018venhola, 2022venhola}. By defining tidal susceptibility from the cluster potential as the ratio between the half-mass radius and the tidal radius of the dwarf ( $r_{tidal} \propto M^{1/3}$ where $M$ is the total mass of the galaxy), these authors found that higher suspecibility values correspond to more disturbed visual morphologies. \par 
Galaxy-galaxy encounters were shown to be irrelevant because the disruption time-scale of a dwarf after an interaction with a massive elliptical galaxy is much larger than the age of the dwarf galaxies themselves in the Fornax cluster \citep[$\sim$10\,Gyr;][]{2001rakos}. A more recent analysis of Fornax-like clusters in the Illustris simulations (Anetjärvi, M. et al. 2023\footnote{\protect\url{http://jultika.oulu.fi/Record/nbnfioulu-202307302907}}) also demonstrates that the influence of dwarf-cluster potential tidal forces is stronger relative to that from dwarf-other galaxy interactions. The latter are only more or similarly significant to the former on the order of once every 10\,Gyr per dwarf, further supporting the findings in \cite{2022asencio}.  \par 
%We cross-matched the sample of dwarfs analysed in \cite{2022asencio} and our 

 \par 

\textbf{The removal of stars via tidal forces:} To consider the impact of cluster tides on our sample, we use the published \cite{2022venhola} classifications to check for visual deformations or signs of tidal features in the outskirts which could lead to the removal of stars. We find that the Fornax sample studied here are 81\% `regular' or `undisturbed' (i.e. galaxies without obvious distortions or tidal tails), 15\% `slightly disturbed' (hints of irregular features in the outskirts) and only 2\% each of class `disturbed' (very clear tidal features or distortions) and `unclear' (difficult to classify due to surrounding contaminants such as neighbouring sources or bright objects). The higher fraction of `regular' shaped or `undisturbed' galaxies is understood because we visually verify that our assumption of elliptical symmetry is an average representation of the data when deriving the radial profiles to identify edges. Such galaxies are consistently more regular in shape (but see Appendix \ref{app:asymmetry} for the more massive asymmetric cases in \citet{2019raj}). \par 

Given these results, \emph{it is reasonable to conclude that the removal of stars via tidal forces may not be the main physical process causing the Fornax edges to be smaller and denser}. However, we cannot rule out the possibility that not \emph{all} stellar tidal features are visible in FDS, as for example, recently demonstrated in the study of the  visibility of tidal features and debris across a range of environments and stellar masses in Rubin-LSST mock imaging from NewHorizon \citep{2022martin} or the FIRE hydrodynamical simulation study of disrupting satellite galaxies \citep{2023shipp}. Nevertheless, from current optical observations of the sample used here, tidal forces from both the cluster and galaxy interactions cannot be directly responsible for the edge properties reported here.\par 

Moreover, this conclusion is further supported if we consider the expected impact of mass loss (including dark matter) in a galaxy, regardless of the process that caused it. When mass is lost, the depth of the potential well of the galaxy is decreased. Assuming that the velocity dispersion of the stars 1) remains unchanged or 2) increased by adding kinetic energy through the interactions, then stars will move at the same speed or faster after the mass is removed. In either case, stars will move into higher orbits because the potential well is shallower \citep{2008bt}. Therefore, we would expect a reduction of the density of stars in the outskirts. However, we find higher edge densities, further supporting our conclusion that mass loss cannot be the process responsible for our results. We discuss how this conclusion compares with that from previous work using the effective radii or fixed iso-density measures in Sect. \ref{sub-sec:compare}. \par 

%\todo{TO DO: Think of how to connect this comparison also to the satellite galaxy sample considered here?}

\begin{figure}[h!]
    \centering
    \includegraphics[width=0.49\textwidth]{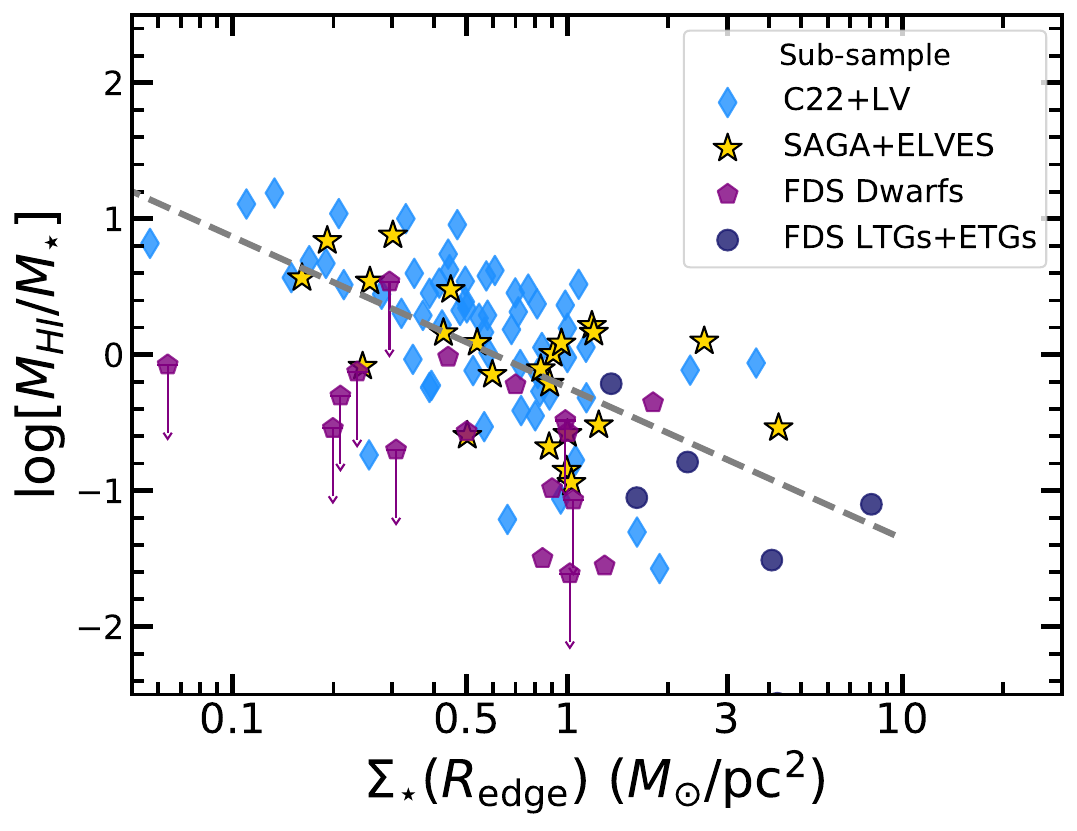}
    \caption{Ratio of \HI{}-to-stellar masses of a sub-sample of local volume field galaxies (blue diamonds, 64 studied here out of the 96 with available DeCaLs imaging from \cite{2013karachentsev}), 9 SAGA and 15 ELVES satellites (yellow stars, matched with the \cite{2020durbala} and \cite{2023zhu} measurements, respectively) and the Fornax galaxies (purple pentagons, 17 studied here out of the 39 \HI{} masses or upper limits reported in \cite{2023kleiner}) as a function of edge surface stellar mass density. The \HI{} masses of the Fornax LTGs and ETGs are taken from \cite{2021loni} when available (dark blue points). Higher edge densities are found when the \HI{}-to-stellar masses ratios are lower. A best fit to the data points is shown as a dashed line to highlight this trend.}
    \label{fig:HI-stellar-ratios}
\end{figure}

\textbf{Ram pressure takes it all:} To consider the influence of ram pressure in edge properties, we compile the known total $\HI{}$ masses of the galaxies in our sample from the catalogues published in \cite{2013karachentsev}, \cite{2020durbala}, \cite{2023zhu} and \cite{2023kleiner}. We plot the ratio between these \HI{} masses and the stellar masses computed here i.e. \HI{} fraction as a function of edge density in Fig. \ref{fig:HI-stellar-ratios}. The standard relations between \HI{} and its ratio with stellar mass as a function of stellar mass \citep[e.g.][]{2012papastergis, 2015maddox} for comparison with ALFALFA data \citep{2020durbala} are shown in Appendix \ref{app:Hi-relations}. From the samples studied here, Fig. \ref{fig:HI-stellar-ratios} shows that galaxies with higher edge densities are those with lower \HI{}-to-stellar masses ratios (r$=-0.23$). 

%\todo{TO DO: Check how all the masses are computed. Compare with ALFALFA sample for reference.?} 
%\todo{TO consider: Estimating a timescale for gas removal by ram pressure and 
%comparing it to the cluster crossing timescale?}

%\todo{TO DO: Need to check HI sensitivity limits in the compiled numbers used, maybe include uncertainties? Needs a companion plot with the MHI/M* - M* plot and overplot the ALFALFA survey parameter space for reference e.g. Paperstergis + 2012. Need to check these masses and why sample if below ALFALFA region before interpretation.}

This result can be connected with the recent finding that most of the Fornax dwarf galaxies are \HI{} deficient \citep[][]{2023kleiner}. This deficiency is thought to have been caused by the removal of \HI{} via tidal forces and ram pressure. 
Tidal forces are required to explain their observations for two main reasons. The first is that $\sim$ half of the \HI{} detected dwarfs have asymmetric stellar bodies, which is a signature of tidal forces acting on these galaxies. \par 
The second reason is that all the detected \HI{} rich dwarfs `avoid' the highly dense regions of the cluster. The densest regions are those populated by the most massive potential wells in the core. The lack of \HI{} rich dwarfs in those regions imply that the dwarfs in the core (see Fig. \ref{fig:FDS-footprint}) have completely lost their \HI{} in the cluster. \par 
Ram pressure is needed to explain the observations mainly because of the configuration of the detected \HI{} tidal tails  in these dwarfs. As ram pressure depends on the density of the intracluster medium \citep{1972gunn}, the tails are shaped by the X-ray emitting sources in the more dense core of the cluster \citep[][see their Fig. 12]{2005scharf}. .\par 

Therefore, while we have established that the removal of stars via tidal forces cannot result in high density edges, the forces can still remove the gas component in a galaxy. Considering that the distribution of \HI{} is several times more extended than the optical counterpart of galaxies \citep[e.g.][]{2016wang}, the detection of \HI{} tidal tails suggests that the \HI{} in the outskirts is likely removed first. Initially, the removal occurs via tidal forces and subsequently ram pressure may continue to remove the loosely bound atomic or molecular gas. This removal may prevent further inflow of cold gas and lowers the respective gas densities. \par 
%These processes may occur multiple times 

The lowering of both \HI{} and \Htwo{} gas density as these components are removed has indeed been reported in the \cite{2023watts} study of the radial stellar and gas (\HI{} and \Htwo) density distributions of Virgo cluster galaxies. These authors find that while the impact on \HI{} is considerably stronger than molecular gas, the \Htwo{} density can be reduced by a factor of three in the cluster galaxies. They additionally show that it is the lowest density molecular gas, preferentially found in low stellar density regions i.e. larger radii, which is removed first (see their Fig. 7). \par 
This finding is consistent with earlier work on the comparison between the optical $R$-band radial distribution and that of a star formation rate indicator, H$_{\alpha}$, in Virgo galaxies by \cite{2004koopmann}. These authors report the existence of extremely truncated galaxies in Virgo where there is no star formation in the outer 60-70\% regions of their disks \citep{2001koopmann}. In a later study considering a larger sample, \citep{2004koopmann} show that the star formation rate in these outer regions has in fact been reduced by a median factor of 9 compared to more isolated galaxies. In other words, star formation is truncated outside-in \citep{2009bekki, 2012zhang}. \par 

Although Fornax is a lower mass cluster compared to Virgo and at a different stage of evolution, there are several pieces of evidence which support the idea that a similar phenomenon might be found in Fornax. For instance, when \Htwo{} deficiency is defined as the difference between the observed \Htwo{} mass and the statistically expected \Htwo{} mass from galaxies of the same stellar mass in the field, \citet{2019zabel} has shown that the \Htwo{} deficiencies of Fornax galaxies are even higher than in Virgo (down to $\sim$-1.1\,dex compared to $\sim$-0.3\,dex). The higher reduction of \HI{} compared to \Htwo{} in Fornax has also been reported in \cite{2022morokuma}. Therefore, we can already anticipate that lower gas densities will also be removed first in Fornax cluster galaxies. \par 
\citet{2022morokuma} additionally report that the lower cold gas fractions (defined as the total \HI{} and \Htwo{} mass as a fraction of stellar mass) due to gas removal is the main cause for the lower star formation activity in Fornax galaxies. If the edges we identify are indeed related to a gas density threshold needed for star formation, then lower gas densities in the outskirts will prevent the formation of stars in those regions if they do not meet the necessary threshold, thus truncating star formation outside-in \citep[see e.g.][]{2009bekki}. \par 

Our finding that Fornax galaxy edges are denser can thus be understood with the scenario that the removal of low density \HI{} and molecular gas via cluster tides and ram pressure exerted on these galaxies during in-fall or as they cross the cluster likely truncated star formation outside-in, preventing the formation of lower density edges and larger sizes. Given that not all galaxy clusters are at the same stage of assembly, have the same mass or interstellar medium \citep[for example the density of hot halos in the Virgo cluster is a factor $\sim$4 higher than in Fornax; see the comaprison in][]{2019zabel} it would be interesting to explore this scenario in other clusters. \bigskip 

%Also found in recent analysis of Fornax-like clusters in Illustris simulations (Venhola, private communication).

%pressure = Sp * rho * v^2  Sp = degrees of freedom. close to half. estimate the stripping time scale of Fornax cluster. F = P/unit area, 

%ram pressure stripping of the galaxy, what is the time scale for this?

%ram pressure from X-ray gas, hot intra-cluster medium. dwarfs characteristics could be imprinted even before they enter the cluster. how to estimate ram pressure??

%In the visual inspection of FDS dwarf galaxies performed by \citet{2022venhola}, the tidal morphology of the dwarfs are categorised as `undisturbed', `possibly/mildly disturbed', `very disturbed' or `unclear'. 

\noindent \textbf{GROUP/SATELLITE GALAXIES} %\todo{\textbf{Tidal forces and ram pressure in group/satellite galaxies:} 
Satellites of massive galaxies may also be influenced by tidal forces and ram pressure. \cite{2009penarrubia} for example study how the shape of the radial profile changes as satellite dwarfs lose stars from tidal forces. They show that as galaxies lose stars, their outer stellar surface density radial profile up-bends, and eventually approaches a $R^{1/4}$ law. By applying their model to observations of Local Group satellites, these authors provide the predictions for the location of the ``break'' radii in these galaxies. \par 

A ``break'' in \cite{2009penarrubia} is defined as the location where the profile up-bends because of tidal forces. The authors show that the predicted ``break'' radii from these models are well beyond the observed breaks in all of the Local Group dwarf galaxies studied except the Sagittarius dwarf, a galaxy with very clear tidal perturbations. \par 
In fact, the predicted radii are also well beyond the edges we find here for satellite galaxies of similar stellar mass. We find that the predicted break radii from \cite{2009penarrubia} are located at $\sim 2-35\,R_{\rm edge}$ depending on the stellar mass of the dwarf. These results indicate that the majority of edges we find here are likely not of tidal origin, similar to the case of Fornax cluster dwarfs. \par 
The comparison of the shapes of profiles themselves is beyond the scope of our work and is highly complicated as it requires the deconvolution of the PSF. Locating the edges themselves do not require this step, see Fig. 12 in \cite{2016trujillo} or Appendix C in \cite{2022chamba}. We leave such an analysis for when extended PSFs for DECam have been established. \par 

More recently, \cite{2021putman} have analysed the impact of the gaseous halo medium of the Milky Way and Andromeda on Local Group dwarfs to study the role of ram pressure and its responsibility in the removal of gas in satellites. These authors show that the minimum halo density to completely remove the gas in the satellites via ram pressure is reached well within the distances to their respective hosts (whichever one is closer). This finding explains the lack of gas-rich galaxies in the Local Group within the virial radii of their hosts. Similar to our study, many of their galaxies do not show signs of tidal distortions or tails and they argue that the impact of tidal forces is likely negligible in comparison to ram pressure. \par 

These results have been further supported by observations of satellite systems in nearby spiral galaxies. In the absence of \HI{} measurements for all the ELVES and SAGA satellites in ALFALFA, \cite{2023zhu} used the late-type morphologies, the presence of H$_{\alpha}$ or both as an indicator for the presence of gas in these galaxies. Using ELVES and SAGA in combination with their own sample of satellite galaxies selected using ALFALFA, \cite{2023zhu} report a similarly low number of gas rich satellites as that in the Local Group. The authors find $\sim$ 0$-$3 gas-rich satellites within the virial radius of the host galaxy (see their Fig. 11). Given this similarity, they argue that ram pressure from the host medium can also explain the lack of gas-rich satellites hosted by Milky Way analogues in the nearby Universe as in the Local Group studied by \cite{2021putman} \citep[see also][]{2023greene}. \par 
Our results that the ELVES and SAGA satellites also have high density edges and smaller sizes compared to the nearly isolated sample is broadly consistent with these findings. In fact, the satellites studied here seem to populate an intermediate regime in terms of size and edge density (Fig. \ref{fig:edge-histograms}) between the Fornax and the field samples. This result \emph{potentially indicates that the impact of ram pressure alone from a single host halo on edge properties is weaker than the cluster tidal+ram pressure scenario discussed above for the case of Fornax galaxies.} It would be interesting to compare these results in future work with the predictions of hydrodynamical simulations that can resolve dwarf galaxies in low and high density environments. \par \bigskip 

%\todo{-Check also \cite{2021putman, 2023zhu} for the current consensus on the gas content of Local Group and simialr satellites outside the Local Group.} \par \bigskip 

%\subsection{Have the cluster galaxies formed at an earlier epoch compared to those in the field?}

%which ones are the cluster galaxies in the plot??

%\textbf{Dwarf galaxies:} We have found that at a fixed stellar mass, bluer galaxies are younger and larger and the redder Fornax cluster galaxies are older and smaller. This result suggests that the Fornax galaxies reached that present-day size at an earlier epoch than field galaxies. \todo{TO DO: Perhaps put this in dialogue with the deviation from the 1/3 slope? All dwarfs follow a higher 0.4 slope so does that mean they in generally under went a different evolution compared to massive galaxies that follow the 1/3 slope? Galaxies can be red because they quenched when in-falling in to the cluster or they really formed at an earlier epoch. Link to effective radii-mass evolution at higher redshift?)}

%\todo{reionization??}

%%\todo{Compare values from Raj et al. papers of break radii, to show they are different?}

\noindent \textbf{MASSIVE GALAXIES} While much of the discussion above has been specific to dwarf galaxies, here we make a few brief remarks on the edge properties of the more massive galaxies in our sample. Our final sample consists of only five massive ETGs from \cite{2019iodice} and 12 massive LTGs from \cite{2019raj, 2020raj} who studied their inner ``break'' radii. For comparison, the break radii reported in \cite{2019raj, 2020raj} are at $0.3-0.7\,R_{\rm edge}$ and at a compatible location only in one galaxy FCC013. There could be potential physical connections between the break and edge radii and may have a common origin as suggested in \cite{2019cristina}, however, exploring this connection is beyond the scope of the current work. \par 

Unfortunately, given the low number of galaxies we also cannot perform a statistical analysis in the stellar mass range of these LTGs and ETGs as we did for galaxies in the lower mass regime. Nevertheless, we point out a few obvious differences in the sample of galaxies studied here and their edge properties. Contrary to the dwarf galaxies in our sample, most of the Fornax LTGs studied here have signs of galaxy interactions (FCC number 312), have slightly asymmetric stellar discs (FCC 113, 121, 285) or transitioning to S0 type (FCC 290). \par 
As discussed in \cite{2019raj, 2020raj}, minor mergers or a recent fly-by galaxy-galaxy interaction may impact the stellar distribution of these galaxies by truncating their star formation outside-in due to the removal of gas. While we cannot use the edge properties to pinpoint the exact formation channel, we find that the densities of the edges are a factor of four times higher compared to the mean expectation from galaxies of similar stellar mass in the field. This result is indicative that galaxy interactions may have a much stronger impact on the properties of edges than cluster tides or ram pressure alone. \par 
Many of the Fornax A sub-group galaxies are also recent or intermediate infallers \cite[FCC numbers 013, 029, 033, 035, 062 from the galaxies in our sample; see the phase space diagram Fig. 7 in][]{2020raj}. While we do not report any significant differences between the edge properties of infallers and galaxies within the main cluster (see Sect. \ref{sub-sec:compare}), there are clear differences between the edge properties of the LTGs and ETGs. The edge densities of ETGs are typically even higher than those of the LTGs by a factor $\sim$1.5 within the Fornax environment (this work), while in the field they are greater than a factor of three \citep{2022chamba}. We discuss the differences between galaxy types in the next section. \par

\subsection{What drives the scatter of the size--stellar mass plane?} 

Differences in the physical properties between galaxy types is expected given that galaxies may undergo varied star formation histories, regardless of their present-day local environment. For instance, massive elliptical galaxies show highly efficient star formation at early times \citep[e.g.][]{2015jaskot}, in stark contrast to the more inefficient formation found in dwarfs \citep[e.g.][]{2012huang}. The median higher edge densities $\sim 5\,M_{\odot}/$pc$^2$ for the massive ETGs, $\sim 3 \,M_{\odot}/$pc$^2$ for the massive LTGs and $\sim 1 \,M_{\odot}/$pc$^2$ for the full dwarf sample we find in the Fornax Cluster is potentially reflective of this intrinsic difference. In the field, the values reported in \cite{2022chamba} were $>3\,M_{\odot}$/pc$^2$ for massive ETGs, $\sim 1\,M_{\odot}/$pc$^2$ for massive LTGs and $\sim 0.6\,M_{\odot}/$pc$^2$ for dwarf galaxies. \par 

Interestingly, by further separating the dwarf galaxies in our sample in to early and late-types, we have additionally shown that the early-types are systematically smaller than the late-types in the Fornax cluster at a given stellar mass. Both dwarf types are also smaller in the Fornax cluster compared to similar galaxies in the field. These findings are potentially linked with the colour and age bifurcation we find in the edge scaling relations shown in Fig. \ref{fig:size-mass-relations}. \par 

Collectively, they suggest that Fornax early-type galaxies reached their present-day sizes earlier than the late-type galaxies. In fact, all but three of the gas-rich dwarfs in Fornax plotted in Fig. \ref{fig:HI-stellar-ratios} are of late-type morphology. Most of the early-types likely lost their gas earlier and the cluster gas-rich galaxies are more recent infallers still evolving in the new environment \citep{2023kleiner}. The colour, age and morphology bifurcations we find in the size--stellar mass relations are thus likely driven by the environment of galaxies. In other words, the scatter of the size--stellar mass plane can be physically explained by the differences in morphology and environment of galaxies. Future work using much larger samples of galaxies, including gas-rich early-type galaxies \citep[e.g.][]{2012serra}, are welcome to explore this issue in more detail and verify our findings. 

\subsection{The dependence of galaxy sizes on cluster-centric distance}
\label{sub-sec:compare}

%\textbf{Dependancy on cluster-centric distance}:
As discussed in \cite{2019venhola} in particular for the Fornax Cluster, the cluster-centric distance defined as the distance of the galaxies with respect to the brightest cluster galaxy NGC~1399, can be used as a robust proxy for galaxy environment (see their Fig. 4 showing its strong correlation with projected galaxy density). These authors find that for the brightest dwarfs, their effective radii decrease as a function of cluster-centric distance. However, the most diffuse, low surface brightness dwarfs with large effective radii are systematically found in the central regions of the cluster compared to those with smaller effective radii \citep{2022venhola}. \par 

%Such an observation is consistent with the morphology-density relation found in other clusters like Virgo \citep{2007lisker} which is expected to impact the physical properties of galaxies, including their radii. \par 

However, in our work, we do not currently find an obvious dependence of the Fornax edge radii or densities on cluster-centric distance. This result does not change even if we compare the edge properties within the in-falling Fornax A sub-group and the main cluster, the locations of which are shown in Fig. \ref{fig:FDS-footprint}. Our main explanation for this result is as follows. \par 
For a dwarf galaxy with stellar mass $\sim10^8\,M_{\odot}$, the gas removal time scale is $\sim 240$\,Myr \citep{2023kleiner}. For comparison, if we consider the simplest estimate of the cluster crossing time of a galaxy using a relative velocity dispersion of 370\,km/s \citep{2001drinkwater} and the virial radius of 700\,kpc of Fornax, the crossing time scale is approximately 1-2 Gyr \citep{2021loni}. %These timescales suggest that gas removal may occur over multiple episodes during the lifetime of a dwarf in Fornax as the galaxies have sufficient time to cross the cluster several times. 
These timescales suggest that gas removal is possible over an extended period that encompasses multiple cluster crossings \citep[see][]{2023kleiner}, implying that a dwarf could be subjected to tidal forces and ram pressure at different stages of its lifetime \citep[e.g.][]{2018jaffe, 2019yun}.\par
%The cluster crossing time is of the order of $1-2$\,Gyr. This time scale implies that galaxies in the cluster have sufficient time to cross the cluster several times.  %As discussed earlier, multiple episodes of gas removal via cluster tides or ram pressure over a $\sim$240 Myr time scale become feasible while the galaxy crosses the cluster. 
%then any dependence of their edge properties with cluster centric distance can also be removed. 
%As the outskirts are more vulnerable compared to the radii traced by the effective radius, it is then reasonable to expect that any dependence on distance can be more easily erased. 
However, it is unknown at which location in the galaxy's trajectory was the gas in the outskirts significantly removed or lowered in order to impact their size. In other words, the galaxy can reach its present-day size by undergoing the aforementioned cluster processes and still continue to cross the cluster.  The galaxy's present-day location may thus not necessarily reflect the location where its edge was most impacted. \par % In this sense,  \par
%The second explanation is that the sizes of galaxies are already set before or soon after reaching the main cluster. 
The rapid quenching of dwarf galaxies due to gas removal in observations and simulations as soon as these galaxies enter the cluster environment has already been shown in earlier work \citep[e.g.][]{2014boselli}. After such quenching, the dwarfs may then not significantly change their size no matter the number of times they could cross the cluster or be impacted by ram pressure at their current location. As the outskirts are more vulnerable compared to the radii traced by the effective radius, it is then reasonable to expect that any dependence on distance can be more easily erased. %his process likely prevents the formation of more extended sizes than when the galaxies first entered. In contrast, those galaxies in lower density environments or in the field may continue to grow via gas accretion, explaining their larger sizes. \par 

%No matter the number of times that the dwarfs ... or wherever they are located in, the dwarfs will not significantly change their sizes?

While the above argument could apply for most of the dwarfs in Fornax, we point out that our sample does not include the faintest, low surface brightness dwarfs studied in \cite{2022venhola}. Future work characterising the edges of such galaxies specifically may provide a more complete picture to the study performed here. Nevertheless, the lack of a dependence on cluster-centric distance we find here compared to previous works using the effective radii highlights that our edge measure is much more sensitive to local environmental processes possible all over the cluster region and is independent of inner structure or light concentration. In Appendix \ref{dis:othersizes} we compare and discuss the use of other size indicators in previous work. \par \bigskip 

%time = d/c 

\section{Conclusions}

The impact of the environment on galaxy size has long been elusive. Previous work addressing this issue by using the effective radius, a measure which traces light concentration, have ignored the fact that the definition is often biased against the outskirts of galaxies. Addressing this bias is crucial because the outskirts of galaxies habour key signatures of environmental processes such as ram pressure or galaxy-galaxy interactions which may transform the surrounding interstellar medium and consequently the galaxy's outer growth. \par 
After several decades of research where the effective radius has shown inconclusive or ambiguous results, we demonstrate the significant impact of the environment on galaxy size when using a physically motivated size definition which is more representative of the in situ, stellar boundaries or ``edges'' of galaxies \citep{2020tck, 2020ctk, 2022chamba}. Our main results are as follows:

\begin{itemize}
    \item The edge radii--stellar mass relation for all galaxies within $10^6-10^{10}\,M_{\odot}$ is $R_{\rm edge} \sim M_{\star}^{0.42}$ with a very low intrinsic scatter $\sim 0.06$\,dex. Compared to the effective radius, the dispersion of this size--stellar mass plane is reduced by a factor more than two.

    % \item The colour bifurcation reflects the bifurcation in morphology where late-type galaxies are larger than the early-types. 
    
    \item The scatter in the edge radii-stellar mass plane is physically driven by the morphology and environment of galaxies. Fornax cluster galaxies can have up to $\sim 50\%$ smaller edge radii than those found in nearly isolated or field environments. The stellar surface density at these edges are fifty percent more dense compared to the field.

     \item When considering morphology, at a fixed stellar mass the redder, older early-type galaxies in the cluster are smaller by $\sim 20\%$ compared to the late-types across the full range studied. A similar difference between galaxy types is found in the field. %extending the results previously reported in \cite{2020tck} and \cite{2022chamba} towards lower mass galaxies. 
     %This result is driven by the environment of galaxies.

     \item Confirmed satellites follow similar trends when compared to the nearly isolated galaxies. However, their sizes are still larger than the cluster galaxies. 

    %\item Both the early- and late-type galaxies in Fornax are smaller than similar galaxies in the field. However, the early-types are even smaller than the late-types, complementing our results at fixed stellar mass. This bifurcation in morphology is driven by the environment of galaxies.

    \item Galaxies with lower \HI{} fractions have edges with higher stellar surface density.
    
\end{itemize}

These results are consistent with recent \HI{} observations of the Fornax Cluster which show that this gas component in member galaxies have been rapidly removed as a consequence of cluster tides and ram pressure \citep{2021loni, 2023kleiner}. These environmental processes truncate galaxies outside-in earlier than in the field, which implies that cluster galaxies may have also reached their present-day size earlier. In this scenario, the formation of larger sizes and low density edges in a cluster environment becomes more difficult. \par 
Finally, our work highlights the importance of deep imaging surveys to characterise the outer growth of the faintest galaxies, a regime which would have otherwise remained invisible. Faint galaxies are not only important for understanding how environmental processes affect the dark matter content of galaxies \citep{2021jackson, 2021rodriquez, 2022mao, 2022asencio, 2022mau, 2023lucie, 2023montes} but also for the future interpretation of deep, multi-band observations upcoming from the Vera Rubin Observatory \cite[][Tsiane et al. in prep]{2022martin, 2023shipp}. Our findings thus offer a unique approach for the future detection of the low surface brightness imprints on galaxy evolution and environment.

\begin{acknowledgements}

We thank the anonymous referee for their comments which improved the quality of our manuscript. This paper has undergone internal review in the LSST Dark Energy Science Collaboration. The internal reviewers were Jeff Carlin, Yuanyuan Zhang and Ian Dell'Antonio. % Optional but recommended
% Standard papers only: author contribution statements. For examples, see http://blogs.nature.com/nautilus/2007/11/post_12.html
% This work used TBD kindly provided by Not-A-DESC Member and benefitted from comments by Another Non-DESC person.
% Standard papers only: A.B.C. acknowledges support from grant 1234 from ... 

This work used the catalogues kindly provided by Aku Venhola (FDS) and Yao-Yuan Mao (SAGA) and benefited from comments by Reynier Peletier, Maria Angela Raj, Aaron Watkins, Maria Carmen Toribio, Kelley Hess, Dane Kleiner and Am\'elie Saintonge.

\vspace{2pt} \\ 
%\textit{Author contributions}:

NC performed the analysis and wrote the paper.
MH discussed the results, implications and commented on the manuscript at all stages. 

\vspace{2pt} \\

NC and MH acknowledge support from the research project grant ‘Understanding the Dynamic Universe’ funded by the Knut and Alice Wallenberg Foundation under Dnr KAW 2018.0067.
The DESC acknowledges ongoing support from the Institut National de 
Physique Nucl\'eaire et de Physique des Particules in France; the 
Science \& Technology Facilities Council in the United Kingdom; and the
Department of Energy, the National Science Foundation, and the LSST 
Corporation in the United States.  DESC uses resources of the IN2P3 
Computing Center (CC-IN2P3--Lyon/Villeurbanne - France) funded by the 
Centre National de la Recherche Scientifique; the National Energy 
Research Scientific Computing Center, a DOE Office of Science User 
Facility supported by the Office of Science of the U.S.\ Department of
Energy under Contract No.\ DE-AC02-05CH11231; STFC DiRAC HPC Facilities, 
funded by UK BEIS National E-infrastructure capital grants; and the UK 
particle physics grid, supported by the GridPP Collaboration.  This 
work was performed in part under DOE Contract DE-AC02-76SF00515.

\vspace{2pt} \\ 
%\textit{Data}: 

The FDS data were produced by the FDS collaboration and the data products are available via the \href{https://archive.eso.org/scienceportal/home?publ_date=2020-08-26}{ESO Science Portal}. See \citet{2020peletier} for a full description of this data release. 

\vspace{2pt} \\ 

Funding for the Sloan Digital Sky Survey V has been provided by the Alfred P. Sloan Foundation, the Heising-Simons Foundation, the National Science Foundation, and the Participating Institutions. SDSS acknowledges support and resources from the Center for High-Performance Computing at the University of Utah. The SDSS web site is \url{www.sdss.org}.

SDSS is managed by the Astrophysical Research Consortium for the Participating Institutions of the SDSS Collaboration, including the Carnegie Institution for Science, Chilean National Time Allocation Committee (CNTAC) ratified researchers, the Gotham Participation Group, Harvard University, Heidelberg University, The Johns Hopkins University, L'Ecole polytechnique f\'{e}d\'{e}rale de Lausanne (EPFL), Leibniz-Institut f\''{u}r Astrophysik Potsdam (AIP), Max-Planck-Institut f\''{u}r Astronomie (MPIA Heidelberg), Max-Planck-Institut f\''{u}r Extraterrestrische Physik (MPE), Nanjing University, National Astronomical Observatories of China (NAOC), New Mexico State University, The Ohio State University, Pennsylvania State University, Smithsonian Astrophysical Observatory, Space Telescope Science Institute (STScI), the Stellar Astrophysics Participation Group, Universidad Nacional Aut\'{o}noma de M\'{e}xico, University of Arizona, University of Colorado Boulder, University of Illinois at Urbana-Champaign, University of Toronto, University of Utah, University of Virginia, Yale University, and Yunnan University.

\vspace{2pt} \\ 

The Legacy Surveys consist of three individual and complementary projects: the Dark Energy Camera Legacy Survey (DECaLS; Proposal ID 2014B-0404; PIs: David Schlegel and Arjun Dey), the Beijing-Arizona Sky Survey (BASS; NOAO Prop. ID 2015A-0801; PIs: Zhou Xu and Xiaohui Fan), and the Mayall z-band Legacy Survey (MzLS; Prop. ID 2016A-0453; PI: Arjun Dey). DECaLS, BASS and MzLS together include data obtained, respectively, at the Blanco telescope, Cerro Tololo Inter-American Observatory, NSF’s NOIRLab; the Bok telescope, Steward Observatory, University of Arizona; and the Mayall telescope, Kitt Peak National Observatory, NOIRLab. Pipeline processing and analyses of the data were supported by NOIRLab and the Lawrence Berkeley National Laboratory (LBNL). The Legacy Surveys project is honored to be permitted to conduct astronomical research on Iolkam Du’ag (Kitt Peak), a mountain with particular significance to the Tohono O’odham Nation.

NOIRLab is operated by the Association of Universities for Research in Astronomy (AURA) under a cooperative agreement with the National Science Foundation. LBNL is managed by the Regents of the University of California under contract to the U.S. Department of Energy. \\

This project used data obtained with the Dark Energy Camera (DECam), which was constructed by the Dark Energy Survey (DES) collaboration. Funding for the DES Projects has been provided by the U.S. Department of Energy, the U.S. National Science Foundation, the Ministry of Science and Education of Spain, the Science and Technology Facilities Council of the United Kingdom, the Higher Education Funding Council for England, the National Center for Supercomputing Applications at the University of Illinois at Urbana-Champaign, the Kavli Institute of Cosmological Physics at the University of Chicago, Center for Cosmology and Astro-Particle Physics at the Ohio State University, the Mitchell Institute for Fundamental Physics and Astronomy at Texas A\&M University, Financiadora de Estudos e Projetos, Fundacao Carlos Chagas Filho de Amparo, Financiadora de Estudos e Projetos, Fundacao Carlos Chagas Filho de Amparo a Pesquisa do Estado do Rio de Janeiro, Conselho Nacional de Desenvolvimento Cientifico e Tecnologico and the Ministerio da Ciencia, Tecnologia e Inovacao, the Deutsche Forschungsgemeinschaft and the Collaborating Institutions in the Dark Energy Survey. The Collaborating Institutions are Argonne National Laboratory, the University of California at Santa Cruz, the University of Cambridge, Centro de Investigaciones Energeticas, Medioambientales y Tecnologicas-Madrid, the University of Chicago, University College London, the DES-Brazil Consortium, the University of Edinburgh, the Eidgenossische Technische Hochschule (ETH) Zurich, Fermi National Accelerator Laboratory, the University of Illinois at Urbana-Champaign, the Institut de Ciencies de l'Espai (IEEC/CSIC), the Institut de Fisica d’Altes Energies, Lawrence Berkeley National Laboratory, the Ludwig Maximilians Universitat Munchen and the associated Excellence Cluster Universe, the University of Michigan, NSF’s NOIRLab, the University of Nottingham, the Ohio State University, the University of Pennsylvania, the University of Portsmouth, SLAC National Accelerator Laboratory, Stanford University, the University of Sussex, and Texas A\&M University.
\\

BASS is a key project of the Telescope Access Program (TAP), which has been funded by the National Astronomical Observatories of China, the Chinese Academy of Sciences (the Strategic Priority Research Program ``The Emergence of Cosmological Structures'' Grant XDB09000000), and the Special Fund for Astronomy from the Ministry of Finance. The BASS is also supported by the External Cooperation Program of Chinese Academy of Sciences (Grant 114A11KYSB20160057), and Chinese National Natural Science Foundation (Grant 12120101003, 11433005). \\

The Legacy Survey team makes use of data products from the Near-Earth Object Wide-field Infrared Survey Explorer (NEOWISE), which is a project of the Jet Propulsion Laboratory/California Institute of Technology. NEOWISE is funded by the National Aeronautics and Space Administration.

The Legacy Surveys imaging of the DESI footprint is supported by the Director, Office of Science, Office of High Energy Physics of the U.S. Department of Energy under Contract No. DE-AC02-05CH1123, by the National Energy Research Scientific Computing Center, a DOE Office of Science User Facility under the same contract; and by the U.S. National Science Foundation, Division of Astronomical Sciences under Contract No. AST-0950945 to NOAO.

\vspace{2pt} \\

\textit{Software}: \texttt{Astropy},\footnote{http://www.astropy.org} a community-developed core \texttt{Python} package for Astronomy \citep{robitaille2013astropy, price2018astropy}; \texttt{SciPy} \citep{scipy}; \texttt{NumPy} \citep{numpy, doi:10.1109/MCSE.2011.37}; \texttt{Matplotlib} \citep{Hunter:2007}; \texttt{Jupyter Notebooks} \citep{Kluyver:2016aa}; \texttt{TOPCAT} \citep{2005ASPC..347...29T}; \texttt{Sourcerer/MTObjects} \citep{2016mto}; \textit{Matrioska} \citep[mascara function;][]{2018borlaff} \texttt{SWarp} \citep{swarp}; and \texttt{SAO Image DS9} \citep{ds9}.
\end{acknowledgements}

 
\bibliographystyle{aa} % style aa.bst
\bibliography{bib} % your references Yourfile.bib

\appendix

\section{Radial profile binning}
\label{app:profile_bins}

\begin{figure*}[t!]
    \centering
    \includegraphics[width=1.0\textwidth]{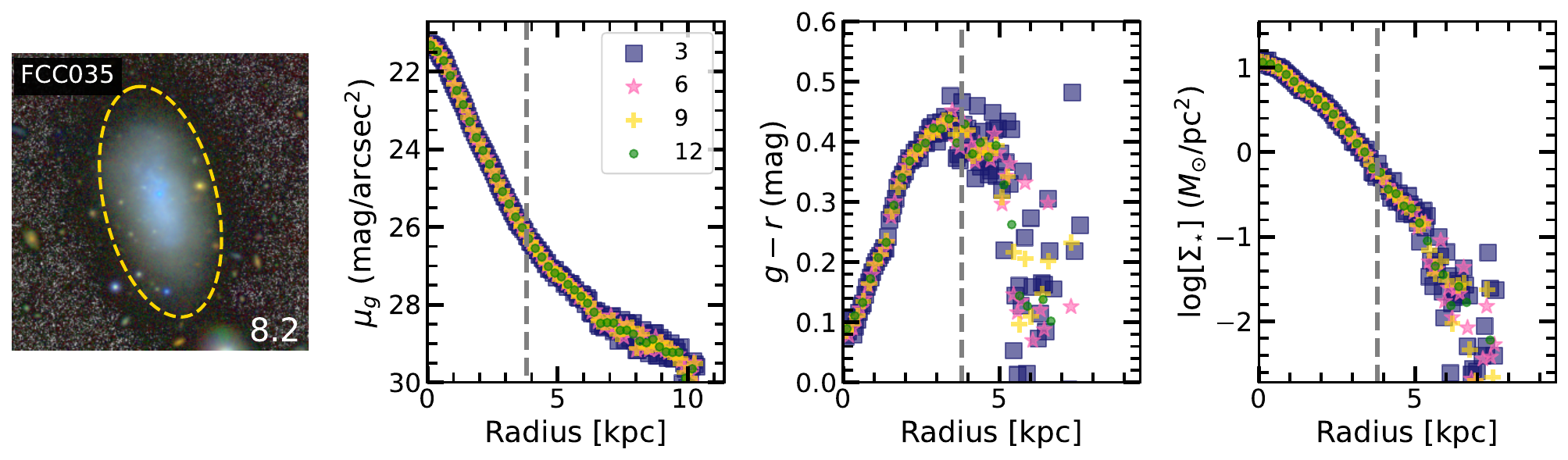}
    \caption{Radial profiles of the dwarf galaxy FCC035 using different bin sizes. The bin sizes are in pixel units. While increasing the bin size reduces the noise in the profiles as expected, the shape of the profiles and the feature we identify as the location of the edge do not change.}
    \label{fig:different_bins}
\end{figure*}

To test the dependence of our edge identification on the bin size we use to derive radial profiles, we show the profiles of a bright dwarf galaxy in the Fornax cluster, FCC035, using different bin sizes in Fig. \ref{app:profile_bins}. The legend indicates the bin size used to derive each profile in pixel units. For the FDS images, 1\,pixel = 0.21\,arcsec. All other steps are the same (see Sect. \ref{sect:methods}). \par 
As expected, using larger bin sizes reduces the fluctuation between data points or smooths the profiles. This smoothing occurs because we are averaging flux over a larger area per elliptical annuli quantified by the bin size. Regardless of this smoothing effect, the location of the edge feature (vertical dashed line) in the $g-r$ colour profile does not change. \par 

\section{Profiles of asymmetric galaxies}
\label{app:asymmetry}

\begin{figure*}
    \centering
    \includegraphics[width=0.9\textwidth]{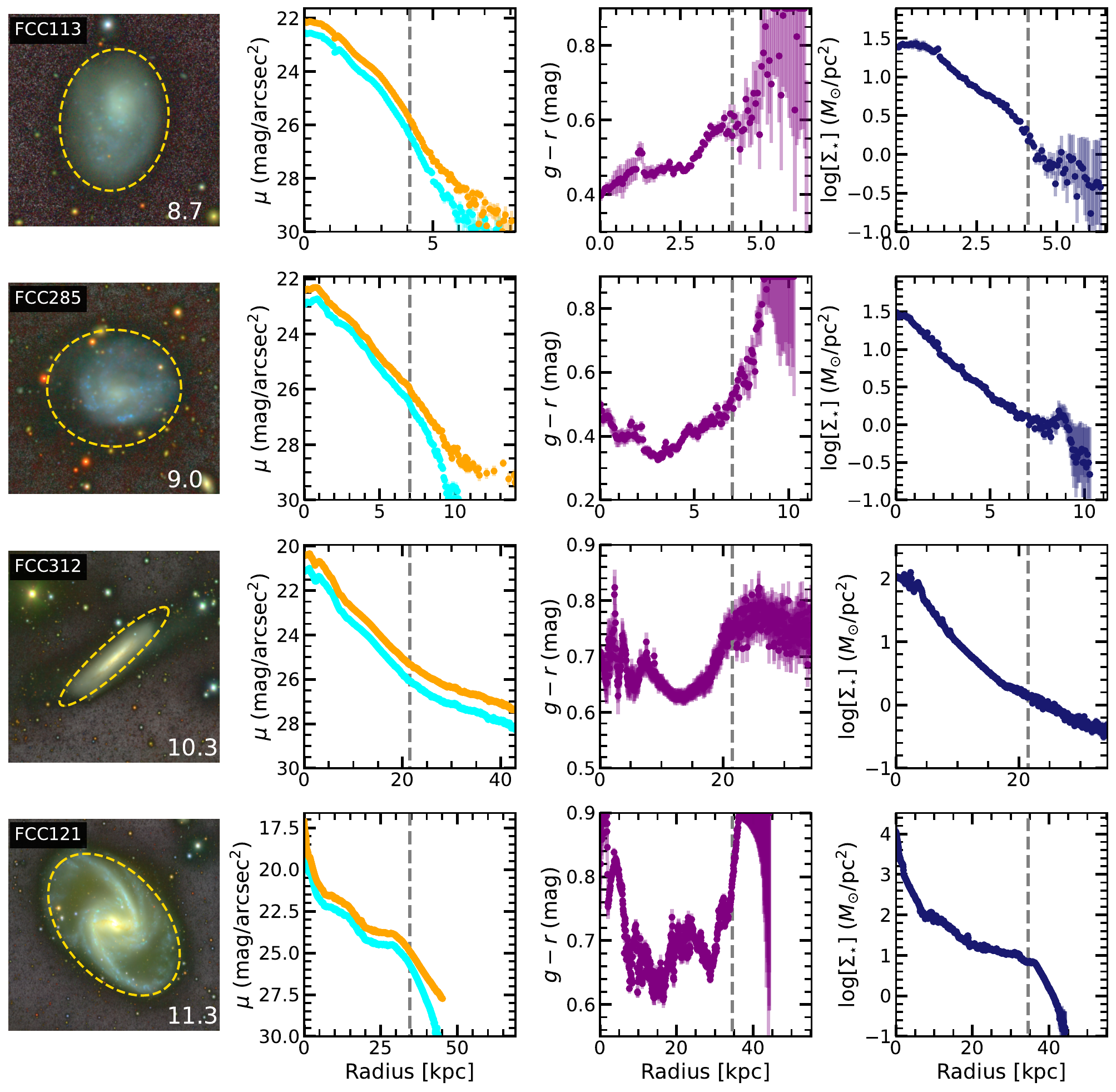}
    \caption{Radial profiles of asymmetric galaxies from the \citet{2019raj} LTG sample. These galaxies are within the virial radius of the main Fornax cluster and are clearly asymmetric in some form (see text for details). In all cases, our default radial profile derivation permits the identification of an edge feature in the colour or density profiles (vertical dashed line). However, as we are using fixed elliptical annuli , our profiles should be considered as an average representation of the data.}
    \label{fig:asymmetry_gals}
\end{figure*}

As we discuss in Sect. \ref{sect:discussion}, the LTG sample from \citet{2019raj} consists of a few asymmetric galaxies. We did not develop any special ad hoc methods to deal with these galaxies and used the same profile and edge derivation method as described in Sect. \ref{sect:methods}. We show the profiles of four of the extreme cases in Fig. \ref{fig:asymmetry_gals}. \par 
All these galaxies are within the virial radius of the main cluster. From the upper to lower panels, FCC113 is a galaxy with a displaced central region, FCC285 is a jellyfish-like galaxy, FCC312 has large tidal features in its outer region and FCC121 is more truncated on one side of its disk. In all cases, our default radial profile derivation permits the identification of an edge feature in the colour or density profiles (vertical dashed line). However, we remark that it would be interesting in future work to explore how changing the axis ratio in the profile derivation step would impact the identification of the edge using model galaxies. We did not work on this idea for our sample as we did not want to report edges which could be artificially caused by a change in position angle or axis ratio (see Sect. \ref{sect:methods}). But this does not mean that the elliptical parameters could change or be different near the location of the edge. Therefore, for these galaxies our profiles should be considered as an average measurement. \par 
We also point out that in the case of FCC312 which has large tidal features in its outer part, the feature we identify as the edge is capable of separating the main body of the galaxy (mainly in situ) and its tidal feature (predominantly ex situ). This idea was previously shown in \citet{2020tck} (see their Fig. 3) and used to define the stellar halo \citep[see][]{2021trujillo}. Future work analysing the stellar halo component of massive galaxies may consider using the edge as a physically motivated approach to define these structures.

%As we analyse in Sect. \ref{sect:discussion}, only 2\% of our FDS sample are considered highly disturbed according to \citet{2022venhola}. 

\section{Using the effective radii from \cite{2021su}}
\label{app:reff_difference}

\begin{figure}[h!]
    \centering
    \includegraphics[width=0.5\textwidth]{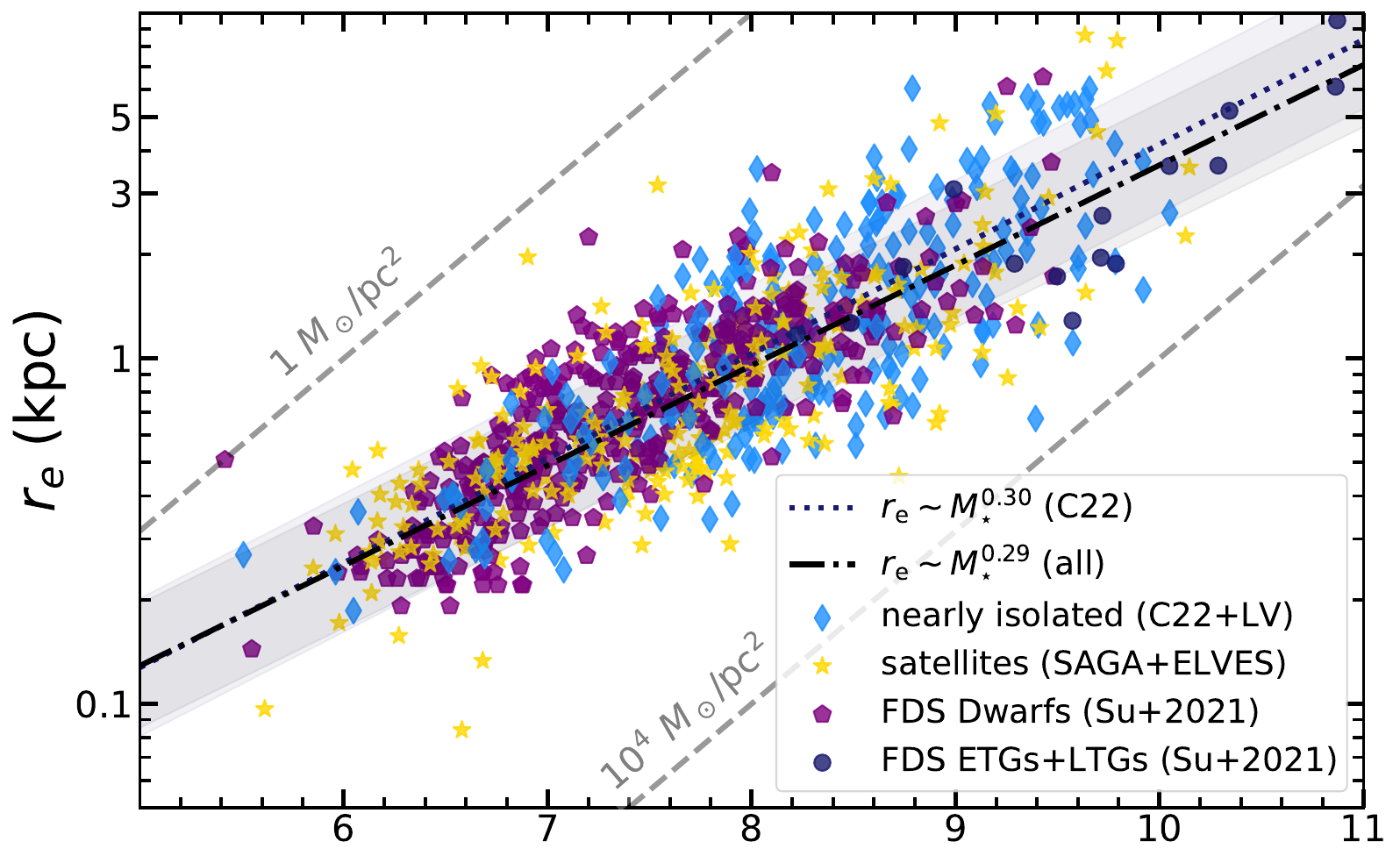}
    \includegraphics[width=0.5\textwidth]{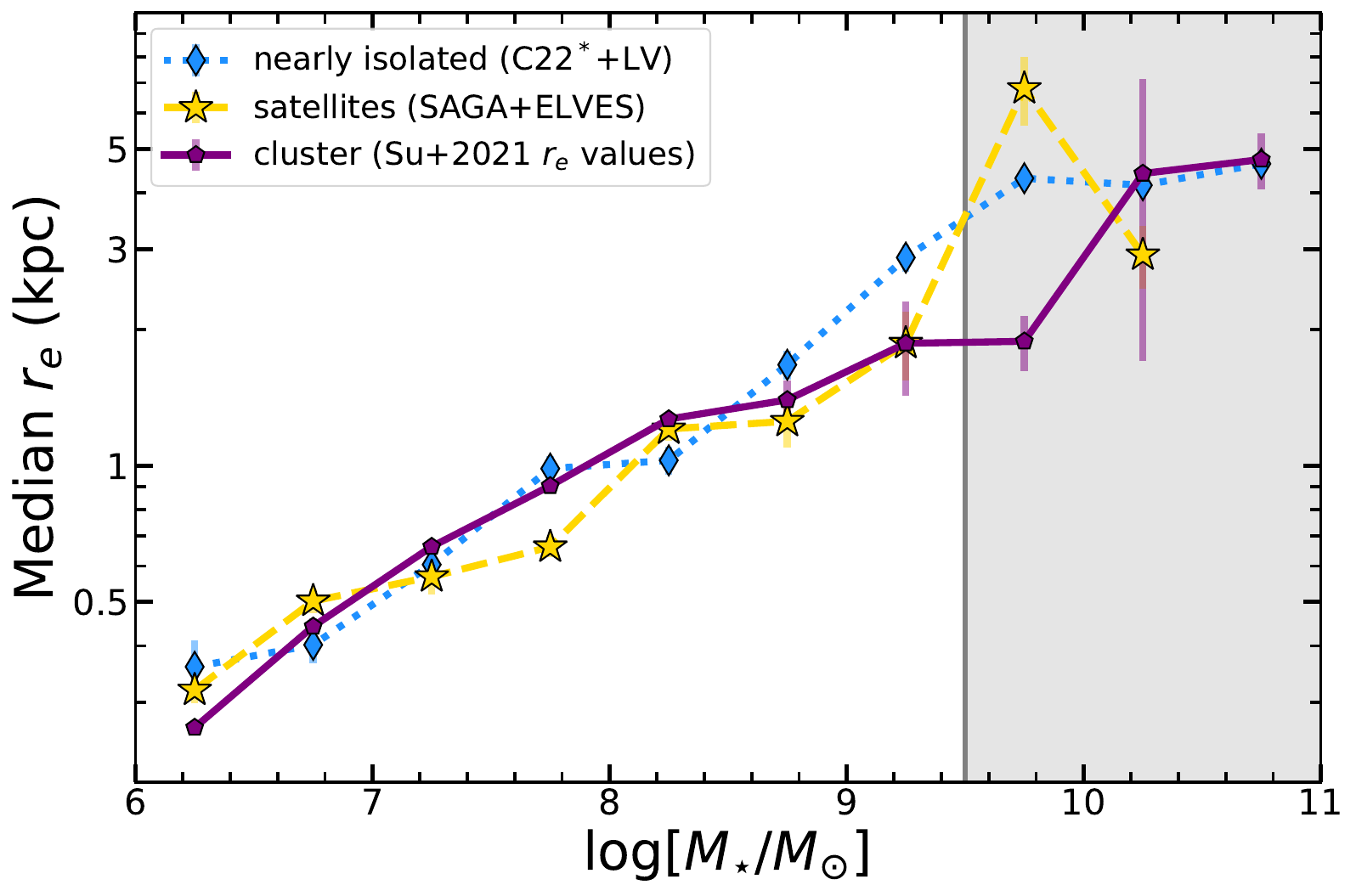}
    \includegraphics[width=0.5\textwidth]{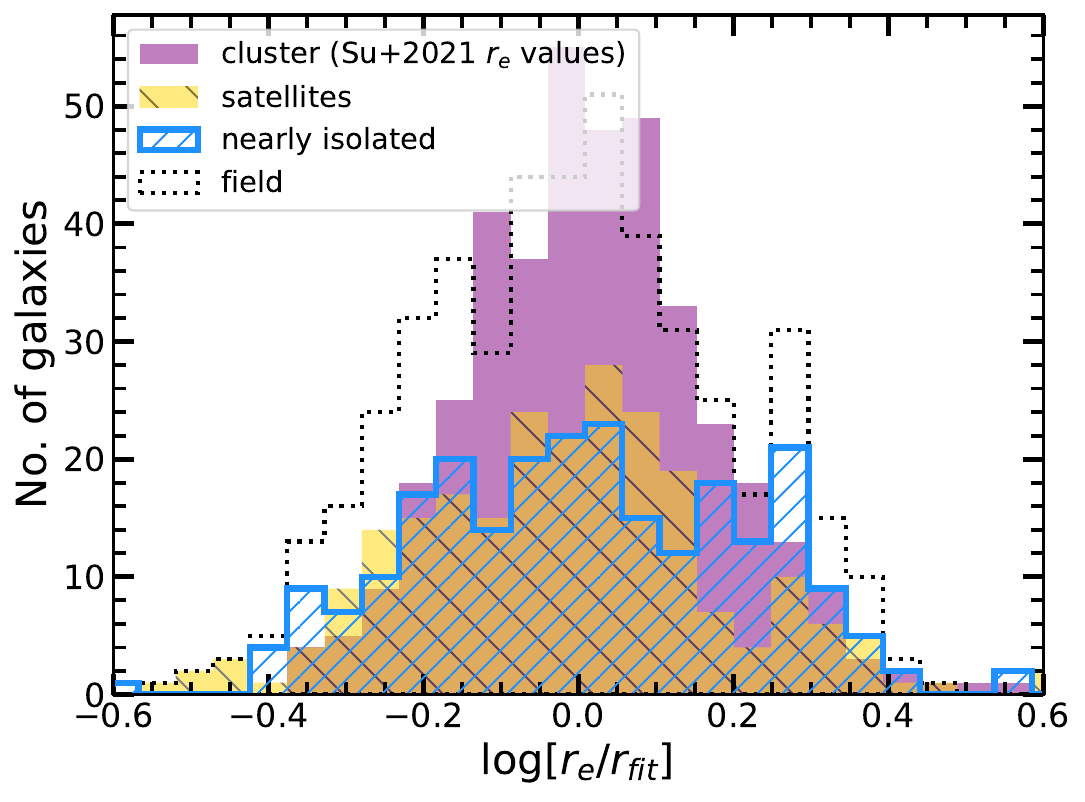}
    \caption{Reproducing the scaling relations and histograms as in Fig. \ref{fig:effective-radius-mass-plane}, \ref{fig:re-histogram} and \ref{fig:comparison-wre-medians} using the modelled  effective radii values from \citet{2021su} for Fornax galaxies. Our main conclusion that the effective radii of Fornax galaxies are not significantly different from those found in the field for low stellar mass galaxies $< 10^{9}\,M_{\odot}$ remains unchanged.}
    \label{fig:reff-su21}
\end{figure}

To confirm that the main conclusions of our paper regarding the use of the effective radius are independent of the method we used to compute the radii, we plot the effective radii--stellar mass scaling relation using the \cite{2021su} modelled values for Fornax galaxies in the upper panel of Fig. \ref{fig:reff-su21}. The middle and lower panels are as in those shown in Fig. \ref{fig:re-histogram} and \ref{fig:comparison-wre-medians} and only show slight differences. \par 
Compared to the upper panel of \ref{fig:comparison-wre-medians}, the median radii value in the lowest stellar mass bin is slightly lower and higher in the highest mass bins, although the latter data point suffers from poor statistics as mentioned earlier. The best fit slope of the relation using the \cite{2021su} values is slightly higher (0.29$\pm$0.02) but well within the uncertainties of the fit using our measurements (see Table \ref{table:bestfit}). Despite these slight differences, our main conclusion that the effective radii of Fornax galaxies are not significantly different from those found in the field for low stellar mass galaxies $< 10^{9}\,M_{\odot}$ remains unchanged.

\section{Alternative metallicity values in the size--stellar mass relation}
\label{app:metallicity}

\begin{figure}[h!]
    \centering
    \includegraphics[width=0.48\textwidth]{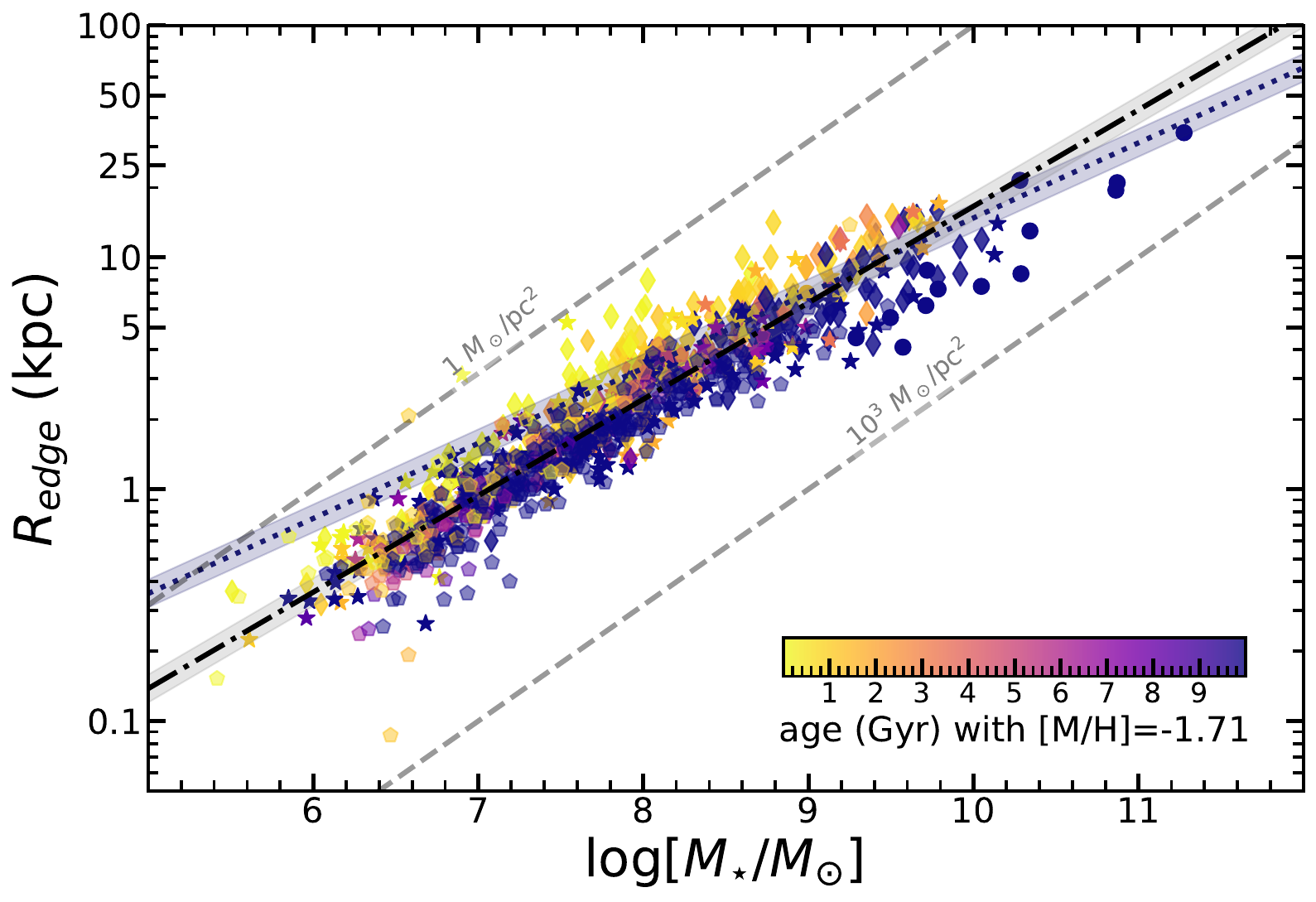}
    \includegraphics[width=0.48\textwidth]{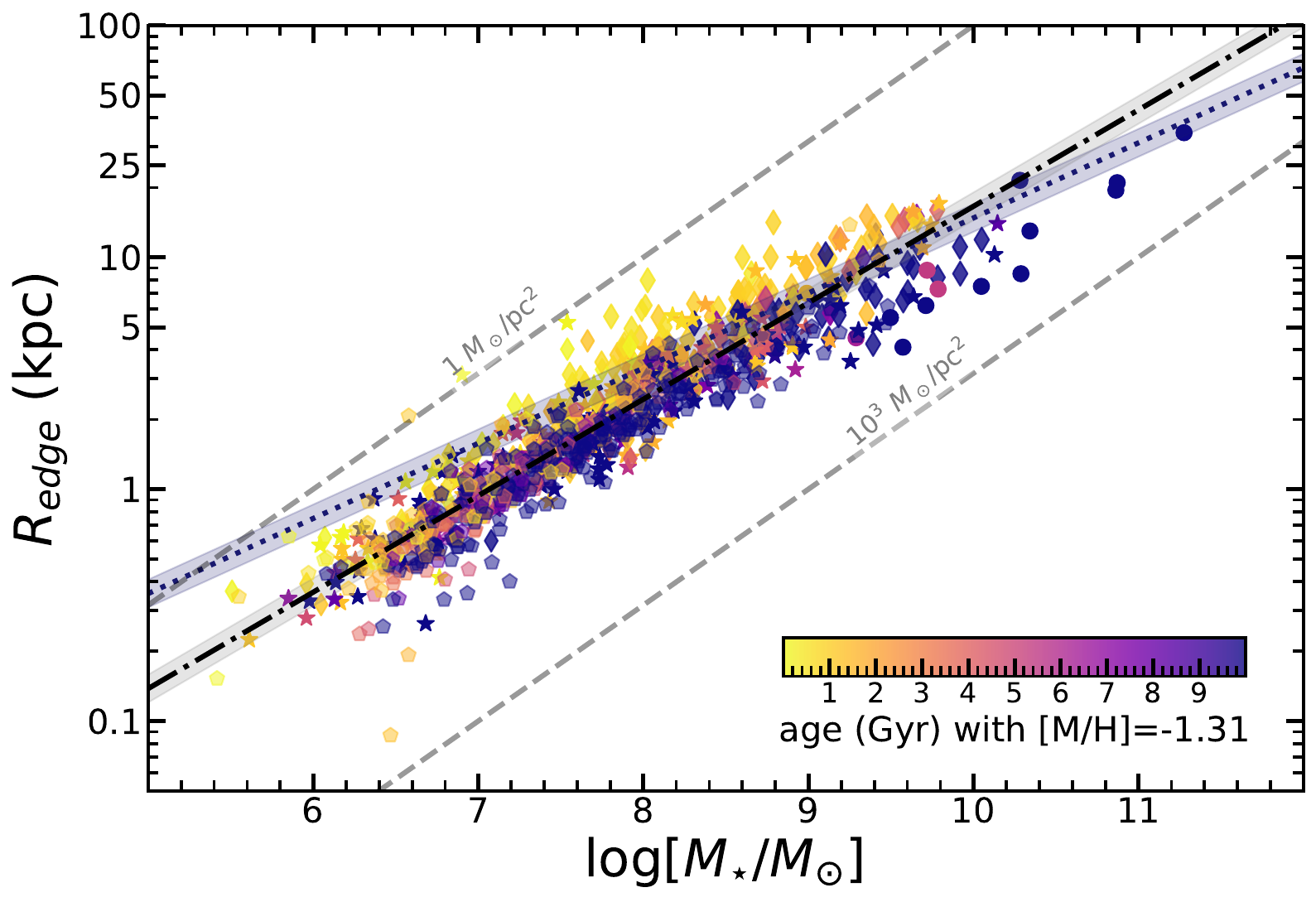}
    \includegraphics[width=0.48\textwidth]{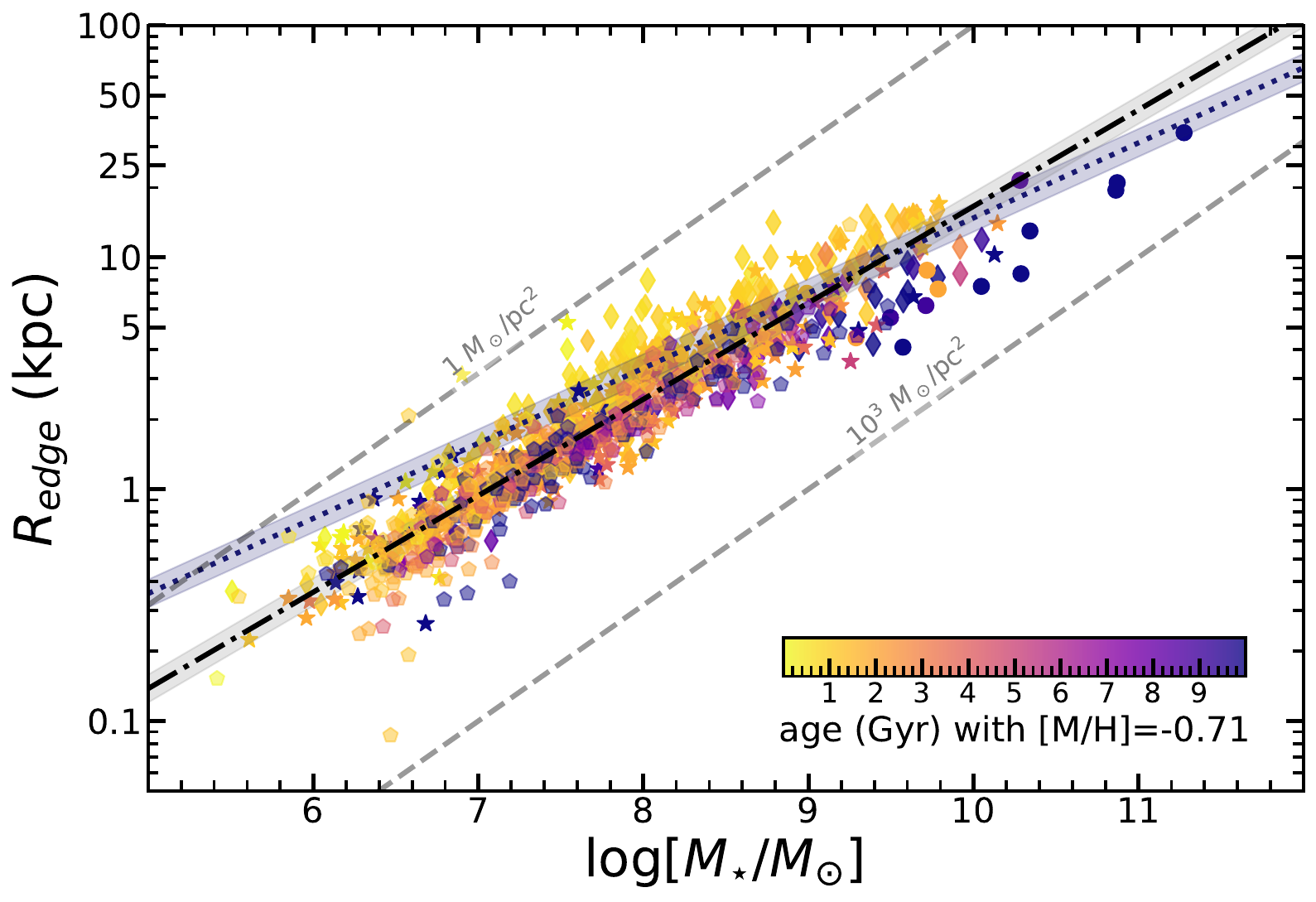}
    
    \caption{Size--stellar mass relations as in Fig. \ref{fig:size-mass-relations} but colour coded according to their age as estimated using the $g-r$ colour at the fixed metallicity value stated in the colour bar. While the differences between the relations using -1.71 and -1.31 are minor, there are larger differences in individual points with the -0.71 relation. However, given that we are interested in relative age distributions, the bifurcation we report in this work remains unchanged when considering either of those relations separately (see text for discussion).}
    \label{fig:diff-metallicity-vals}
\end{figure}

In Fig. \ref{fig:size-mass-relations}, we use a fixed low metallicity value [M/H] = -0.71 taken from the E-MILES models to estimate the ages of galaxies, accounting for the stellar populations in their outskirts. As an alternative estimation method, we can consider the \citet{2013kirby} average stellar mass-metallicity relation for local dwarf galaxies where [M/H] $\sim M_{\star}^{0.3}$ if we assume solar abundance ratios. The scatter in the \citet{2013kirby} relation is $\sim$ 0.17\,dex. While the \citet{2013kirby} relation strictly holds for galaxies $<10^9\,M_{\odot}$, their Fig. 9 shows that we can extend the relation to galaxies with $M_{\star} < 10^{10}\,M{\odot}$ as the average relation for these higher mass galaxies shown in previous work is still within the scatter of the \citet{2013kirby} extrapolation  \citep[see a more recent study by][]{2023dominguez}. \par 
If we assume the \citet{2013kirby} relation, which is applicable
for $10^6\,M_{\odot}< M_{\star} < 10^{10}M_{\odot}$, then the relevant stellar metalicities are
%The relevant metallicity values from E-MILES if we strictly assume the \citet{2013kirby} relation and which are applicable for stellar masses $\sim 10^6\,M_{\odot}$, $10^{7.5}\,M_{\odot}$ and $10^{9.5}\,M_{\odot}$ are then
-1.71, -1.31 and -0.71. %Fig. \ref{fig:size-mass-relations} already shows the relation using -0.71. 
We show the relations using these values below in Fig. \ref{fig:diff-metallicity-vals}. While the differences between the relations using -1.71 and -1.31 are minor, there are larger differences in individual points with the -0.71 relation we presented previously in Fig. \ref{fig:size-mass-relations}. The ratio between the age estimated using either -1.71 or -1.31 and -0.71 deviate by more than a factor of 1.5 for half our sample.
%This behaviour is not unexpected as the \citet{2013kirby} relation scales linearly with stellar mass. 
But given that we are interested in relative age distributions, the bifurcation we report in this work remains unchanged when considering any of these relations separately. \par 
Further examination using the \citet{2013kirby} relation suggests that the average metallicity value at our median stellar mass $\sim 10^8\,M_{\odot}$ is closer to -1.00 and could likely be more representative of our sample. Unfortunately, as the E-MILES predictions for SDSS filters are not already available at this metallicity value, we do not currently have the models to present them here. Therefore, we adopted the same value as in \citet{2022chamba}, i.e. -0.71. However, we point out that using the alternative metallicity value close to -1.00, i.e. -1.31, would only strengthen the age and colour bifurcation we report in this work. We conclude that using either value would thus not change our results significantly. \par

% \section{RGB panels}

% \begin{figure*}[h!]
%     \centering
%     %\includegraphics[width=0.49\textwidth]{figs/fds_dwarfs_examples_2.png}
%     % \includegraphics[width=0.9\textwidth]{figs/fds_examples_faint.png}
%     \includegraphics[width=0.95\textwidth]{figs/examples_fornax_atlas_orange.pdf}
%     \caption{Visualising the edges of galaxies in the Fornax Cluster. Each panel is the $gri$-colour composite image using FDS images \citep{2018venhola}, overlaid on a grey scale background to highlight the low surface brightness boundaries of these galaxies. White in this scale indicates non-detections or background pixels in the data. FDS or FCC names for each galaxy from \citet{2021su}, the estimated stellar mass in units of log$M_{\star}/M_{\odot}$ (left) and edge locations in kpc (right, dotted ellipses) are indicated in the corners of each panel. Galaxies are ordered by increasing stellar mass from the upper to lower panels. }
%     \label{fig:fds_examples}
% \end{figure*}

% \begin{figure*}[h!]
%     \centering
%     %\includegraphics[width=0.49\textwidth]{figs/fds_dwarfs_examples_2.png}
%     % \includegraphics[width=0.9\textwidth]{figs/fds_examples_faint.png}
%     \includegraphics[width=0.95\textwidth]{figs/examples_elves_atlas_yellow-v1.pdf}
%     \caption{Similar to Fig. \ref{fig:fds_examples} but for satellite galaxies from ELVES \cite{2021carlsten} within DECaLs \citep{2019dey}.}
%     \label{fig:elves_examples}
% \end{figure*}

\section{\HI{} properties and scaling relations}
\label{app:Hi-relations}

\begin{figure}[h!]
    \centering
    \includegraphics[width=0.5\textwidth]{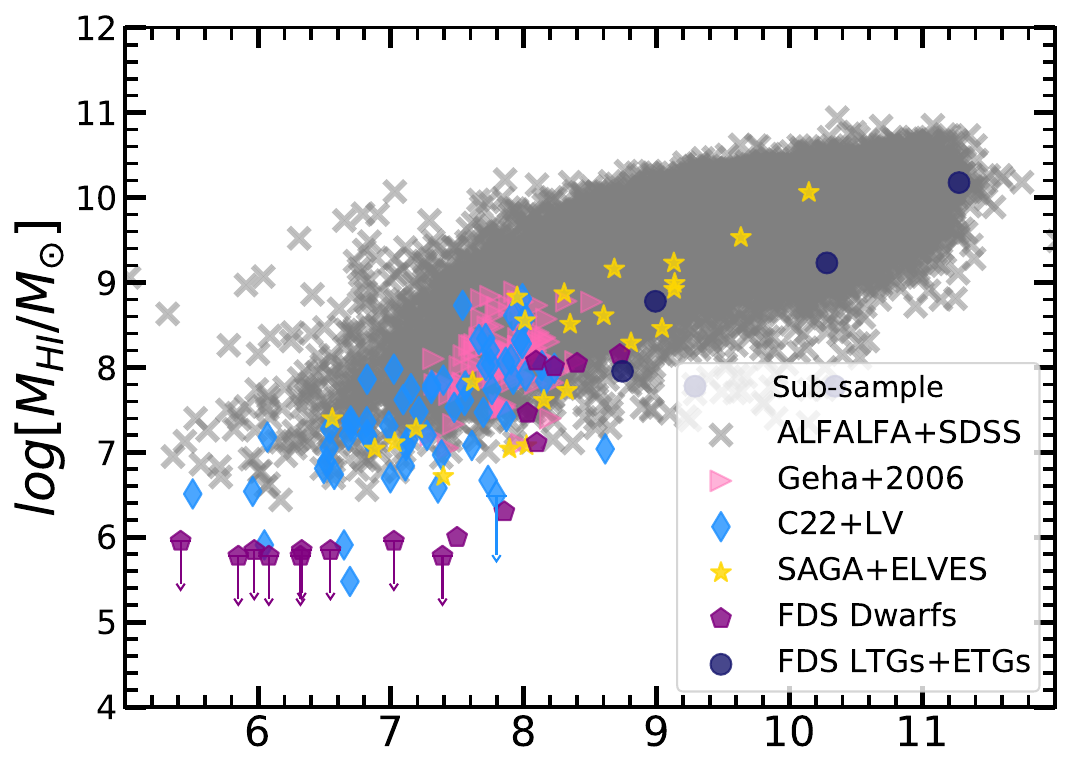}
    \includegraphics[width=0.5\textwidth]{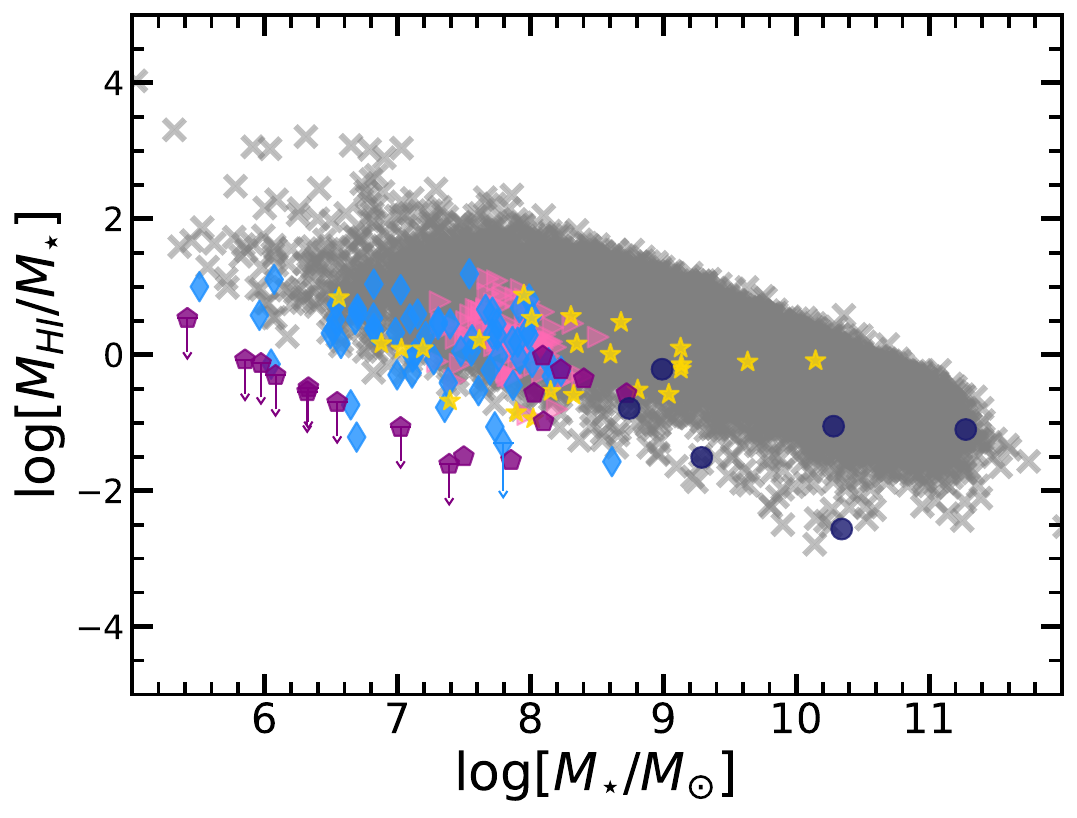}
    \caption{\HI{}-stellar mass relation (upper) and the \HI{}-stellar mass ratio as a function of stellar mass (lower) for a sub-sample of Fornax, satellites and field galaxies in our work as labelled. Arecibo Observatory and Green Bank Telescope targeted observations by \cite{2006geha} of optically selected dwarfs and ALFALFA detections (both for SDSS galaxies) from \cite{2020durbala} are also included for reference. The offsets between the datapoints in our sub-samples, the \cite{2006geha} work and ALFALFA is expected when galaxies are optically selected (see text for details). Arrows show \HI{} upper limits where reported in the respective catalogues.}
    \label{fig:maddox-paper}
\end{figure}

In this section, we show that our result from Fig. \ref{fig:HI-stellar-ratios} where galaxies with lower \HI{} fractions possess higher edge stellar densities, is likely not a selection effect. Complementing Fig. \ref{fig:HI-stellar-ratios} is Fig. \ref{fig:maddox-paper} where we plot the \HI{}-stellar mass relation in the upper panel \citep[as in][]{2015maddox} and the \HI{}-stellar mass ratio as a function of stellar mass \citep{2012papastergis} in the lower panel for a sub-sample of galaxies in our work as labelled. The ALFALFA detections for SDSS galaxies from \cite{2020durbala} as well as Arecibo Observatory and Green Bank Telescope targeted observations of SDSS dwarfs by \cite{2006geha} are also included for reference.  \par 

In the \HI{}-stellar mass plane, our low mass galaxies populate the lower $M_{\star} < 10^9\,M_{\odot}$ end of the ALFALFA+SDSS sample. As discussed in \cite{2015maddox}, this regime is generally populated by low mass, low metallicity, irregular galaxies that are \HI{}-rich. And the few that lie above $M_{\star} > 10^9\,M_{\odot}$ are in the regime populated by extended \HI{} disks and massive galaxies where star formation is efficient. The bending or break of the relation at a limiting \HI{} fraction is currently thought to be set by an upper limit in the spin parameter of the halo \citep{2015maddox}. While we do not attempt to estimate the spin of the galaxies in our sample, the fact that the most \HI{} poor galaxies in our work (Fornax) lie even below this relation in both regimes is expected because ALFALFA consistently detects the brighter \HI{}-rich galaxies given the shallower depth of the data compared to MeerKAT observations for Fornax. \par

The systematic offset between the ALFALFA and all our other sub-sample datapoints is also seen in the lower panel. As discussed and also shown by \citet[][in their Fig. 19]{2012papastergis}, because ALFALFA is a blind \HI{} survey where the integrated time of observations per pointing is fixed, the distribution of ALFALFA detections in the \HI{}-stellar mass ratio vs. stellar mass plane is expected to lie above of those from targeted \HI{} observations of optically selected galaxies. We reproduce this systematic offset as our sub-samples were also optically selected, although with very different criteria. \par 
Our result from Fig. \ref{fig:HI-stellar-ratios} that lower \HI{} fractions could potentially lead to higher edge stellar surface densities can thus not be one due to a selection effect. It would certainly be worthwhile in future work to populate Fig. \ref{fig:HI-stellar-ratios} with \HI{}-rich ETGs \citep[e.g.][]{2012serra} to confirm this result.

%\section{\cite{2009penarrubia} predictions of ``break'' radii from tidal forces}
%\label{app:tidal-break-radii}

\section{The use of scale lengths, fixed surface brightness or isodensity radii measures as galaxy size indicators}
\label{dis:othersizes}
Core to this paper is an attempt to overcome the limitations of past size indicators and locate a common ``edge'' feature in galaxies of a large range in stellar mass and morphoplogy in the Fornax Cluster and other environments, connecting size to the outermost limits of star formation in a galaxy \citep[see][]{2020tck, 2020chamba, 2022chamba}. As a reference for the reader, the median effective radii derived here for Fornax dwarfs are at 0.6$\pm$0.2\,$R_{\rm edge}$. The scale lengths derived by \cite{2021su} via decomposition modelling of the disk (see their equations 5 and 6) of the same sample are also located close to this value. Therefore, it is already clear that the edge radii found here are inherently different from these past measures. \par 
Isophotal radius measures have also been frequently used in the literature. For example, in the \cite{2001koopmann} result mentioned earlier on the truncation of H$_{\alpha}$ radial profiles in Virgo galaxies, it is interesting to point out that these authors normalised their sample using an isophotal radius in the $R$-band at 24\,mag/arcsec$^2$ ($r_{24}$). They also explored the use of disk scale lengths derived by first creating a $r^{1/4}$ bulge + exponential disk decomposition model and then locating the radius where the flux with respect to the centre has fallen by 1/$e$. However, they caution the reader about the difficultly in fitting such models to galaxies: \par 
``\textit{Given the systematic effects of differing surface brightnesses on isophotal radii} [referring to the intrinsic variations between galaxies and its general sensitivity to background subtraction] \textit{and the difficulties of model fitting and unknown extinction values on disk scale length, it is not obvious which radial normalizer is a better indicator of galaxy/disk size.}'' -- \cite{2001koopmann} \par 
In other words, both fixed isophotal radii and scale lengths have their own limitations when considered as galaxy size indicators. While they find that their main conclusions are independent on which radii measure was used, they justify the selection of $r_{24}$ ``because more galaxies could be included in the analysis'' considering ``the more uncertain nature of disk scale length fitting'', i.e. a pragmatic choice given the goals of their work. \par 
We refer the interested reader to \cite{2020tck} for an in-depth comparison between different size definitions, including the half-mass and \cite{1958holmberg} isophotal radii. Here we report that the median distance of the isophote at 24\,mag/arcsec$^2$ in the $r$-band (close to $r_{24}$) is 0.4$\pm0.2\,R_{\rm edge}$ and at 26\,mag/arcsec$^2$ in the $g$-band is 0.7$\pm0.2\,R_{\rm edge}$ for the Fornax dwarfs studied. In future work, it would be interesting to study the H$_{\alpha}$ radial profiles of Fornax galaxies for comparison with the optical edges reported here and the analysis performed in \cite{2001koopmann}. \par 

\cite{2023watkins} has also recently used the fixed iso-density contour at 3.63 $M_{\odot}$/pc$^2$ surface stellar density as a radius measure for FDS dwarfs. These authors find that the radius - mass relation significantly curves/drops towards smaller radii at lower masses $M_{\star} < 10^7\,M_{\odot}$ (see their Fig. 2). The outliers in this relation are those dwarfs with very low central densities. The authors argue that the outliers are low surface brightness galaxies which are more susceptible to tidal forces as they are found in the denser regions of the cluster. However, we point out that the visual morphologies of these outliers have not been reported or analysed in that work. \par 
In contrast, we do not find the same drop in the edge radii-mass relation but a constant slope of 0.4 (Fig. \ref{fig:size-mass-relations}). As addressed throughout \cite{2023watkins}, the density value of 3.63$M_{\odot}$/pc$^2$ reaches the central regions of several dwarfs in their sample, corresponding to smaller or zero radii (see their Fig. 2). This fact implies that the contour is no longer representative of the outskirts of these galaxies or any structures at larger radii. For the sample of FDS dwarfs studied here, the 3.63$M_{\odot}$/pc$^2$ iso-density radius when non-zero occurs at a median distance of 0.5$\pm$0.2\,$R_{\rm edge}$. This value is even slightly less than that achieved with the effective radii or scale lengths as discussed earlier, reflecting that the measure is more representative of the central or inner regions of these galaxies.\par 
 
\emph{Similar to radius measurements dependent on light concentration, any fixed threshold will thus not reflect the same regions of galaxies across stellar mass as the variation of their inner structure is broad. This fact makes it even more difficult to decipher the environmental influence on scaling relations based on such measurements.} However, if the reader is specifically interested in selecting the lowest density or morphologically diffuse dwarfs, such relations could perhaps present to be a useful guide. \par

\end{document}